# Distinguishing between Communicating Transactions

Vasileios Koutavas[a,*], Maciej Gazda[a], Matthew Hennessy[a]

[a]*School of Computer Science and Statistics, Trinity College Dublin, Dublin, Ireland*

**Abstract**

Communicating transactions is a form of distributed, non-isolated transactions which provides a simple construct for building concurrent systems. In this paper we develop a logical framework to express properties of the observable behaviour of such systems. This comprises three nominal modal logics which share standard communication modalities but have distinct past and future modalities involving transactional commits. All three logics have the same distinguishing power over systems because their associated weak bisimulations coincide with contextual equivalence. Furthermore, they are equally expressive because there are semantics-preserving translations between their formulae. Using the logics we can clearly exhibit subtle example inequivalences. This work presents the first property logics for non-isolated transactions.

## 1. Introduction

Transactional constructs without the isolation principle have been proposed as useful building blocks of distributed systems (e.g., [10, 11, 17, 23, 3, 6]). *Communicating transactions* is such a construct, equipped with a rich theory providing techniques for proving behavioural equivalence of transactional systems [8, 9, 16]. To develop useful verification tools, however, it is also essential to have techniques for exhibiting the *in*-equivalence of systems, rather than relying on the absence of equivalence proofs.

Numerous existing verification tools accept two formal descriptions of computing systems and determine whether or not they are behaviourally equivalent (e.g., [13, 2, 5]), and, crucially, provide coherent explanations as to why two descriptions are behaviourally distinguishable. Perhaps the most widely cited

☆This work was supported by the Science Foundation Ireland grant 13/RC/2094 and co-funded under the European Regional Development Fund through the Southern & Eastern Regional Operational Programme to Lero – the Irish Software Research Centre.

*Corresponding author.

*Email addresses:* `Vasileios.Koutavas@scss.tcd.ie` (Vasileios Koutavas), `gazdam@scss.tcd.ie` (Maciej Gazda), `Matthew.Hennessy@scss.tcd.ie` (Matthew Hennessy)

example is the relationship between the property language HML and the behavioural equivalence called *bisimulation equivalence* for processes written in the language *CCS*, [18]. Two processes are not equivalent, $P \not\approx Q$, if and only if there is an HML property $\phi$ which $P$ enjoys and $Q$ does not, [18, 14]. Thus $\phi$ can be considered an explanation as to why $P$ and $Q$ have different behaviour. Indeed an algorithm has been proposed by Cleveland [4] and implemented in the *concurrency workbench* [5] which, when presented with descriptions of two finite state processes, either calculates a bisimulation, a formal justification for their behavioural equivalence, or returns a distinguishing HML formula.

For example consider the CCS process $P_0 = a.(b.\mathbb{0} + c.\mathbb{0})$, which performs an $a$-action, followed by offering a choice between a $b$- and a $c$-action, after which it terminates. According to the definition of bisimulation equivalence, $P_0 \not\approx Q_0$, where $Q_0$ denotes the slightly different process $a.(b.\mathbb{0} + c.\mathbb{0}) + a.b.\mathbb{0}$. Intuitively $p_1$ satisfies the property: *whenever it performs an $a$-action it must be subsequently able to perform a $c$-action*; whereas $p_2$ does not. In HML this property is captured by using modality operators, $[a]$ for necessity and $\langle a \rangle$ for possibility. Thus the property distinguishing $P_0$ from $Q_0$ is written formally as $[a]\,\langle c \rangle\, \texttt{true}$.

The purpose of this paper is to develop similar property logics which characterise contextual equivalence for communicating transactions. As a formalism we use the abstract language *TCCS*$^m$, for which a natural contextual equivalence has been defined and characterised using a form of weak bisimulation over *configurations*, run-time entities recording the current state of the transactional system, together with information on historical interactions with its environment [16].

The transactional language *TCCS*$^m$ is obtained by adding to *CCS* constructs for describing transactions. For example $P_1 = [\![a.b.\,\mathsf{co} \rhd_l d.\mathbb{0}]\!]$ describes a transaction named $l$ which can either perform the sequence of actions $a$, $b$ in its entirety, or else fails and performs the action $d$. The transaction $Q_1 = [\![a.(b.\,\mathsf{co} + c.\mathbb{0}) \rhd_l d.\mathbb{0}]\!]$ is a slight variation in which there is an apparent possibility of performing a $c$-action after $a$. However if this $c$-action is performed then the transaction can never commit (i.e., perform a $\mathsf{co}$-action) and therefore the presence of this potential $c$-action is superfluous. According to *TCCS*$^m$ reduction barbed equivalence theory [16] these two transactions are behaviourally equivalent. Consequently an extension of HML we propose should not be able to distinguish them, despite the fact that $Q_1$ can apparently perform a $c$-action or at least attempt to do so.

The notion of weak bisimulation developed in [16] for transactions contains constraints on the actions which transactions may perform—the standard *transfer property*. This in effect compares the future behaviour of transactions. But the definition of bisimulation also contains constraints on past behaviour, as encoded in configurations. For example, $P_2 = [\![a.\,\mathsf{co} \rhd_{k_1} \mathbb{0}]\!] \mid [\![b.\,\mathsf{co} \rhd_{k_2} \mathbb{0}]\!]$ can perform actions $a$ and $b$ and reach a state where these actions can be committed independently. On the other hand $Q_2 = \nu p.\,[\![a.p.\,\mathsf{co} + a.\,\mathsf{co} \rhd_{k_1} \mathbb{0}]\!] \mid [\![b.\overline{p}.\,\mathsf{co} + b.\,\mathsf{co} \rhd_{k_2} \mathbb{0}]\!]$ can perform the same actions and then, through the internal communication on $p$, can only commit the past actions $a$ and $b$ si-multaneously. Thus two configurations reachable starting from $P_2$ and $Q_2$ are $\mathcal{C}_1 = \langle l_1(a), l_2(b) \bullet [\![\mathsf{co} \rhd_{l_1} \mathbb{0}]\!] \mid [\![\mathsf{co} \rhd_{l_2} \mathbb{0}]\!]\rangle$ and $\mathcal{C}_2 = \langle l(a), l(b) \bullet [\![\mathsf{co} \rhd_{l} \mathbb{0}]\!] \mid [\![\mathsf{co} \rhd_{l} \mathbb{0}]\!]\rangle$, respectively. In the latter configuration, the two transactions have been *merged* and thus obtained the same name $l$. After a single commit, $C_1$ becomes $\langle a, l_2(b) \bullet \mathbb{0} \mid [\![\mathsf{co} \rhd_{l_2} \mathbb{0}]\!]\rangle$ where only past action $a$ is committed. Because there is no matching future configuration of $\mathcal{C}_2$, weak bisimulation from [16] distinguishes the two processes. Thus, to distinguish $P_2$ from $Q_2$, one would expect a property logic for transactions containing, in addition to standard future-oriented modal operators discussed above, operators for examining past behaviour.

In this paper we provide two such property logics, with different past operators. We also provide a property logic with *no* past operators; instead a richer collection of future-oriented operators are used. In the example of $P_2$ and $Q_2$ above, the first logic, $\mathcal{L}_{\mathsf{Hasco}}$, using only an additional "has committed" predicate on past actions ($\mathtt{Hasco}(k)$) can express the inequivalence as the following rather involved formula satisfied by $Q_2$:

$$\langle x(a)\rangle \, \langle y(b)\rangle \, \big(\neg\mathtt{Hasco}(x) \;\wedge\; \langle\tau\rangle\, \mathtt{Hasco}(x) \;\wedge\; ([\tau](\mathtt{Hasco}(x) \leftrightarrow \mathtt{Hasco}(y)))\big)$$

This states that the process can perform an $a$-action followed by a $b$-action and reach a state where: (1) the $a$-action has not been committed yet; (2) the $a$-action can be committed after some internal ($\tau$) transitions; and (3) in any future configurations reachable by $\tau$-transitions, the past $a$- and $b$-actions are either both committed or both aborted. Note that $\leftrightarrow$ is double implication, and $x$ and $y$ are bound variables representing the transactions performing $a$ and $b$, respectively.

The second logic, $\mathcal{L}_{\mathsf{Eq}}$, distinguishes $P_2$ from $Q_2$ by the significantly simpler formula $\langle x(a)\rangle \, \langle y(b)\rangle (x =_{\mathsf{co}} y)$ which expresses the possibility of performing actions $a$ and $b$, reaching a state where both have been committed by a single transaction, possibly as a result of transactional merging. The last logic, $\mathcal{L}_{\mathsf{Canco}}$, distinguishes the same processes with the formula $\langle x(a)\rangle \, \langle y(b)\rangle \, \langle \mathsf{co}(\{x, y\})\rangle \, \mathtt{true}$, expressing the possibility of performing $a$, then $b$, and then committing both actions simultaneously.

The main results of the paper include:

- Three property logics for $TCCS^m$, and their natural associated bisimulation relations. The first logic encapsulates the intuitions on observable past actions from [16]; the second encodes a more powerful predicate on past actions which we use to write more succinct formulas; the third logic uses only future action modalities giving rise to standard bisimulation equivalence. All logics include nominal [21, 12] versions of the standard HML modal operators, and are based on a novel labelled transition system for $TCCS^m$.

- Proofs that all logics have the same distinguishing power over $TCCS^m$ terms. In effect each of their associated bisimulations precisely coincides with the natural contextual equivalence.

- Proofs that each of the logics are *equally expressive*. We provide translations between the formulas of the three logics and show that any property definable in one logic is also expressible in each of the other two.

The remainder of the paper is organised as follows. First in Section 2 we recall the theory of *TCCS*$^m$ from [16], in particular recalling the definition of bisimulation over *configurations* which characterises the natural contextual equivalence for transactions. Then in Section 3 we first explain the expressive deficiencies in this notion of configuration; that is the limited access it gives to past behaviour. We then propose a more expressive notion of *extended configuration*, together with a new notion of bisimulation, `Hasco`-bisimulation, over these extended configurations. This is similar in style to that in [16]; a transfer condition between possible actions puts requirements on the future behaviour of processes, while predicates on the configurations enforce requirements on past behaviour. Our new notion of bisimulation equivalence over extended configurations, $\approx_{\mathsf{Hasco}}$, generates the same behavioural equivalence over transactions as that defined in [16].

This is followed in Section 4 by an exposition of the natural property language $\mathcal{L}_{\mathsf{Hasco}}$ which characterises the new bisimulation equivalence $\approx_{\mathsf{Hasco}}$ over transactions. It has nominal [21, 12] versions of the standard modal operators from HML for examining future behaviour, but also contains predicates over extended configurations for examining past behaviour. This section also contains an exposition of the alternative property logics $\mathcal{L}_{\mathsf{Eq}}$ and $\mathcal{L}_{\mathsf{Canco}}$ already alluded to. The latter only contains operators for examining the future behaviour, but the absence of past operators is compensated for by a richer collection of future-oriented modal operators.

The final two sections of the paper are devoted to comparing the three property logics, $\mathcal{L}_{\mathsf{Hasco}}, \mathcal{L}_{\mathsf{Eq}}$ and $\mathcal{L}_{\mathsf{Canco}}$. In Section 5 we show that all three are equally powerful from the point of view of being able to distinguish between transactions. This is proved by comparing their associated bisimulation relations.

Then in Section 6 we show the stronger result that all three logics are equally expressive; there are translations between the logics which preserve satisfiability. This means that for any property expressible in one logic there is an exact corresponding property in each of the other two. The paper ends with Section 7 containing a brief conclusion, including remarks on future work.

## 2. A review of *TCCS*$^m$

In this section we recall the language *TCCS*$^m$ from [16] and its behaviour theory. In this first sub-section we give the syntax, together with its reduction semantics. This is used in the following sub-section where we recall a version of contextual equivalence, called *reduction barbed equivalence.* Then we recall the bisimulation equivalence for *TCCS*$^m$, which is based on an LTS (labelled transition system) between *configurations.*

**$TCCS^m$ Syntax**

$$P, Q, R \quad ::= \quad \textstyle\sum_{i \in I} \mu_i.P_i,\ I \text{ countable} \;\big|\; P \mid Q \;\big|\; \nu a.P \;\big|\; X \;\big|\; \mathtt{rec}X.P$$
$$\big|\; [\![P \rhd_k Q]\!] \;\big|\; \mathtt{co} \;\big|\; [\![P \blacktriangleright Q]\!]$$

**CCS Transitions**

$$\textsc{Sum}\ \frac{}{\Sigma\mu_i.P_i \xrightarrow{\mu_i}_{\varepsilon} P_i} \qquad \textsc{Sync}\ \frac{P \xrightarrow{a}_{\varepsilon} P' \qquad Q \xrightarrow{\overline{a}}_{\varepsilon} Q'}{P \mid Q \xrightarrow{\tau}_{\varepsilon} P' \mid Q'} \qquad \textsc{Rec}\ \frac{}{\mathtt{rec}X.P \xrightarrow{\tau}_{\varepsilon} P[\mathtt{rec}X.P/X]}$$

**Transactional Transitions**

$$\textsc{TrSum}\ \frac{}{\Sigma\mu_i.P_i \xrightarrow{k(a)}_{\varepsilon\mapsto k} [\![P_j \mid \mathtt{co} \rhd_k \Sigma\mu_i.P_i]\!]}\ \mu_j = a \qquad \textsc{TrMu}\ \frac{P \xrightarrow{\mu}_{\varepsilon} P'}{[\![P \rhd_l Q]\!] \xrightarrow{k(\mu)}_{l\mapsto k} [\![P' \rhd_k Q]\!]}\ k \sharp l$$

$$\textsc{TrSync}\ \frac{P \xrightarrow{k(a)}_{\sigma_1} P' \qquad Q \xrightarrow{k(\overline{a})}_{\sigma_2} Q'}{P \mid Q \xrightarrow{k(\tau)}_{(\widetilde{l}_1,\widetilde{l}_2)\mapsto k} P'\sigma_2 \mid Q'\sigma_1}\ \begin{array}{l}\sigma_1 = \widetilde{l}_1 \mapsto k\\ \sigma_2 = \widetilde{l}_2 \mapsto k\end{array}$$

**Propagation Transitions**

$$\textsc{ParL}\ \frac{P \xrightarrow{\alpha}_{\sigma} P'}{P \mid Q \xrightarrow{\alpha}_{\sigma} P' \mid Q\sigma}\ \mathit{range}(\sigma) \sharp Q \qquad \textsc{Restr}\ \frac{P \xrightarrow{\alpha}_{\sigma} P'}{\nu a.P \xrightarrow{\alpha}_{\sigma} \nu a.P'}\ a \notin \alpha$$

**Transactional Reconfiguration Transitions**

$$\textsc{TrCo}\ \frac{P \equiv \mathtt{co} \mid P'' \qquad P \rightsquigarrow_{\mathtt{co}} P'}{[\![P \rhd_k Q]\!] \xrightarrow{\mathtt{co}\,k} P'} \qquad \textsc{TrNew}\ \frac{}{[\![P \blacktriangleright Q]\!] \xrightarrow{\mathtt{new}\,k} [\![P \rhd_k Q]\!]} \qquad \textsc{TrAb}\ \frac{}{[\![P \rhd_k Q]\!] \xrightarrow{\mathtt{ab}\,k} Q}$$

$$\textsc{TrBCast}\ \frac{P \xrightarrow{\beta} P' \qquad Q \xrightarrow{\beta} Q'}{P \mid Q \xrightarrow{\beta} P' \mid Q'}\ \beta = \mathtt{co}\,k, \mathtt{ab}\,k \qquad \textsc{TrIgnore}\ \frac{P \xrightarrow{\beta} P'}{P \mid Q \xrightarrow{\beta} P' \mid Q}\ \beta \sharp Q \qquad \textsc{TrRestr}\ \frac{P \xrightarrow{\beta} P'}{\nu a.P \xrightarrow{\beta} \nu a.P'}$$

Figure 1: *TCCS*$^m$ syntax and transitions (omitting symmetric rules).

### *2.1. Reduction semantics*

Here we recall the language *TCCS*$^m$, its reduction semantics, and the standard notion of bisimulation equivalence from [16].

We assume an infinite set of action names $\mathsf{Act}$, and an additional special action $\tau$ which represents internal computation; we use $a, b, \ldots$ to range over $\mathsf{Act}$, while $\mu$ ranges over the disjoint union $\mathsf{Act} \uplus \{\tau\}$. We use a standard collection of operators from CCS, including a parallel operator, action restriction, recursion and an infinite version of summation $\sum_{i \in I} \mu_i.P_i$, where $I$ is an arbitrary countable indexing set. When $I$ is a singleton set we have the standard prefixing operator, $\mu.P$, while when it is empty we get the deadlocked or inert process which we abbreviate to $\mathbb{0}$. Moreover we systematically use the standard convention of omitting trailing occurrences of $\mathbb{0}$; for example rendering $a.\mathbb{0}$ as $a$.

Our language $TCCS^m$ is obtained by extending this version of CCS with the form $[\![P \rhd_k Q]\!]$ denoting a transaction running $P$—the *default* part of the transaction—which may abort and become $Q$—the *alternative* part of the transaction. The name $k \in \mathsf{TrN}$ will be used in the operational semantics to identify a transaction. Multiple transactions may share the same name, thus forming a *distributed transaction*. The intention is that such a distributed transaction commits when all its default components simultaneously execute the special command co, another addition to the language.

The syntax of $TCCS^m$ is given in Figure 1 together with rules for deriving judgements of the form

$$P \xrightarrow{\alpha}_{\sigma} Q$$

where $\alpha$ ranges over $\{a, \tau, k(\tau), k(a) \mid k \in \mathsf{TrN}, a \in \mathsf{Act}\}$, and $\sigma$ is a name substitution of the form $\tilde{k} \mapsto l$. Intuitively this means that in performing the action $\alpha$ the transaction names $\tilde{k}$ in $P$ are all renamed to $l$, which is always fresh. In derived transitions, the domain of a $\sigma$ substitution, $\tilde{k}$, can have zero, one or two elements. If it contains two elements this means that performing $\alpha$ results in the *merging* of previously independent transactions. Thus we have distributed transactions, in that we have transactions with the same name executing in parallel.

The rule TrMu allows the default part of a transaction to perform transitions, representing actions. Two of these transitions can be combined in the rule TrSync, thereby allowing communication or synchronisation between two transactions; the effect of the substitutions on the transitions is that the communication transactions now have the same (fresh) name. On the other hand, the use in TrSync of a transition obtained from an application of TrSum allows communication between a process and a transaction.

In Figure 1 we also give rules for deriving judgements of the form

$$P \xrightarrow{\beta} Q$$

where $\beta$ ranges over $\{\mathtt{ab}\, k,\ \mathtt{co}\, k,\ \mathtt{new}\, k\}$, encoding commit/abort transitions and the creation of new named transactions. The rule TrAb allows an individual component of transaction $k$ to abort at any time. But TrBCast together with TrIgnore obliges all components of a distributed $k$ transaction to abort together. Thus *aborts* are *broadcast actions*.

In a similar manner *commits* are *broadcast actions*, with the rule TrCo allowing the execution of a top-level commit inside the default part of a transaction. It uses an auxiliary deterministic relation $P \rightsquigarrow_{\mathtt{co}} P'$ which intuitively means the elimination of all co outside of dormant ▶-transactions, prefixes and recursion. Formally, $P \rightsquigarrow_{\mathtt{co}} Q$ is defined by the rules of Figure 2; note that the rule for sums has a particularly useful instance: $\mathbb{0} \rightsquigarrow_{\mathtt{co}} \mathbb{0}$.

We use $ftn(o)$ to denote the set of transaction names which occur in the syntactic object $o$, and write $o \sharp o'$ when no transaction name in $o$ appears in $o'$.

The presence of infinite sums in the language means that in general $ftn(P)$ may be infinite. Here we restrict attention to terms containing a finite number of

$$\frac{}{\mathtt{co} \rightsquigarrow_{\mathsf{co}} \mathbf{0}} \qquad \frac{}{\sum_{i\in I} \mu_i.P_i \rightsquigarrow_{\mathsf{co}} \sum_{i\in I} \mu_i.P_i} \qquad \frac{}{X \rightsquigarrow_{\mathsf{co}} X} \qquad \frac{}{\mathtt{rec}X.P \rightsquigarrow_{\mathsf{co}} \mathtt{rec}X.P}$$

$$\frac{Q \rightsquigarrow_{\mathsf{co}} Q'}{\nu a.Q \rightsquigarrow_{\mathsf{co}} \nu a.Q'} \qquad \frac{Q_1 \rightsquigarrow_{\mathsf{co}} Q_1' \qquad Q_2 \rightsquigarrow_{\mathsf{co}} Q_2'}{Q_1 \mid Q_2 \rightsquigarrow_{\mathsf{co}} Q_1' \mid Q_2'} \qquad \frac{}{[\![P \blacktriangleright Q]\!] \rightsquigarrow_{\mathsf{co}} [\![P \blacktriangleright Q]\!]}$$

Figure 2: Elimination of commits.

transaction names. Moreover, we only consider terms in which running transactions are top-level, and dormant transactions do not appear in the default parts of running transactions.

**Definition 2.1** (Well-formed terms)**.** Term $P$ is *well-formed* when

1. $P$ is closed; i.e., all occurrences of process variables $X$ are bound by enclosing occurrences of the binder $\mathtt{rec}X.-$
2. $\mathit{ftn}(P)$ is finite;

and for all *sub*terms of $P$ of the form $[\![P_1 \rhd_k P_1']\!]$, $[\![P_2 \blacktriangleright P_2']\!]$, $\sum \mu_i.Q_i$ and $\mathtt{rec}X.R$:

3. $\mathit{ftn}(P_1, P_2, P_1', P_2', Q_i, R) = \emptyset$; i.e., these subterms do not contain name transactions of the form $[\![- \rhd_{_} -]\!]$;
4. $P_1$, $P_2$ and $Q_i$ do not contain dormant transactions of the form $[\![- \blacktriangleright -]\!]$;
5. $P_1$ and $P_2$ are closed. ◊

The first condition of this definition is standard; the second condition ensures that, although the language contains infinite sums, there are always enough fresh names to be used by reductions and process renaming. The third condition in effect disallows transaction nesting, simplifying the complexity of the language. The last two conditions ensure that dormant transactions do not appear in the default part of any running transaction, thus disallowing another form of transaction nesting. We refer to terms satisfying these well-formedness restrictions as *processes.*

### *2.2. Contextual equivalence*

We now recall the contextual equivalence for $\mathit{TCCS}^m$, $\cong_{\mathsf{rbe}}$, from [16]. It follows a standard formulation, originally from [15], which requires

- a reduction relation $\rightarrow$ over processes in $\mathit{TCCS}^m$
- a notion of *barb*, formalising a primitive notion of observation.

For the former, following [16], we let $P \rightarrow Q$ whenever

(1) $P \xrightarrow{\tau}_{\varepsilon} Q$, or

(2) $P \xrightarrow{\beta} Q$, or

(3) $P \xrightarrow{k(\tau)}_{\sigma} Q$.

For the latter we use a version of the language in which the set of actions now takes the form $\mathsf{Act} \uplus \Omega$, where $\Omega$ is a distinct infinite set of special actions ranged over by $\omega$. Intuitively observations will be formulated as contexts which use the special actions from $\Omega$, and observations will be deemed successful when a designated collection of these special actions can be performed at the top-level. Formally we write $Q{\Downarrow}\omega$ whenever $Q \rightarrow^* Q'$, where $Q' \equiv Q'_1 \mid \omega.Q'_2$; here $\equiv$ denotes the standard structural equivalence for CCS, [18], generalised in the obvious manner to handle transactions. We will see an instance of the use of barbs in Example 2.3.

**Definition 2.2** (Reduction barbed equivalence ). ($\cong_{\mathsf{rbe}}$) is the largest relation over processes for which $P \cong_{\mathsf{rbe}} Q$ when:

(1) for every $\omega \in \Omega$, $P{\Downarrow}\omega$ iff $Q{\Downarrow}\omega$,

(2) if $P \rightarrow P'$ then there exists $Q'$ such that $Q \rightarrow^* Q'$ and $P' \cong_{\mathsf{rbe}} Q'$,

(3) if $Q \rightarrow Q'$ then there exists $P'$ such that $P \rightarrow^* P'$ and $P' \cong_{\mathsf{rbe}} Q'$,

(4) $P \mid R \cong_{\mathsf{rbe}} Q \mid R$ for any $R$ with $R \sharp P, Q$. $\Diamond$

We refer the reader to [16] for a discussion of why this provides a natural notion of contextual equivalence for $TCCS^m$. Instead we give two examples.

**Example 2.3.** Consider the processes

$$P_2 = [\![a.\,\mathsf{co} \rhd_{k_1} \mathbb{0}]\!] \mid [\![b.\,\mathsf{co} \rhd_{k_2} \mathbb{0}]\!]$$
$$Q_2 = \nu p.\,[\![a.p.\,\mathsf{co} + a.\,\mathsf{co} \rhd_{k_1} \mathbb{0}]\!] \mid [\![b.\overline{p}.\,\mathsf{co} + b.\,\mathsf{co} \rhd_{k_2} \mathbb{0}]\!]$$

from the introduction. We show that $P_2 \not\cong_{\mathsf{rbe}} Q_2$. This follows by showing that $P_2 \mid O \not\cong_{\mathsf{rbe}} Q_2 \mid O$, where $O$ is the observing process $[\![\overline{a}.(\mathsf{co} \mid \omega_1) \rhd_{l_1} \mathbb{0}]\!] \mid [\![\overline{b}.(\mathsf{co} \mid \omega_2) \rhd_{l_2} \mathbb{0}]\!]$.

In turn this follows by considering the reduction

$$Q_2 \mid O \rightarrow^* [\![\mathsf{co} \rhd_m \mathbb{0}]\!] \mid [\![\mathsf{co} \rhd_m \mathbb{0}]\!] \mid [\![\omega_1 \mid \mathsf{co} \rhd_m \mathbb{0}]\!] \mid [\![\omega_2 \mid \mathsf{co} \rhd_m \mathbb{0}]\!] \;=\; R$$

consisting of three communications on $a$, $b$ and $p$. In $R$ there is a single (distributed) transaction called $m$; thus rule TrBCast from Figure 1 ensures that either all individual components of this single transaction commit together, or none do. This has important consequences for all residuals of $R$: For any $R'$ such that $R \rightarrow^* R'$,

- either $R'\not\Downarrow\omega_1$ and $R'\not\Downarrow\omega_2$, when the transaction has aborted
- or $R'{\Downarrow}\omega_1$ and $R'{\Downarrow}\omega_2$, otherwise.

For this reason there is no corresponding reduction from $P_2 \mid O \to^* S$ such that $S \cong_{\mathsf{rbe}} R$.

There are many candidates for the process $S$. Perhaps the most interesting is

$$P_2 \mid O \to^* [\![\mathsf{co} \rhd_{m_1} \mathbb{0}]\!] \mid [\![\omega_1 \mid \mathsf{co} \rhd_{m_1} \mathbb{0}]\!] \mid [\![\mathsf{co} \rhd_{m_2} \mathbb{0}]\!] \mid [\![\omega_2 \mid \mathsf{co} \rhd_{m_2} \mathbb{0}]\!] \; = \; S_1$$

Here there are two independent transactions $m_1, m_2$ and for this reason one can show that $S_1 \not\cong_{\mathsf{rbe}} R$. This is because $S_1 \to \mathbb{0} \mid \mathbb{0} \mid [\![\mathsf{co} \rhd_{m_2} \mathbb{0}]\!] \mid [\![\omega_2 \mid \mathsf{co} \rhd_{m_2} \mathbb{0}]\!] = S_1'$ with $S_1' \Downarrow \omega_2$ and $S_1' \not\Downarrow \omega_1$, obtained by aborting the first transaction, cannot be matched by a reduction $R \to^* R'$ in such a way that $R' \cong_{\mathsf{rbe}} S_1'$. All other candidates for $S$ can be eliminated in a similar fashion. $\Diamond$

**Example 2.4.** Transactional communications that cannot commit are not observable. Consider $P_1 = [\![a.\,\mathsf{co} \rhd_l \mathbb{0}]\!]$ and $P_2 = [\![a.\,\mathsf{co} + b.\mathbb{0} \rhd_l \mathbb{0}]\!]$. In order to show that these processes are distinguished by reduction barbed congruence we need to provide a context that probes the $b$-action of $P_2$ and only then releases an $\omega$-barb. However, after communication on $b$, such a context would become merged with transaction $l$, which would not be able to commit to release the barb. Indeed it will follow from Example 2.9 and Theorem 2.10 that $P_1 \cong_{\mathsf{rbe}} P_2$. $\Diamond$

*2.3. The LTS and bisimulation equivalence*

Bisimulation equivalence is defined over a Labelled Transition System (LTS) semantics of *configurations* $\langle H \bullet P \rangle$, ranged over by $\mathcal{C}$. The *history* $H$ records the dependencies of previous actions of $P$ on transactions yet to be committed.

**Definition 2.5** (History)**.** A *history* $H$ is a finite partial function from objects $i$ of a countable set $I$ to the set $\{a, \star, k(a), k(\star), \mathsf{ab} \mid a \in \mathsf{Act}\}$. The domain of $H$ is defined to be $\{\, i \in I \mid H(i) \text{ is defined} \,\}$. $\Diamond$

**Example 2.6.** Consider the configuration $\langle H \bullet P \rangle$ where

$$H = (i_1 \mapsto k(a)), (i_2 \mapsto k(b)), (i_3 \mapsto l(c)), (i_4 \mapsto a), (i_5 \mapsto \mathsf{ab}), (i_6 \mapsto l(\star))$$

This history records the *past communications* of $P$ with its environment. In particular, $P$ has performed six communications with its environment, in an unspecified order, identified by $i_1, \dots, i_6$. Communications $i_1, i_2, i_3$ are tentative, which means that they may be committed or aborted, depending on whether the corresponding transaction commits or aborts. For example, $i_1$ and $i_2$ will become permanent (or be aborted) once transaction $k$ commits (or respectively aborts); similarly for communication $i_3$ and transaction $l$.

In addition, $H$ records that communication $i_4$ has been committed and is now a permanent $a$-action. On the other hand, $i_5$ has been aborted and thus the original tentative action performed is not important.

Communication $i_6$ records a *degenerate communication* of $P$ with its environment. As we will see, the bisimulation allows to match a tentative communication $k(a)$ with a degenerate $k(\star)$ provided that these communications are never committed. $\Diamond$

$$
\begin{array}{llll}
\langle H \bullet P\rangle \xrightarrow{\tau} \langle H \bullet Q\rangle & \text{if} & P \xrightarrow{\tau}_{\varepsilon} Q & (\text{LTS}\tau) \\
\langle H \bullet P\rangle \xrightarrow{\tau} \langle \sigma(H) \bullet Q\rangle & \text{if} & P \xrightarrow{k(\tau)}_{\sigma} Q \text{ and } k \sharp H & (\text{LTS}k(\tau)) \\
\langle H \bullet P\rangle \xrightarrow{\tau} \langle H \bullet Q\rangle & \text{if} & P \xrightarrow{\mathtt{new}\,k} Q \text{ and } k \sharp H & (\text{LTS}\mathtt{new}) \\
\langle H \bullet P\rangle \xrightarrow{\tau} \langle H \setminus_{\mathsf{co}} k \bullet Q\rangle & \text{if} & P \xrightarrow{\mathsf{co}\,k} Q & (\text{LTS}\mathtt{co}) \\
\langle H \bullet P\rangle \xrightarrow{\tau} \langle H \setminus_{\mathsf{ab}} k \bullet Q\rangle & \text{if} & P \xrightarrow{\mathsf{ab}\,k} Q & (\text{LTS}\mathtt{ab}) \\
\langle H \bullet P\rangle \xrightarrow{k} \langle \sigma(H), k(a) \bullet Q\rangle & \text{if} & P \xrightarrow{k(a)}_{\sigma} Q \text{ and } k \sharp H & (\text{LTS}k(a)) \\
\langle H \bullet P\rangle \xrightarrow{k} \langle H, k(\star) \bullet P\rangle & \text{if} & k \sharp H, P & (\text{LTS}\star)
\end{array}
$$

We define $\overset{\zeta}{\Rightarrow}$ to be $(\xrightarrow{\tau})^*$ when $\zeta$ is $\tau$ , and $\overset{\tau}{\Rightarrow}\xrightarrow{\zeta}\overset{\tau}{\Rightarrow}$ otherwise.

Figure 3: Standard transitions.

This LTS semantics for $TCCS^m$ has judgements of the form

$$\langle H \bullet P\rangle \xrightarrow{\zeta} \langle H' \bullet P'\rangle$$

where $\zeta$ ranges over $\{\tau, k\}$. The rules for inferring these judgements are in Figure 3.

The first five rules in the above definition encode the $TCCS^m$ reduction semantics, updating the history of the configurations accordingly: LTS$\tau$ and LTS$k(\tau)$ respectively encode non-transactional and transactional internal moves; LTS`new` represents the initiation of a new transaction; LTS`co` and LTS`ab` respectively encode the commitment and abortion of a transaction. The functions $H \setminus_{\mathsf{co}} k$ and $H \setminus_{\mathsf{ab}} k$ are the lifting to sets of the operations:

$$
\begin{array}{llll l}
(i \mapsto k(\hat{a})) \setminus_{\mathsf{co}} k & = (i \mapsto \hat{a}) & (i \mapsto k(\hat{a})) \setminus_{\mathsf{ab}} k & = (i \mapsto \mathsf{ab}) & \\
(i \mapsto l(\hat{a})) \setminus_{\mathsf{co}} k & = (i \mapsto l(\hat{a})) & (i \mapsto l(\hat{a})) \setminus_{\mathsf{ab}} k & = (i \mapsto l(\hat{a})) & \text{when } k \sharp l \\
(i \mapsto \hat{a}) \setminus_{\mathsf{co}} k & = (i \mapsto \hat{a}) & (i \mapsto \hat{a}) \setminus_{\mathsf{ab}} k & = (i \mapsto \hat{a}) & \\
(i \mapsto \mathsf{ab}) \setminus_{\mathsf{co}} k & = (i \mapsto \mathsf{ab}) & (i \mapsto \mathsf{ab}) \setminus_{\mathsf{ab}} k & = (i \mapsto \mathsf{ab}) &
\end{array}
$$

Here $\hat{a}$ ranges over actions $a$ and $\star$.

Rule LTS$k(a)$ encodes the synchronisation between a transaction in the process and its environment, yielding a fresh transaction $k$; this tentative action is recorded in the history. Note that the reduction semantics (Figure 1) rename the transactions involved in $k(a)$ and $k(\tau)$ transitions using some substitution $\sigma$. This substitution is applied to the history in rules LTS$k(\tau)$ and LTS$k(a)$ for consistency of $H$ and $P$.

Rule LTS$\star$ allows an arbitrary configuration to execute a degenerate $k$ transition; for their usefulness see Example 2.9 below. If the transition $\mathcal{C}_1 \xrightarrow{\zeta} \mathcal{C}_1'$ can be inferred without using this rule it is called a *challenger move*.

The definition of weak bisimulation in [16] requires matching (up to $\tau$-steps) future challenger transitions of related processes, but also matching past actions recorded in related histories, expressed with the notion of *consistent histories*.

**Definition 2.7** (Consistency)**.** Two histories $H_1$ and $H_2$ are *consistent* when they have the same domain, and for all $i \in I$, $a \in \mathsf{Act}$: $H_1(i) = a$ iff $H_2(i) = a$. Two configurations are *consistent* if their history components are. $\Diamond$

**Definition 2.8** (Weak Bisimulation)**.** A binary relation $\mathcal{R}$ over configurations is a (weak) bisimulation when for all $\mathcal{C}_1 \mathrel{\mathcal{R}} \mathcal{C}_2$:

1. (Consistency) $\mathcal{C}_1$ and $\mathcal{C}_2$ are consistent,
2. (Transfer condition) for every $\zeta \in \{\tau, k\}$,
   - (i) if $\mathcal{C}_1 \xrightarrow{\zeta} \mathcal{C}_1'$ is a challenger move where $\zeta \sharp \mathcal{C}_2$, then $\mathcal{C}_2 \stackrel{\hat{\zeta}}{\Rightarrow} \mathcal{C}_2'$ for some $\mathcal{C}_2'$ such that $\mathcal{C}_1' \mathrel{\mathcal{R}} \mathcal{C}_2'$,
   - (ii) the converse of the preceding condition.

Bisimilarity ($\approx$) is the largest (weak) bisimulation over configurations, and extends to processes $P \approx Q$ meaning $\langle \varepsilon \bullet P \rangle \approx \langle \varepsilon \bullet Q \rangle$. ◊

Challenger moves in this definition are for easing proofs of equivalence, avoiding the trivial requirement that $k(\star)$-moves are matched in the transfer condition. A $k(\star)$ history annotation can thus appear only as a response to a challenger $k(a)$-move; committing such moves would violate consistency.

The condition that $\zeta$ is fresh from $\mathcal{C}_2$ guarantees that the choice of fresh transaction names in $\zeta$ does not hinder the transition from $\mathcal{C}_2$. This is in line with the matching of bound outputs in the pi-calculus [22, Sec. 2.2.1].

The following example shows the use of the $k(\star)$-transitions.

**Example 2.9.** Continuing from Example 2.4, we would expect

$$P_1 \approx P_2$$

where $P_1 = [\![a.\,\mathtt{co} \rhd_l \mathbb{0}]\!]$ and $P_2 = [\![a.\,\mathtt{co} + b.\mathbb{0} \rhd_l \mathbb{0}]\!]$. For this to be true we need the transition

$$\langle \varepsilon \bullet P_2 \rangle \xrightarrow{k} \langle k(b) \bullet [\![\mathbb{0} \rhd_l \mathbb{0}]\!] \rangle$$

to be matched in some way by one from $P_1$. Using the rule LTS$\star$ we can infer the degenerate transition, followed by an application of LTS$\mathtt{ab}$

$$\langle \varepsilon \bullet P_1 \rangle \xrightarrow{k} \langle k(\star) \bullet P_1 \rangle \xrightarrow{\tau} \langle k(\star) \bullet \mathbb{0} \rangle \tag{1}$$

which will supply the required matching move in the bisimulation game that establishes $\langle \varepsilon \bullet P_1 \rangle \approx \langle \varepsilon \bullet P_2 \rangle$.

Let $\mathcal{R}$ be the relation over configurations which contains the following pairs:

$$\begin{array}{rcll} \langle \varepsilon \bullet P_1 \rangle & \leftrightarrow & \langle \varepsilon \bullet P_2 \rangle & \\ \langle k(\star) \bullet \mathbb{0} \rangle & \leftrightarrow & \langle k(b) \bullet [\![\mathbb{0} \rhd_l \mathbb{0}]\!] \rangle & \\ \langle k(\star) \bullet \mathbb{0} \rangle & \leftrightarrow & \langle \mathtt{ab} \bullet \mathbb{0} \rangle & \text{for all } k \end{array}$$

Then the use of the transition (1) above is the only non-trivial part of showing that $\mathcal{R} \cup \mathsf{Id}$, where $\mathsf{Id}$ is the identity relation, satisfies the requirements of Definition 2.8. ◊

**Theorem 2.10** (Full-abstraction for bisimulations)**.** *$P \cong_{\mathsf{rbe}} Q$ if and only if $P \approx Q$.*

*Proof.* See Theorem 5.1 and Theorem 5.4 of [16]. □

In the sequel we will introduce alternative bisimulation equivalences for transactions. All will coincide with the contextual equivalence $\cong_{\mathsf{rbe}}$; in each case this will be established by linking the equivalence to the standard bisimulation equivalence $\approx$ over processes.

## 3. More expressive LTSs

### *3.1. Intuition*

The transitions in Figure 3 endow the set of configurations with the structure of an LTS over which there is a standard interpretation of the modal logic HML. However this logic does not characterise bisimilarity from Definition 2.8. The reason is the extra requirement, in condition (1), that the histories be consistent. This means that equivalence between configurations depends not only on their ability to perform interactions with the environment but also on their past behaviour.

One way forward would be to augment standard HML with primitive predicates which take into account past behaviour by interrogating the internal state of configurations. But here we argue that the existing notion of configuration cannot support primitives which give meaningful information about the past behaviour of transactions.

Consider again the processes $P_2$ and $Q_2$ from Example 2.3, which we have shown not to be contextually equivalent. In terms of bisimulations, intuitively $Q_2$ can simulate $P_2$ but also has some additional possible behaviour. Namely it can perform tentative actions $a$ followed by $b$ which are executed in independent transactions, say $m_1$ and $m_2$, but still arrive at a state in which thereafter either both transactions, or none, are committed. This possibility is captured by the configuration $\mathcal{C}_Q$ in the following derivation; no comparable configuration is reachable from $\langle \varepsilon \bullet P_2 \rangle$.

$$\begin{array}{lll} \langle \varepsilon \bullet Q_2 \rangle & \xrightarrow{m_1} & \dots \\ & \xrightarrow{m_2} & \langle m_1(a), m_2(b) \bullet \nu p.\, [\![ p.\, \mathsf{co} \rhd_{m_1} \mathbb{0} ]\!] \mid [\![ \overline{p}.\, \mathsf{co} \rhd_{m_2} \mathbb{0} ]\!] \rangle \; = \mathcal{C}_Q \\ & \xrightarrow{\tau} & \langle m(a), m(b) \bullet \nu p.\, [\![ \mathsf{co} \rhd_{m} \mathbb{0} ]\!] \mid [\![ \mathsf{co} \rhd_{m} \mathbb{0} ]\!] \rangle \; = \mathcal{C}'_Q \end{array}$$

The intuitive characteristic property of $\mathcal{C}_Q$ is that in any future state the commitment status of the two tentative actions $m_1, m_2$ is identical. But formally this cannot be expressed as a property of the successor configurations of $\mathcal{C}_Q$. For example in $\mathcal{C}'_Q$ the transaction names $m_1, m_2$ no longer occur, and so it does not make sense to make an assertion for $\mathcal{C}'_Q$ about them.

We could extend the modal logic with some *past operators*, as in [7], with which $\mathcal{C}$ could gain access to the transaction names used in its past. Instead we take a simpler approach, augmenting the notion of configuration so that the knowledge of all previously used transaction names is retained. This is the topic of the next subsection. We also show that a version of Theorem 2.10 can be established with this version of augmented configurations.

### *3.2. New Transition Semantics*

The main intuition in this new semantics is that configurations contain both a history $H$ and an equivalence relation $E$ between transaction names. Configuration transitions may extend $H$ and $E$, but never apply substitutions to $H$. Thus any $k(a)$ that appears in the history $H$ of a configuration will remain unchanged in all subsequent configurations. The merging of transactions, resulting from communications between transactions, will instead be captured by increasing the equivalence relation $E$.

Here we divide the set of transaction names $\mathsf{TrN}$ into two disjoint sets; $\mathsf{TrN} = \mathsf{inTrN} \uplus \mathsf{exTrN}$. We do this for clarity: $\mathsf{inTrN}$ contains *internal names* used for transactions confined inside configurations, whereas $\mathsf{exTrN}$ contains *external names* used in transactions shared between a configuration and its environment. The latter are the names appearing in histories; we write $\mathit{eftn}(o)$ to mean the external transaction names in syntactic object $o$.

Histories are now finite partial functions $H$ from the countable (unordered) index set $I$ to the set $\{\, k(a), k(\mathsf{ab}), k(\mathsf{co}) \mid k \in \mathsf{exTrN},\ a \in \mathsf{Act}\,\}$. For $i \in \mathit{dom}(H)$ we use $H^{\mathtt{trn}}(i)$ to denote the transaction name used in $H(i)$, while $H^{\mathtt{trn}}$ denotes the set $\{\, H^{\mathtt{trn}}(i) \mid i \in \mathit{dom}(H)\,\}$. Moreover,

- $\mathtt{IsR}(H) = \{\, k \in \mathsf{exTrN} \mid H(i) = k(a),$ for some $i \in \mathit{dom}(H),\ a \in \mathsf{Act}\,\}$; these are the active transactions with tentative actions in the history;
- $\mathtt{Hasab}(H) = \{\, k \in \mathsf{exTrN} \mid H(i) = k(\mathsf{ab})$, for some $i \in \mathit{dom}(H)\,\}$; these are the aborted transactions;
- $\mathtt{Hasco}(H) = \{\, k \in \mathsf{exTrN} \mid H(i) = k(\mathsf{co})$, for some $i \in \mathit{dom}(H)\,\}$; the committed transactions.

**Definition 3.1** (Extended configurations)**.** An *extended history* consists of a pair $E; H$ where $H$ is a partial function as above, satisfying

(i) $H^{\mathtt{trn}}(i) = H^{\mathtt{trn}}(j)$ implies $i = j$,

and $E$ is an equivalence relation over a finite subset of $\mathsf{TrN}$ satisfying

(ii) for every $k \in \mathsf{exTrN}$:

- $k\,E\,k$ if and only if $k \in H^{\mathtt{trn}}$
- if $k\,E\,k'$ then $k \in \mathtt{Hasco}(H)$ if and only if $k' \in \mathtt{Hasco}(H)$.

An *extended configuration* is a pair $\langle \Delta \bullet P \rangle$ where $P$ is a well-formed process and $\Delta = E; H$ is an extended history satisfying[1]

[1] Condition (iv) is used in proof of Lemma B.3; (iii) is used to ensure Property (4) in Lemma 3.3; (iii) also appears to be true for $\mathtt{Hasab}(H)$ but is not needed.

$$\langle \Delta \bullet P\rangle \stackrel{\tau}{\twoheadrightarrow} \langle \Delta \bullet Q\rangle \quad \text{if } P \xrightarrow{\tau}_{\varepsilon} Q \qquad (\text{eLTS}\tau)$$

$$\langle E; H \bullet P\rangle \stackrel{\tau}{\twoheadrightarrow} \langle \sigma(E); H \bullet Q\rangle \quad \text{if } P \xrightarrow{k(\tau)}_{\sigma} Q,\; k \sharp E, H,\; k \in \mathsf{inTrN} \qquad (\text{eLTS}k(\tau))$$

$$\langle \Delta \bullet P\rangle \stackrel{\tau}{\twoheadrightarrow} \langle \Delta \bullet Q\rangle \quad \text{if } P \xrightarrow{\mathtt{new}\,k} Q,\; k \sharp \Delta,\; k \in \mathsf{inTrN} \qquad (\text{eLTS}\mathtt{new})$$

$$\langle \Delta \bullet P\rangle \stackrel{\tau}{\twoheadrightarrow} \langle \Delta \setminus_{\mathsf{co}} k \bullet Q\rangle \quad \text{if } P \xrightarrow{\mathsf{co}\,k} Q \qquad (\text{eLTS}\mathtt{co})$$

$$\langle \Delta \bullet P\rangle \stackrel{\tau}{\twoheadrightarrow} \langle \Delta \setminus_{\mathsf{ab}} k \bullet Q\rangle \quad \text{if } P \xrightarrow{\mathsf{ab}\,k} Q \qquad (\text{eLTS}\mathtt{ab})$$

$$\langle E; H \bullet P\rangle \stackrel{k(a)}{\twoheadrightarrow} \langle \sigma(E); H, k(a) \bullet Q\rangle \quad \text{if } P \xrightarrow{k(a)}_{\sigma} Q,\; k \sharp E, H,\; k \in \mathsf{exTrN} \qquad (\text{eLTS}k(a))$$

$$\langle E; H \bullet P\rangle \stackrel{k(a)}{\twoheadrightarrow} \langle E, (k,k); H, k(\mathsf{ab}) \bullet P\rangle \quad \text{if } k \sharp E, H, P,\; k \in \mathsf{exTrN} \qquad (\text{eLTS}\star)$$

We define $\stackrel{\zeta}{\Rightarrow}$ to be $(\stackrel{\tau}{\twoheadrightarrow})^*$ when $\zeta$ is $\tau$ , and $\stackrel{\tau}{\Rightarrow} \stackrel{\zeta}{\twoheadrightarrow} \stackrel{\tau}{\Rightarrow}$ otherwise.

Figure 4: Extended transitions.

(iii) $ftn(P) \cap \mathtt{Hasco}(H) = \emptyset$

(iv) $k\,E\,k'$ and $k, k' \in ftn(P)$ implies $k = k'$.

We extend functions such as $\mathtt{Hasco}(-)$, originally defined on the partial functions $H$, to extended configurations in the obvious manner. For example $\mathtt{IsR}(\langle E; H \bullet P\rangle)$ refers to $\mathtt{IsR}(H)$. ◊

The intuition here is that $H(i)$ records the transaction name, from $\mathsf{exTrN}$, used in the $i^{\text{th}}$ interaction with the environment, although we do not need to impose an order on $i$'s in the domain of $H$. As the history grows, new fresh transaction names are used. These names never change, hence the restriction that $H^{\mathtt{trn}}(i) = H^{\mathtt{trn}}(j)$ implies $i = j$. The identifications introduced by the merging of transactions, caused by inter-transactional communication, are tabulated by the equivalence relation $E$, which develops dynamically as a computation proceeds. Note that in the reduction semantics from Figure 1, internal communication automatically leads to the fresh renaming of transaction names. In our revised transitions these internal communications use names from $\mathsf{inTrN}$, and thus in general the equivalence relation $E$ may contain these fresh names, from $\mathsf{inTrN}$, which do not appear in $H^{\mathtt{trn}}$. See Example 3.2 for an instance.

A further difference is that $E$ also retains all the historical names used, including those committed or aborted. But the major significance of the equivalence classes is that there may be two distinct transaction names $k_1, k_2 \in H^{\mathtt{trn}}$ satisfying $k_1\,E\,k_2$, encoding the fact that $k_1$ and $k_2$ have been merged into a single transaction.

We use a number of operations on extended histories to record the effects of transitions such as commits, aborts and internal communications. These are more or less inherited from [16], and are defined presently. They are used in the definition of extended transitions $\mathcal{C} \stackrel{\tau}{\twoheadrightarrow} \mathcal{C}'$ derived from the rules of Figure 4.

These rules use the following operations on extended histories:

(a) Renaming: For any substitution $\sigma = \tilde{k} \mapsto l$, $\sigma(E)$ is the smallest equivalence relation $F$ satisfying

$$E \subseteq F \qquad\qquad l\,F\,l \qquad\qquad k\,F\,l \text{ for every } k \in \mathit{dom}(\sigma)$$

(b) Committing: $E; H \setminus_{\mathtt{co}} k$ leaves $E$ unchanged but, for every $k'\,E\,k$, changes every occurrence of $k'(a)$ in $H$ into $k'(\mathtt{co})$.

(c) Aborting: $E; H \setminus_{\mathtt{ab}} k$ again leaves $E$ unchanged but, for every $k'\,E\,k$, changes every occurrence of $k'(a)$ in $H$ into $k'(\mathtt{ab})$.

We also use $\mathtt{Hasco}\langle E; H \bullet P\rangle$ to denote the set $\{k \in \mathsf{exTrN} \mid \exists i.\ H(i) = k(\mathtt{co})\}$.

**Example 3.2.** Consider again the processes $P_2$ and $Q_2$ from Example 2.3. Using the extended transitions in Figure 4 we have the following computation from $\langle \emptyset; \varepsilon \bullet Q_2\rangle$:

$$\begin{aligned}\langle \emptyset; \varepsilon \bullet Q_2\rangle &\xrightarrow{m_1(a)}\!\!\!\!\rightarrow \ \ldots \\ &\xrightarrow{m_2(b)}\!\!\!\!\rightarrow \ \langle \mathrm{Id}; m_1(a), m_2(b) \bullet \nu p.[\![p.\,\mathtt{co} \rhd_{m_1} \mathbb{0}]\!] \mid [\![\overline{p}.\,\mathtt{co} \rhd_{m_2} \mathbb{0}]\!]\rangle \\ &\xrightarrow{\tau}\!\!\!\!\rightarrow \ \langle E; m_1(a), m_2(b) \bullet \nu p.[\![\mathtt{co} \rhd_{m_3} \mathbb{0}]\!] \mid [\![\mathtt{co} \rhd_{m_3} \mathbb{0}]\!]\rangle \ = \mathcal{C}_Q\end{aligned}$$

where Id is the identity relation and $E$ contains the single equivalence class $\{m_1, m_2, m_3, k_1, k_2\}$. Because of the rules from Figure 4 used to infer these transitions we know that $m_1, m_2 \in \mathsf{exTrN}$ whereas $m_3 \in \mathsf{inTrN}$.

Here we have retained the historical transaction names $m_1, m_2$ in the residual configuration $\mathcal{C}_Q$. In a logic endowed with suitable primitives we will be able to assert that the future commitments of $m_1$ and $m_2$ will coincide; even in configurations in which these names are absent from the processes; see Example 4.3. On the other hand the only comparable computation from $\langle \emptyset; \varepsilon \bullet P_2\rangle$ is

$$\begin{aligned}\langle \emptyset; \varepsilon \bullet P_2\rangle &\xrightarrow{m_1(a)}\!\!\!\!\rightarrow \ \ldots \\ &\xrightarrow{m_2(b)}\!\!\!\!\rightarrow \ \langle \mathrm{Id}; m_1(a), m_2(b) \bullet [\![\mathtt{co} \rhd_{m_1} \mathbb{0}]\!] \mid [\![\mathtt{co} \rhd_{m_2} \mathbb{0}]\!]\rangle \ = \mathcal{C}_P\end{aligned}$$

In $\mathcal{C}_P$ the historical transaction names $m_1, m_2$ have independent commitment possibilities. ◊

**Lemma 3.3** (Sanity Check 1)**.** *Suppose*

$$\mathcal{C}_1 = \langle E_1; H_1 \bullet P_1\rangle \xrightarrow{\zeta}\!\!\!\!\rightarrow \mathcal{C}_2 = \langle E_2; H_2 \bullet P_2\rangle$$

*and $\mathcal{C}_1$ is an extended configuration. Then*

*(1) $\zeta \sharp \mathcal{C}_1$ and $\mathit{eftn}(\mathcal{C}_2) \subseteq \mathit{eftn}(\mathcal{C}_1) \cup \mathit{ftn}(\zeta)$*

*(2) $\mathcal{C}_2$ is also an extended configuration*

(3) *(Monotonicity)* $E_1 \subseteq E_2$, $\mathtt{IsR}(\mathcal{C}_2) \subseteq \mathtt{IsR}(\mathcal{C}_1) \cup \mathit{ftn}(\zeta)$, *and* $\mathtt{Hasco}(\mathcal{C}_1) \subseteq \mathtt{Hasco}(\mathcal{C}_2) \subseteq \mathtt{Hasco}(\mathcal{C}_1) \cup \mathtt{IsR}(\mathcal{C}_1)$

(4) *if* $l, l' \in \mathtt{Hasco}(\mathcal{C}_1)$ *and* $l\, E_2\, l'$ *then* $l\, E_1\, l'$.

*Proof.* In Figure 4 there are seven ways of inferring the judgement $\mathcal{C} \xrightarrow{\zeta}\!\!\!\!\rightarrow \mathcal{C}'$. The proof proceeds by examining each of these seven cases in turn. In each case property (1) is straightforward.

For property (2) we give one example. Suppose $\zeta = k(a)$ and $H_2 = H_1, k(a)$, $E_2 = \sigma(E_1)$ because $P_1 \xrightarrow{k(a)}_{\sigma} P_2$ for some fresh $k$. We need to show that $\mathcal{C}_1$ satisfies the requirements of Definition 3.1. The only non-trivial requirement is (iv): $l\, \sigma(E_1)\, l'$ and $l, l' \in \mathit{ftn}(P_2)$ implies $l = l'$. As a prelude to the proof of this we first show

$$l \in E_1(\mathit{dom}(\sigma)) \text{ implies } l \notin \mathit{ftn}(P_2) \qquad (\dagger)$$

For, if $l\, E_1\, n$ for some $n \in \mathit{dom}(\sigma)$ then we know $l$ is different than the fresh $k$. Also from Lemma A.2 (1) we know that $n \in \mathit{ftn}(P_1)$ and $n \notin \mathit{ftn}(P_2)$.

There are two possibilities for $l$. If $l \notin \mathit{ftn}(P_1)$ then since it is different from $k$ it is also not in $\mathit{ftn}(P_2)$ by property (1), as required. On the other hand if $l \in \mathit{ftn}(P_1)$ then since $\mathcal{C}$ satisfies the requirements of an extended configuration (Definition 3.1) we have that $l = n$, and therefore is not in $\mathit{ftn}(P_2)$ as required.

Having established (†) above, now let us prove property (iv) of Definition 3.1 for $\mathcal{C}_2$. Suppose $l\, \sigma(E_1)\, l'$ where $l, l' \in \mathit{ftn}(P_2)$. We have to show $l = l'$. Using the characterisation of $\sigma(E_1)$ in Lemma A.1 (2) there are two possibilities.

(i) $l\, E_1\, l'$. Here we know both $l, l'$ are different than $k$ and so by property (4) of Lemma A.2, $l, l' \in \mathit{ftn}(P)$. The fact that $\mathcal{C}$ is a well-formed extended configuration now gives the required $l = l'$.

(ii) $l, l' \in E_1(\mathit{dom}(\sigma)) \cup \mathit{dom}(\sigma) \cup \{k\}$. But now by (†) above we know that $l, l' \notin E_1(\mathit{dom}(\sigma))$, and also by property (1) of Lemma A.2 $l, l' \notin \mathit{dom}(\sigma)$. So the only possibility is that $l, l' = k$, when the required result is immediate.

The proof of (3) is a question of examining the seven cases in Figure 4 and by inspection ensuring that the three subset relations are retained.

The proof of (4) is a little more delicate. But note that in only three of the seven cases in Figure 4 does $E_2$ differ from $E_1$; and in one of these, eLTS$\star$, the proof is trivial. Here we examine one other case, eLTS$k(a)$. So we have $\zeta = k(a)$ and $H_2 = H_1, k(a)$, $E_2 = \sigma(E_1)$ because $P_1 \xrightarrow{k(a)}_{\sigma} P_2$, where $k$ is fresh. Suppose $l\, \sigma(E_1)\, l'$, where $l, l' \in \mathtt{Hasco}(H_1)$. We have to show that $l\, E_1\, l'$. First note that if $l\, E_1\, n$ for any name $n$ then $n \notin \mathit{dom}(\sigma)$. This follows because for any such $n$ we know by Lemma A.2 (1) that $n \in \mathit{ftn}(P_1)$. The well-formedness condition (ii) in Definition 3.1 then ensures that $n \in \mathtt{Hasco}(H_1)$, which in turn contradicts condition (iii) of the same definition.

The same is true of $l'$, and the freshness of $k$ means that neither $l$ nor $l'$ can be $k$. So by the characterisation of $\sigma(E_1)$ in Lemma A.1 (2) the required result follows. □

The transition semantics, both for processes in Figure 1 and for (extended) configurations in Figure 4, make extensive use of transaction names, and their systematic renaming by fresh names. The intention is to handle these names as if they are formally bound, in the same way as names are *scoped* in the pi-calculus, [22]. Intuitively the behaviour of transactions should be independent of their names. This is formally captured by showing that the transitions are preserved by arbitrary permutations of the transaction names.

**Definition 3.4** (Permutations)**.** A permutation $\pi$ is an injective and surjective mapping over the set of names $\mathsf{TrN}$ satisfying

(1) (Finite) $\pi(k) = k$ for almost all $k$

(2) $\pi(k) \in \mathsf{exTrN}$ if and only if $k \in \mathsf{exTrN}$. ♢

We use $\pi(P)$ to denote the result of replacing all occurrences of a transaction name $k$ in $P$ with $\pi(k)$. For an extended history $E; H$ we let $\pi(E; H)$ denote $E_\pi; (\pi \cdot H)$, where $E_\pi$ is shorthand for $\pi^{-1} \cdot E \cdot \pi^{-1}$; that is $k \, E_\pi \, k'$ when $\pi^{-1}(k) \, E \, \pi^{-1}(k')$. If $\langle \Delta \bullet P \rangle$ is an extended configuration then so is $\langle \pi(\Delta) \bullet \pi(P) \rangle$; we refer to the latter as $\pi(\langle \Delta \bullet P \rangle)$.

**Proposition 3.5** (Transition permutations)**.** *For any permutation $\pi$, $\mathcal{C} \xrightarrow{\zeta}\!\!\!\!\rightarrow \mathcal{C}'$ implies $\pi(\mathcal{C}) \xrightarrow{\pi(\zeta)}\!\!\!\!\rightarrow \pi(\mathcal{C}')$.*

*Proof.* In Figure 4 there are seven ways of inferring the judgement $\mathcal{C} \xrightarrow{\zeta}\!\!\!\!\rightarrow \mathcal{C}'$. The proof proceeds by examining each of these seven cases in turn. We give two example cases.

(a) Suppose $\mathcal{C} = \langle E; H \bullet P \rangle \xrightarrow{\tau}\!\!\!\!\rightarrow \langle \sigma(E); H \bullet Q \rangle = \mathcal{C}'$ because $P \xrightarrow{k(\tau)}_\sigma Q$, where $k$ is fresh.

From Lemma A.2 $\pi(P) \xrightarrow{\pi(k)(\tau)}_{\sigma_\pi} \pi(Q)$, and therefore employing the rule $\textsc{eLTS}k(\tau)$ we have $\pi(\langle E; H \bullet P \rangle) \xrightarrow{\tau}\!\!\!\!\rightarrow \langle \sigma_\pi(E_\pi); \pi \cdot H \bullet \pi(Q) \rangle$. The result now follows since by Lemma A.1 (3) we know that $\sigma_\pi(E_\pi) = (\sigma(E))_\pi$.

(b) Suppose $\mathcal{C} = \langle E; H \bullet P \rangle \xrightarrow{\tau}\!\!\!\!\rightarrow \langle E; H \setminus_{\mathsf{co}} k \bullet Q \rangle = \mathcal{C}'$ because $P \xrightarrow{\mathsf{co}\, k} Q$.

By Lemma A.3 we have $\pi(P) \xrightarrow{\mathsf{co}\, \pi(k)} \pi(Q)$, and so employing the rule $\textsc{eLTS}k(\tau)$ we have $\pi(\langle E; H \bullet P \rangle) \xrightarrow{\tau}\!\!\!\!\rightarrow \langle E_\pi; (\pi \cdot H) \setminus_{\mathsf{co}} \pi(k) \bullet \pi(Q) \rangle$. The result now follows since $(\pi \cdot H) \setminus_{\mathsf{co}} \pi(k)$ is the same as $\pi \cdot (H \setminus_{\mathsf{co}} k)$. □

This result implies that the LTS determined by the transitions in Figure 4 constitutes a *nominal transition system* as given in Definition 1 of [20].

*3.3. New Bisimulation Equivalence*

We now adapt the definition of bisimulation from Definition 2.8 to extended configurations and show that the two bisimulations coincide for source-level configurations. This new definition requires the matching of future actions and past behaviour, as recorded in (extended) histories. But the comparison of these past actions is somewhat different, and is captured in the following definition.

**Definition 3.6** (Commit consistent configurations)**.** We say that $\mathcal{C}_1, \mathcal{C}_2$ are *commit consistent*, written $\mathcal{C}_1 \, \texttt{Hasco} \, \mathcal{C}_2$, whenever $\texttt{Hasco}(\mathcal{C}_1) = \texttt{Hasco}(\mathcal{C}_2)$. $\diamondsuit$

Here we only require that two commit consistent configurations have the same set of committed transaction names. This is in contrast to the notion of *consistent* (histories) in Definition 2.7, where the comparison is index-based. It is possible for two configurations with extended histories satisfying

$$E_1; H_1, (i \mapsto k(a)) \, \texttt{Hasco} \, E_2; H_2, (i \mapsto k(b))$$

to commit transaction $k$ without violating commit consistency. This would violate a consistency in the sense of Definition 2.7, where committed actions must be equal.

**Definition 3.7** ($\texttt{Hasco}$-Bisimulations)**.** A binary relation $\mathcal{R}$ over (extended) configurations is a (weak) $\texttt{Hasco}$-bisimulation when for all $\mathcal{C}_1 \; \mathcal{R} \; \mathcal{C}_2$

1. $\mathcal{C}_1 \; \texttt{Hasco} \; \mathcal{C}_2$
2. Transfer property:
   (i) if $\mathcal{C}_1 \stackrel{\zeta}{\Rrightarrow} \mathcal{C}_1'$ where $\zeta \sharp \mathcal{C}_2$ then $\mathcal{C}_2 \stackrel{\zeta}{\Rrightarrow} \mathcal{C}_2'$ for some $\mathcal{C}_2'$ such that $\mathcal{C}_1' \; \mathcal{R} \; \mathcal{C}_2'$,
   (ii) the converse of condition (i)

$\texttt{Hasco}$-bisimilarity ($\approx_{\mathsf{Hasco}}$) is the largest $\texttt{Hasco}$-bisimulation over configurations, and extends to processes: $P \approx_{\mathsf{Hasco}} Q$ if $\langle \emptyset; \varepsilon \bullet P \rangle \approx_{\mathsf{Hasco}} \langle \emptyset; \varepsilon \bullet Q \rangle$. $\diamondsuit$

In this definition we use a weak move $\mathcal{C}_1 \stackrel{\zeta}{\Rrightarrow} \mathcal{C}_1'$ instead of a strong challenger one as in Definition 2.8. This simplifies proofs when comparing versions of bisimulation to logical equivalences in the following section.

**Proposition 3.8.** *For any permutation $\pi$, $\mathcal{C} \approx_{\mathsf{Hasco}} \pi(\mathcal{C})$.*

*Proof.* The relation $\mathcal{R} = \{ (\mathcal{C}, \pi(\mathcal{C})) \mid \pi \text{ a permutation} \}$ can be shown to be a $\texttt{Hasco}$-bisimulation by the systematic application of Proposition 3.5. $\square$

The following theorem justifies our new form of bisimulation, $\approx_{\mathsf{Hasco}}$, by showing that the process equivalence it denotes coincides with the original bisimulation of [16], $\approx$, recalled in Definition 2.8.

**Theorem 3.9.** *$P \approx Q$ if and only if $P \approx_{\mathsf{Hasco}} Q$.*

*Proof.* See Appendix B. $\square$

From the above it follows that $\approx_{\mathsf{Hasco}}$ coincides with contextual equivalence.

**Theorem 3.10.** *$P \cong_{\mathsf{rbe}} Q$ if and only if $P \approx_{\mathsf{Hasco}} Q$.*

*Proof.* Immediate from Theorems 2.10 and 3.9. □

**Remark 3.11.** In the following sections we will see two more bisimulation equivalences. All three relations will be shown equivalent, each giving rise to an interesting modal logic. ♢

## 4. HML for transactions

Here we return to the topic of property logics for transactions. We first explain the natural logic $\mathcal{L}_{\mathsf{Hasco}}$ associated with the bisimulation equivalence $\approx_{\mathsf{Hasco}}$. Recall that in the definition of these bisimulations there is a requirement on the freshness of transaction names. To capture this we use a nominal interpretation of the standard modal operators from HML, [18]. In addition we have operators for interrogating past behaviour, as encoded in extended configurations.

There are natural variations possible on these predicates which examine the past behaviour of processes. One instance is given in the following section, $\mathcal{L}_{\mathsf{Eq}}$. Then in Section 4.3 we give an alternative logic $\mathcal{L}_{\mathsf{Canco}}$, in which past behaviour remains unexamined. This requires yet a new transition semantics which makes commitments externally visible.

In the following section we will show that all three logics are equally powerful in their ability to distinguish processes. They can distinguish, and only distinguish, all processes differentiated by the contextual equivalence $\cong_{\mathsf{rbe}}$.

### *4.1. The Property Logic $\mathcal{L}_{\mathsf{Hasco}}$*

Here we design a property logic $\mathcal{L}_{\mathsf{Hasco}}$ which captures the bisimulation equivalence $\approx_{\mathsf{Hasco}}$. As with the original HML [18, p. 88] it uses modal formulae to capture future actions of processes. But we also need formulae for capturing past behaviour; in this case a collection of predicates will suffice. However there is a complication with the use of modal operators to capture future actions. In the definition of Hasco-bisimulations (Definition 3.7) the transaction names in the actions used to interrogate configurations are required to be fresh. As already remarked, this is common for calculi which manipulate bound names, [22, 20]. To mimic this freshness, we use a separate syntactic category of variables, Var, which is a countable set distinct from exTrN. These variables are used in the definition of the modal constructs, and semantically are interpreted *nominally* [12, 21]; on the other hand the names in exTrN are treated as constants.

$$
\begin{aligned}
\text{Properties: } \phi \in \mathcal{L}_{\mathsf{Hasco}} \quad &::= \quad \langle \tau \rangle\, \phi \mid \langle x(a) \rangle\, \phi \quad \text{when } x \in \mathsf{Var} \\
&\qquad \mid \neg \phi \mid \textstyle\bigwedge_{\{i \in I\}} \phi_i \\
&\qquad \mid \mathtt{Hasco}(v) \\
v \in \mathsf{Values} \quad &::= \quad l \in \mathsf{exTrN} \mid x \in \mathsf{Var}
\end{aligned}
$$

As usual true is encoded with an empty conjunction and false with $\neg$true. In the formula $\langle x(a)\rangle\, \phi$ all the occurrences of the name $x$ in $\phi$ are bound. We thus have the standard notion of free and bound occurrences of variables. We are interested in *closed* formulae, that is those containing no free variables.

We use $ftn(\phi)$ to denote the set of names from exTrN in $\phi$, and assume a standard notion of applying a permutation $\pi$ to a formula $\phi$, written $\pi(\phi)$; this substitutes names for names, and can be defined in a straightforward manner by structural induction. We also require the notion of substitution of a name $k$ for all free occurrences of a variable $x$, written $\phi[k/x]$; again this can be defined straightforwardly by structural induction on $\phi$, and does not require any notion of *alpha-equivalence*, as in [20]. Finally note that unlike papers such as [19, 20, 1] we tolerate formulae containing an infinite number of names.

**Definition 4.1.** The satisfaction relation $\mathcal{C} \models_{\mathsf{hc}} \phi$, defined for closed $\phi$, is:

(1) $\mathcal{C} \models_{\mathsf{hc}} \mathtt{Hasco}(k)$ whenever $k \in \mathtt{Hasco}(\mathcal{C})$

(2) $\mathcal{C} \models_{\mathsf{hc}} \langle\tau\rangle\, \phi'$ if $\mathcal{C} \overset{\tau}{\Rrightarrow} \mathcal{C}'$ such that $\mathcal{C}' \models_{\mathsf{hc}} \phi'$

(3) $\mathcal{C} \models_{\mathsf{hc}} \langle x(a)\rangle\, \phi'$ if there exists cofinite $N \subseteq \mathsf{exTrN}$ such that for all $l \in N$, there exists $\mathcal{C}'$ such that $\mathcal{C} \overset{l(a)}{\Longrightarrow} \mathcal{C}'$ and $\mathcal{C}' \models_{\mathsf{hc}} \phi'[l/x]$

(4) $\mathcal{C} \models_{\mathsf{hc}} \wedge_{\{i\in I\}}\phi_i$ if for each $i \in I$, $\mathcal{C} \models_{\mathsf{hc}} \phi_i$.

We let $\mathcal{L}_{\mathsf{Hasco}}(\mathcal{C}) = \{\, \phi \mid \mathcal{C} \models_{\mathsf{hc}} \phi \,\}$ and abbreviate $\mathcal{L}_{\mathsf{Hasco}}\langle\emptyset; \varepsilon \bullet P\rangle$ to $\mathcal{L}_{\mathsf{Hasco}}(P)$. ◊

**Lemma 4.2.** *Let $\pi$ be permutation of* exTrN. *If $\mathcal{C} \models_{\mathsf{hc}} \phi$ then $\pi(\mathcal{C}) \models_{\mathsf{hc}} \pi(\phi)$.*

*Proof.* By induction on the size of $\phi$. We give one example.

Suppose $\mathcal{C} \models_{\mathsf{hc}} \langle x(a)\rangle\, \phi'$; that is $\mathcal{C} \overset{l(a)}{\Longrightarrow} \mathcal{C}_l$ such that $\mathcal{C}_l \models_{\mathsf{hc}} \phi'[l/x]$ for all $l$ in some cofinite set $N$. We show $\pi(\mathcal{C}) \models_{\mathsf{hc}} \pi(\langle x(a)\rangle\, \phi')$, that is $\pi(\mathcal{C}) \models_{\mathsf{hc}} \langle x(a)\rangle\, \pi(\phi')$.

Using Proposition 3.5 we have that $\pi(\mathcal{C}) \overset{\pi(l)(a)}{\Longrightarrow} \pi(\mathcal{C}_l)$ for all $l \in N$. Also by induction $\pi(\mathcal{C}_l) \models_{\mathsf{hc}} \pi(\phi'[l/x])$; this formula can also be written as $\pi(\phi')[\pi(l)/x]$ since name permutations leave variables unchanged.

Let $K$ denote $\pi(N)$, which is a cofinite set. Then we have shown that $\pi(\mathcal{C}) \overset{k(a)}{\Longrightarrow} C_k$ for some $C_k$ such that $C_k \models_{\mathsf{hc}} \pi(\phi')[k/x]$, for all $k$ in $K$. By definition this means $\pi(C) \models_{\mathsf{hc}} \langle x(a)\rangle\, \pi(\phi')$. □

**Example 4.3.** Consider again the processes from Examples 2.3 and 3.2,

$$P_2 = [\![a.\,\mathsf{co} \rhd_{k_1} \mathbb{0}]\!] \mid [\![b.\,\mathsf{co} \rhd_{k_2} \mathbb{0}]\!]$$
$$Q_2 = \nu p.\, [\![a.p.\,\mathsf{co} + a.\,\mathsf{co} \rhd_{k_1} \mathbb{0}]\!] \mid [\![b.\overline{p}.\,\mathsf{co} + b.\,\mathsf{co} \rhd_{k_2} \mathbb{0}]\!]$$

For convenience let $\phi$ denote the property

$$\langle x(a)\rangle\, \langle y(b)\rangle\, \big(\neg\mathtt{Hasco}(x) \wedge \langle\tau\rangle\, \mathtt{Hasco}(x) \wedge [\tau](\mathtt{Hasco}(x) \leftrightarrow \mathtt{Hasco}(y))\big)$$

where $\leftrightarrow$ is double implication, definable in $\mathcal{L}_{\mathsf{Hasco}}$, and $[\mu]\,\psi$ is the standard shorthand for $\neg\,\langle\mu\rangle\,\neg\psi$.

First we show that $\langle\emptyset;\varepsilon \bullet Q_2\rangle \models_{\mathsf{hc}} \phi$.

To see this consider the derivation in Example 3.2:

$$\langle\emptyset;\varepsilon \bullet Q_2\rangle \xrightarrow{m_1(a)}\!\!\!\!\twoheadrightarrow \xrightarrow{m_2(b)}\!\!\!\!\twoheadrightarrow \xrightarrow{\tau}\!\!\!\!\twoheadrightarrow \mathcal{C}_Q \xrightarrow{\tau}\!\!\!\!\twoheadrightarrow \mathcal{C}'_Q$$

where $\mathcal{C}_Q$, $\mathcal{C}'_Q$ respectively denote

$$\langle E; m_1(a), m_2(b) \bullet \nu p. [\![\mathsf{co} \rhd_{m_3} \mathbb{0}]\!] \mid [\![\mathsf{co} \rhd_{m_3} \mathbb{0}]\!]\rangle \quad \langle E; m_1(\mathsf{co}), m_2(\mathsf{co}) \bullet \nu p. \mathbb{0} \mid \mathbb{0}\rangle$$

Then $\mathcal{C}_Q \models_{\mathsf{hc}} \neg\mathtt{Hasco}(m_1) \wedge \langle\tau\rangle\,\mathtt{Hasco}(m_1)$. Also $\mathcal{C}_Q \models_{\mathsf{hc}} \mathtt{Hasco}(m_1) \leftrightarrow \mathtt{Hasco}(m_2)$, since in $\mathcal{C}_Q$ neither of $m_1, m_2$ are committed, and $\mathcal{C}'_Q \models_{\mathsf{hc}} \mathtt{Hasco}(m_1) \leftrightarrow \mathtt{Hasco}(m_2)$ because both are committed. In fact one can show that $\mathcal{C}' \models_{\mathsf{hc}} \mathtt{Hasco}(m_1) \leftrightarrow \mathtt{Hasco}(m_2)$ whenever $\mathcal{C}_Q \stackrel{\tau}{\Rightarrow}\!\!\!\!\twoheadrightarrow \mathcal{C}'$, and thus $\mathcal{C}_Q \models_{\mathsf{hc}} [\tau](\mathtt{Hasco}(m_1) \leftrightarrow \mathtt{Hasco}(m_2))$.

This argument can be repeated for almost all $m_1, m_2$, which means that

$$Q_2 \models_{\mathsf{hc}} \phi$$

However

$$P_2 \not\models_{\mathsf{hc}} \phi$$

It is not possible to find a derivation $\langle\emptyset;\varepsilon \bullet P_2\rangle \stackrel{m_1(a)}{\Longrightarrow}\!\!\!\!\twoheadrightarrow \stackrel{m_2(b)}{\Longrightarrow}\!\!\!\!\twoheadrightarrow \mathcal{C}_P$ such that $\mathcal{C}_P \models_{\mathsf{hc}} \langle\tau\rangle\,\mathtt{Hasco}(m_1)$ all of whose successors satisfy $\mathtt{Hasco}(m_1) \leftrightarrow \mathtt{Hasco}(m_2)$. $\Diamond$

We now discuss the relationship between the logic $\mathcal{L}_{\mathsf{Hasco}}$ and the bisimulation equivalence Hasco.

**Proposition 4.4.** *If $\mathcal{C}_1 \approx_{\mathsf{Hasco}} \mathcal{C}_2$ then $\mathcal{L}_{\mathsf{Hasco}}(\mathcal{C}_1) = \mathcal{L}_{\mathsf{Hasco}}(\mathcal{C}_2)$.*

*Proof.* The proof relies on the fact that the support of a configuration, that is the set of transaction names occurring in it, is finite; see Definition 2.1.

Suppose that $\mathcal{C}_1 \models_{\mathsf{hc}} \phi$. We prove by induction on the size of $\phi$ that $\mathcal{C}_2 \models_{\mathsf{hc}} \phi$. The cases $\phi = \langle\tau\rangle\,\phi'$ and $\phi = \wedge_{\{i \in I\}}\phi_i$ follow by the induction hypothesis. When $\phi = \mathtt{Hasco}(k)$ the proof follows by the consistency requirement of bisimulation. The only interesting case is the following:

*Case* $\phi = \langle x(a)\rangle\,\phi'$. By Definition 4.1, there exists cofinite set $N$ such that for all $l \in N$, there exists $\mathcal{C}'_1$ such that $\mathcal{C}_1 \stackrel{l(a)}{\Longrightarrow}\!\!\!\!\twoheadrightarrow \mathcal{C}'_1$ and $\mathcal{C}'_1 \models_{\mathsf{hc}} \phi'[l/x]$. If we remove from $N$ the finite support of $C_2$, that is $\mathit{ftn}(C_2)$, we get the cofinite set $N' = N \setminus \mathit{ftn}(C_2)$.

Let $l \in N'$; from the above we derive $\mathcal{C}'_1$ such that $\mathcal{C}_1 \stackrel{l(a)}{\Longrightarrow}\!\!\!\!\twoheadrightarrow \mathcal{C}'_1$ and $\mathcal{C}'_1 \models_{\mathsf{hc}} \phi'[l/x]$. Because $l \sharp C_2$, from the hypothesis and the transfer condition of the bisimulation we get $C'_2$ such that $C_2 \stackrel{l(a)}{\Longrightarrow}\!\!\!\!\twoheadrightarrow C'_2$ and $C'_1 \approx_{\mathsf{Hasco}} C'_2$. By the induction hypothesis we get $C'_2 \models_{\mathsf{hc}} \phi'[l/x]$.

Thus we have established that for all $l \in N'$ there exists $\mathcal{C}'_2$ such that $\mathcal{C}_2 \stackrel{l(a)}{\Longrightarrow}\!\!\!\!\twoheadrightarrow \mathcal{C}'_2$ and $\mathcal{C}'_2 \models_{\mathsf{hc}} \phi'[l/x]$. Therefore, by Definition 4.1 we get $\mathcal{C}_2 \models_{\mathsf{hc}} \langle x(a)\rangle\,\phi'$. $\Box$

**Proposition 4.5.** *If* $\mathcal{L}_{\mathsf{Hasco}}(\mathcal{C}_1) = \mathcal{L}_{\mathsf{Hasco}}(\mathcal{C}_2)$ *then* $\mathcal{C}_1 \approx_{\mathsf{Hasco}} \mathcal{C}_2$.

*Proof.* For the purposes of this proof let $\equiv_{\mathcal{L}}$ be the relation between configurations defined by letting $\mathcal{C}_1 \equiv_{\mathcal{L}} \mathcal{C}_2$ whenever, for all $\phi$ satisfying $\mathit{ftn}(\phi) \subseteq \mathit{ftn}(\mathcal{C}_1) \cup \mathit{ftn}(\mathcal{C}_2)$,

$$\mathcal{C}_1 \models_{\mathsf{hc}} \phi \text{ if and only if } \mathcal{C}_2 \models_{\mathsf{hc}} \phi$$

We show that the relation $\equiv_{\mathcal{L}}$ is a `Hasco`-bisimulation, from which the result follows.

The proof proceeds by contradiction. Note that $\mathcal{C}_1 \equiv_{\mathcal{L}} \mathcal{C}_2$ implies $\mathcal{C}_1$ `Hasco` $\mathcal{C}_2$ and so if $\equiv_{\mathcal{L}}$ is not a `Hasco`-bisimulation it must be that it does not satisfy the transfer condition in Definition 3.7. So without loss of generality suppose that $\mathcal{C}_1 \equiv_{\mathcal{L}} \mathcal{C}_2$ and that for some $\zeta \in \{\tau, l(a)\}, \mathcal{C}_1 \stackrel{\zeta}{\Rightarrow} \mathcal{C}_1'$ such that for every $\mathcal{C}_2 \stackrel{\zeta}{\Rightarrow} \mathcal{C}_2'$, $\mathcal{C}_1' \not\equiv_{\mathcal{L}} \mathcal{C}_2'$. We examine the case when $\zeta$ takes the form $l(a)$; the other case, when $\zeta = \tau$, is similar but simpler.

Let $S = \{\, \mathcal{C}_2^i \mid \mathcal{C}_2 \stackrel{l(a)}{\Longrightarrow} \mathcal{C}_2^i \,\}$. The language $\mathcal{L}_{\mathsf{Hasco}}$ is closed with respect to negation. So we can assume that for every $C_2^i \in S$ there exists some formula $\phi_i$ such that $\mathcal{C}_1' \models_{\mathsf{hc}} \phi_i$ but $\mathcal{C}_2^i \not\models_{\mathsf{hc}} \phi_i$. Moreover $\mathit{ftn}(\phi_i) \subseteq \mathit{ftn}(\mathcal{C}_1') \cup \mathit{ftn}(\mathcal{C}_2^i)$.

We use $I$ to index the configurations in $S$ and the corresponding formulae $\phi_i$. Then let $\phi$ denote the property $\wedge_{i \in I} \phi_i$. We therefore have $\mathcal{C}_1' \models_{\mathsf{hc}} \phi$ and $\mathcal{C}_2^i \not\models_{\mathsf{hc}} \phi$ for every $i \in I$. Also by Lemma 3.3 we know that $\mathit{ftn}(\phi) \subseteq \mathit{ftn}(\mathcal{C}_1) \cup \mathit{ftn}(\mathcal{C}_2) \cup \{l\}$.

Let $N$ denote $\mathsf{exTrN} \setminus (\mathit{ftn}(\mathcal{C}_1) \cup \mathit{ftn}(\mathcal{C}_2 \cup \{l\}))$, which is cofinite. For any $k \in N$ let $\pi_k$ denote the permutation which exchanges $k$ with $l$. Note that $\pi_k(\mathcal{C}_1) = \mathcal{C}_1$, $\pi_k(\mathcal{C}_1') = \mathcal{C}_1'[k/l]$ and $\pi_k(\phi') = \phi'[k/l]$. So by Proposition 3.5 $\mathcal{C}_1 \stackrel{k(a)}{\Longrightarrow} \mathcal{C}_1'[k/l]$ for all $k$ in the cofinite set $N$. Moreover, from Lemma 4.2, $\mathcal{C}_1'[k/l] \models_{\mathsf{hc}} \phi'[k/l]$.

Now choose a variable $x$ which does not appear in $\phi$. This is always possible, by renaming the countable set of variables which occur in $\phi$ if necessary. If $\psi$ denotes the formula which results from replacing all occurrences of $l$ in $\phi$ with $x$, then we have shown that $\mathcal{C}_1 \models_{\mathsf{hc}} \langle x(a) \rangle\, \psi$, since $\psi[k/x] = \phi[k/l]$.

Note that $\mathit{ftn}(\langle x(a) \rangle\, \psi) \subseteq \mathit{ftn}(\mathcal{C}_1) \cup \mathit{ftn}(\mathcal{C}_2)$. Therefore if we show

$$\mathcal{C}_2 \not\models_{\mathsf{hc}} \langle x(a) \rangle\, \psi$$

we will have established a contradiction to the assumption $\mathcal{C}_1 \equiv_{\mathcal{L}} \mathcal{C}_2$.

Let $N'$ be any cofinite set and suppose that for all $k' \in N'$, there exists $\mathcal{C}_2'$ such that $\mathcal{C}_2 \stackrel{k'(a)}{\Longrightarrow} \mathcal{C}_2'$ and $\mathcal{C}_2' \models_{\mathsf{hc}} \phi'[k'/l]$.

Pick some $k \in N'$ such that $k \notin \mathit{ftn}(\mathcal{C}_1) \cup \mathit{ftn}(\mathcal{C}_2)$; this is possible since all this set is finite. Consider the permutation $\pi_k$ defined above. Then $\pi_k(\mathcal{C}_2) = \mathcal{C}_2$, since $l, k \sharp \mathcal{C}_2$ and $\pi_k(\mathcal{C}_2') = \mathcal{C}_2'[l/k]$, since $l \sharp \mathcal{C}_2'$, and so by Proposition 3.5 $\mathcal{C}_2 \stackrel{l(a)}{\Longrightarrow} \mathcal{C}_2'[l/k]$.

Also since $l \notin \mathit{ftn}(\phi[k/l])$ by Lemma 4.2 $\mathcal{C}_2'[l/k] \models_{\mathsf{hc}} (\phi[k/l])[l/k] = \phi$.

However $\mathcal{C}_2'[l/k] \in S$ and we have already established that $\mathcal{C} \not\models_{\mathsf{hc}} \phi'$ whenever $\mathcal{C} \in S$. Therefore we have a contradiction and the proposition holds. $\square$

With these two propositions we have established

$$\mathcal{C}_1 \approx_{\mathsf{Hasco}} \mathcal{C}_2 \text{ if and only if } \mathcal{L}_{\mathsf{Hasco}}(\mathcal{C}_1) = \mathcal{L}_{\mathsf{Hasco}}(\mathcal{C}_2)$$

However because of the proof technique used in the latter we can obtain a slightly stronger result for processes, that is configurations of the form $\langle \emptyset; \varepsilon \bullet P \rangle$. Let $\mathcal{L}^{c}_{\mathsf{Hasco}}(P) = \{ \phi \in \mathcal{L}_{\mathsf{Hasco}}(\langle \emptyset; \varepsilon \bullet P \rangle) \mid \mathit{ftn}(\phi) = \emptyset \}$. Then one can also show that

$$P \approx_{\mathsf{Hasco}} Q \text{ if and only if } \mathcal{L}^{c}_{\mathsf{Hasco}}(P) = \mathcal{L}^{c}_{\mathsf{Hasco}}(Q)$$

*4.2. The Property Logic $\mathcal{L}_{\mathsf{Eq}}$*

The previous logic captured some aspects of past behaviour by interrogating the current state of configurations. But these configurations, in particular the extended histories $E; H$, contain more information about past events. For example the equivalence relation $E$ contains information about names which were originally independent but at some point were merged due to communications between transactions. Our new logic $\mathcal{L}_{\mathsf{Eq}}$ uses this information to interrogate more fully the past behaviour.

$$\begin{array}{rcll} \text{Properties: } \phi \in \mathcal{L}_{\mathsf{Eq}} & ::= & \langle \tau \rangle \, \phi \mid \langle x(a) \rangle \, \phi & \text{when } x \in \mathsf{Var} \\ & & \mid \neg \phi \mid \wedge_{\{i \in I\}} \phi_i & \\ & & \mid v_1 =_{\mathsf{co}} v_2 & \\ v \in \mathsf{Values} & ::= & l \in \mathsf{exTrN} \mid x \in \mathsf{Var} & \end{array}$$

Here we have replaced the predicates `Hasco` with the new $(v_1 =_{\mathsf{co}} v_2)$ stating that these two committed transaction names are *essentially the same*, in that they have been merged sometime in the past by a communication between transactions. However we must be very restrictive in when we make these intensional assertions.

**Definition 4.6** (Name consistent configurations)**.** We write $\Delta \models k =_{\mathsf{co}} k'$, where $\Delta = E; H$, whenever

- $k(\mathsf{co}) \in H$ and $k'(\mathsf{co}) \in H$
- $k \, E \, k'$

Then let $\Delta$ `Eq` $\Delta'$ if for all $k \in \mathsf{exTrN}$, $\Delta \models k =_{\mathsf{co}} k'$ if and only if $\Delta' \models k =_{\mathsf{co}} k'$.

This relation is extended to configurations in the standard manner. ♢

The definition of satisfaction $\mathcal{C} \models_{\mathsf{eq}} \phi$, for $\phi \in \mathcal{L}_{\mathsf{Eq}}$, is essentially the same as in Definition 4.1, with the one change in condition (1):

(1) $\mathcal{C} \models_{\mathsf{eq}} k =_{\mathsf{co}} k'$, whenever $\mathcal{C} = \langle \Delta \bullet P \rangle$ and $\Delta \models k =_{\mathsf{co}} k'$

Thus in the logic we can only assert that transaction names are *essentially* the same when the transactions have committed but not before that.

We use $\mathcal{L}_{\mathsf{Eq}}(\mathcal{C})$ to denote the set $\{ \phi \in \mathcal{L}_{\mathsf{Eq}} \mid \mathcal{C} \models_{\mathsf{eq}} \phi \}$, and $\mathcal{L}_{\mathsf{Eq}}(P)$ to denote the set $\{ \phi \mid \langle \emptyset; \varepsilon \bullet P \rangle \models_{\mathsf{eq}} \phi \}$.

**Example 4.7.** Consider the two processes defined by:

$$P_3 = [\![a.b.\,\mathsf{co} + b.a.\,\mathsf{co} \triangleright_k \mathbb{0}]\!]$$
$$Q_3 = [\![a.\,\mathsf{co} \triangleright_{k_1} \mathbb{0}]\!] \mid [\![b.\,\mathsf{co} \triangleright_{k_2} \mathbb{0}]\!]$$

Both processes can possibly execute an $a$-action followed by a $b$-action, or vice-versa. We can show that $P_3 \not\cong_{\mathsf{rbe}} Q_3$. Intuitively the reason for their different behaviour is quite straightforward. In $P_3$ both actions are within the scope of the same transaction, while in $Q_3$ they are in independent transactions.

We know that there is some formula $\phi \in \mathcal{L}_{\mathsf{Hasco}}$, which explains this difference; $P_3 \models_{\mathsf{hc}} \phi$ and $Q_3 \not\models_{\mathsf{hc}} \phi$. But $\phi$ must formalise this intuitive difference using the predicates which code up the fact that certain transactions have been committed and others have not. In fact this formula is

$$\phi = \langle x(a)\rangle\, \langle y(b)\rangle\, (\neg\, \mathtt{Hasco}(x) \wedge \neg\, \mathtt{Hasco}(y) \wedge [\tau](\mathtt{Hasco}(x) \leftrightarrow \mathtt{Hasco}(y)))$$

This expresses the fact that $P_3$, after performing an $a$- and a $b$-action can reach a state where the two actions are not committed and in any state reachable with internal transitions, the $a$-action is committed if and only if the $b$-action is committed.

However one can express this intuitive difference in a straightforward manner using $\mathcal{L}_{\mathsf{Eq}}$:

$$\langle \emptyset; \varepsilon \bullet P_3\rangle \models_{\mathsf{eq}} \langle x(a)\rangle\, \langle y(b)\rangle\; x =_{\mathsf{co}} y \qquad \langle \emptyset; \varepsilon \bullet Q_3\rangle \not\models_{\mathsf{eq}} \langle x(a)\rangle\, \langle y(b)\rangle\; x =_{\mathsf{co}} y$$

Thus $\mathcal{L}_{\mathsf{Eq}}(P_3) \neq \mathcal{L}_{\mathsf{Eq}}(Q_3)$. ◊

It is easy to see that the logic $\mathcal{L}_{\mathsf{Eq}}$ is at least as powerful as $\mathcal{L}_{\mathsf{Hasco}}$. For the predicate $\mathtt{Hasco}(v)$ from $\mathcal{L}_{\mathsf{Hasco}}$ can be modelled in $\mathcal{L}_{\mathsf{Eq}}$ using the predicate $v =_{\mathsf{co}} v$. In general it is strictly more powerful.

**Example 4.8.** Let $\mathcal{C}_1$, $\mathcal{C}_2$ be the configurations $\langle \mathrm{Id}; H \bullet \mathbb{0}\rangle$, $\langle \mathrm{U}; H \bullet \mathbb{0}\rangle$, where $H$ is the history $k(\mathsf{co}), l(\mathsf{co})$, Id is the identity relation over $\{l, k\}$, and $U$ is the universal relation over $\{l, k\}$.

Then $\mathcal{C}_1\, \mathtt{Hasco}\, \mathcal{C}_2$ because they have exactly the same committed transaction names, $k, l$. Since neither can perform any actions this means that $\mathcal{L}_{\mathsf{Hasco}}(\mathcal{C}_1) = \mathcal{L}_{\mathsf{Hasco}}(\mathcal{C}_2)$.

However $\mathcal{C}_2 \models_{\mathsf{eq}} l =_{\mathsf{co}} k$ whereas $\mathcal{C}_1 \not\models_{\mathsf{eq}} l =_{\mathsf{co}} k$, which means that $\mathcal{L}_{\mathsf{Eq}}(\mathcal{C}_1) \neq \mathcal{L}_{\mathsf{Eq}}(\mathcal{C}_2)$. ◊

We leave a more detailed discussion of the relationship between the two logics $\mathcal{L}_{\mathsf{Hasco}}$ and $\mathcal{L}_{\mathsf{Eq}}$ to Section 5. However we can easily adapt the definition of $\mathtt{Hasco}$-bisimulations so as to obtain one appropriate to $\mathcal{L}_{\mathsf{Eq}}$, by changing the first condition of Definition 3.7 to $\mathcal{C}_1\, \mathtt{Eq}\, \mathcal{C}_2$.

**Definition 4.9** ($\mathtt{Eq}$-Bisimulations)**.** A binary relation $\mathcal{R}$ over (extended) configurations is a (weak) $\mathtt{Eq}$-bisimulation when for all $\mathcal{C}_1\, \mathcal{R}\, \mathcal{C}_2$

1. $\mathcal{C}_1\, \mathtt{Eq}\, \mathcal{C}_2$

$$
\begin{array}{llll}
\langle \Delta \bullet P \rangle & \xrightarrow{\tau}\!\!\!\!\twoheadrightarrow \langle \Delta \bullet Q \rangle & \text{if} & P \xrightarrow{\tau}_{\varepsilon} Q \\
\langle E; H \bullet P \rangle & \xrightarrow{\tau}\!\!\!\!\twoheadrightarrow \langle \sigma(E); H \bullet Q \rangle & \text{if} & P \xrightarrow{k(\tau)}_{\sigma} Q,\ k \sharp E; H,\ k \in \mathsf{inTrN} \\
\langle \Delta \bullet P \rangle & \xrightarrow{\tau}\!\!\!\!\twoheadrightarrow \langle \Delta \bullet Q \rangle & \text{if} & P \xrightarrow{\mathsf{new}\, k} Q,\ k \sharp \Delta,\ k \in \mathsf{inTrN} \\
\langle E; H \bullet P \rangle & \xrightarrow{\mathsf{co}(E_{ext}(k))}\!\!\!\!\twoheadrightarrow \langle (E; H) \setminus_{\mathsf{co}} k \bullet Q \rangle & \text{if} & P \xrightarrow{\mathsf{co}\, k} Q,\ E_{ext}(k) \neq \emptyset \\
\langle E; H \bullet P \rangle & \xrightarrow{\tau}\!\!\!\!\twoheadrightarrow \langle (E; H) \setminus_{\mathsf{co}} k \bullet Q \rangle & \text{if} & P \xrightarrow{\mathsf{co}\, k} Q,\ E_{ext}(k) = \emptyset \\
\langle \Delta \bullet P \rangle & \xrightarrow{\tau}\!\!\!\!\twoheadrightarrow \langle \Delta \setminus_{\mathsf{ab}} k \bullet Q \rangle & \text{if} & P \xrightarrow{\mathsf{ab}\, k} Q \\
\langle E; H \bullet P \rangle & \xrightarrow{k(a)}\!\!\!\!\twoheadrightarrow \langle \sigma(E); H, k(a) \bullet Q \rangle & \text{if} & P \xrightarrow{k(a)}_{\sigma} Q,\ k \sharp E; H,\ k \in \mathsf{exTrN} \\
\langle E; H \bullet P \rangle & \xrightarrow{k(a)}\!\!\!\!\twoheadrightarrow \langle E \cup (k, k); H, k(\mathsf{ab}) \bullet P \rangle & \text{if} & k \sharp E; H,\ k \in \mathsf{exTrN}
\end{array}
$$

We define $\stackrel{\zeta}{\Rrightarrow}$ to be $(\xrightarrow{\tau}\!\!\twoheadrightarrow)^*$ when $\zeta$ is $\tau$ , and $\stackrel{\tau}{\Rrightarrow} \xrightarrow{\zeta}\!\!\twoheadrightarrow \stackrel{\tau}{\Rrightarrow}$ otherwise.

Figure 5: Commit-sensitive transitions.

2. (Transfer property)
   - (i) if $\mathcal{C}_1 \stackrel{\zeta}{\Rightarrow} \mathcal{C}_1'$ where $\zeta \sharp \mathcal{C}_2$ then $\mathcal{C}_2 \stackrel{\zeta}{\Rightarrow} \mathcal{C}_2'$ for some $\mathcal{C}_2'$ such that $\mathcal{C}_1' \mathrel{\mathcal{R}} \mathcal{C}_2'$,
   - (ii) the converse of condition (i)

$\mathsf{Eq}$-Bisimilarity ($\approx_{\mathsf{Eq}}$) is the largest $\mathsf{Eq}$-bisimulation over configurations, and extends to processes in the standard manner. ◊

**Proposition 4.10.** *$P \approx_{\mathsf{Eq}} Q$ if and only if $\mathcal{L}_{\mathsf{Eq}}(P) = \mathcal{L}_{\mathsf{Eq}}(Q)$.*

*Proof.* Virtually identical to that of Propositions 4.4 and 4.5. □

*4.3. The Property Logic $\mathcal{L}_{\mathsf{Canco}}$*

The two logics we have seen already both have mechanisms for interrogating the past behaviour, in addition to the more standard modal operators for predicting future behaviour. We now design a logic with no constructs for interrogating the past, thereby obtaining a more standard property logic over an LTS. Of course this revised LTS needs to be more complicated if we are to retain the distinguishing power of the other logics.

In $\mathcal{L}_{\mathsf{Hasco}}$ there are operators for seeing if particular transactions have been committed sometime in the past. Here these are replaced by operators which interrogate if transactions names can be committed in the present; that is we make commit actions observable.

The new LTS is shown in Figure 5. We let $E_{ext}(k)$ denote the set $\{\, k' \in \mathsf{exTrN} \mid k \mathrel{E} k' \,\}$, where $E$ is an equivalence relation over $\mathsf{TrN}$. In the internal transition $\mathcal{C}_1 \stackrel{\tau}{\Rrightarrow} \mathcal{C}_2$ we are assured that no commits have been made of external names, that is names which are shared with the environment. The essential novelty of the LTS are the *commit* transitions of the form $\mathcal{C}_1 \stackrel{\mathsf{co}(K)}{\Rrightarrow} \mathcal{C}_2$. This may be broken down into $\mathcal{C}_1 \stackrel{\tau}{\Rrightarrow} \mathcal{C}_1' \xrightarrow{\mathsf{co}(K)}\!\!\twoheadrightarrow \mathcal{C}_2' \stackrel{\tau}{\Rrightarrow} \mathcal{C}_2$ where

- in the internal moves from $\mathcal{C}_1$ to $\mathcal{C}_1'$ and from $\mathcal{C}_2'$ to $\mathcal{C}_2$ there can only be commits of *internal* transaction names, those which do not appear in the history components of the configurations

- in the transition from $\mathcal{C}_1'$ to $\mathcal{C}_2'$ there is a single commit move of some active transaction name $k$, inferred from the rule TrCo in Figure 1, where $K$ is the non-empty set of external names which are equivalent to $k$.

**Lemma 4.11** (Sanity Check 2). *Suppose*

$$\mathcal{C}_1 = \langle E_1; H_1 \bullet P_1 \rangle \xrightarrow{\zeta}\!\!\!\!\!\twoheadrightarrow \mathcal{C}_2 = \langle E_2; H_2 \bullet P_2 \rangle$$

*and $\mathcal{C}_1$ is an extended configuration. Then*[2]

*(1) $\mathcal{C}_2$ is also an extended configuration*

*(2) if $\zeta \in \{\tau, k(a)\}$ then*

*(i) $\zeta \sharp \mathcal{C}_1$ and $eftn(\mathcal{C}_2) \subseteq eftn(\mathcal{C}_1) \cup ftn(\zeta)$*

*(ii) $E_1 \subseteq E_2$, $\mathtt{IsR}(\mathcal{C}_2) \subseteq \mathtt{IsR}(\mathcal{C}_1) \cup ftn(\zeta)$, and $\mathtt{Hasco}(\mathcal{C}_2) = \mathtt{Hasco}(\mathcal{C}_1)$*

*(3) if $\zeta = \mathsf{co}\, A$ then*

*(i) $A \subseteq \mathtt{IsR}(\mathcal{C}_1)$ and $A \times A \subseteq E_1$ and $eftn(\mathcal{C}_2) = eftn(\mathcal{C}_1)$*

*(ii) $E_1 = E_2$, $\mathtt{IsR}(\mathcal{C}_2) = \mathtt{IsR}(\mathcal{C}_1) \setminus A$, and $\mathtt{Hasco}(\mathcal{C}_2) = \mathtt{Hasco}(\mathcal{C}_1) \uplus A$*

*(4) if $l, l' \in \mathtt{Hasco}(\mathcal{C}_1)$ and $l\, E_2\, l'$ then $l\, E_1\, l'$.*

*Proof.* As in Lemma 3.3. □

We can now design a variation on our logics for this new LTS, which has no past operators but instead a modal operator for the external commit transitions:

$$\begin{array}{rll} \phi \in \mathcal{L}_{\mathsf{Canco}} ::= & \langle\tau\rangle\,\phi \mid \langle x(a)\rangle\,\phi & \text{when } x \in \mathsf{Var} \\ & \mid \neg\phi \mid \bigwedge_{\{i\in I\}} \phi_i \mid \langle \mathsf{co}(K)\rangle\,\phi & \text{when } \emptyset \neq K \subseteq \mathsf{exTrN} \uplus \mathsf{Var} \end{array}$$

The satisfaction relation for closed formulae, $\mathcal{C} \models_{\mathsf{cc}} \phi$, is defined by adapting Definition 4.1, using the clause

(1) $\mathcal{C} \models_{\mathsf{cc}} \langle \mathsf{co}(K)\rangle\,\phi$ whenever $\mathcal{C} \xrightarrow{\mathsf{co}(K)}\!\!\!\!\!\twoheadrightarrow \mathcal{C}'$ such that $\mathcal{C}' \models_{\mathsf{cc}} \phi'$.

As usual $\mathcal{L}_{\mathsf{Canco}}(\mathcal{C})$ denotes the set $\{\, \phi \in \mathcal{L}_{\mathsf{Canco}} \mid \mathcal{C} \models_{\mathsf{cc}} \phi \,\}$, and $\mathcal{L}_{\mathsf{Canco}}(P)$ abbreviates $\mathcal{L}_{\mathsf{Canco}}(\langle \emptyset; \varepsilon \bullet P \rangle \models_{\mathsf{cc}} \phi)$.

**Example 4.12.** Consider again $P_2, Q_2$ from Example 4.3, and the derivation

$$\langle \emptyset; \varepsilon \bullet Q_2 \rangle \xrightarrow{m_1(a)}\!\!\!\!\!\twoheadrightarrow \xrightarrow{m_2(b)}\!\!\!\!\!\twoheadrightarrow \xrightarrow{\tau}\!\!\!\!\!\twoheadrightarrow \langle E; m_1(a), m_2(b) \bullet \nu p.\, [\![\mathsf{co} \rhd_{m_3} \mathbb{0}]\!] \mid [\![\mathsf{co} \rhd_{m_3} \mathbb{0}]\!] \rangle = \mathcal{C}_Q$$

Because $m_1\, E\, m_2$ we have $\mathcal{C}_Q \models_{\mathsf{cc}} \langle \mathsf{co}(\{m_1, m_2\})\rangle\, \mathtt{true}$. This reasoning holds for any pair of distinct names $m_1, m_2$, thus $Q_2 \models_{\mathsf{cc}} \langle x(a)\rangle\, \langle y(b)\rangle\, \langle \mathsf{co}(\{x, y\})\rangle\, \mathtt{true}$.

[2] Property (2) is used in Theorems 5.3 and 6.4 and Lemma 5.4; property (3) is used in Theorems 5.3, 6.4 and 6.7 and Lemma 5.4; and property (4) is used in Lemma 5.4.

However if $\langle \varepsilon \bullet P_2 \rangle \xrightarrow{m_1(a)}\!\!\!\!\!\!\twoheadrightarrow \xrightarrow{m_2(a)}\!\!\!\!\!\!\twoheadrightarrow \mathcal{C}_P$ then, regardless of the choice of $m_1, m_2$, it will never be the case that $\mathcal{C}_P \models_{\mathsf{cc}} \langle \mathtt{co}(\{m_1, m_2\})\rangle\, \mathtt{true}$. This in turn means that

$$P_2 \not\models_{\mathsf{cc}} \langle x(a)\rangle \langle y(b)\rangle \langle \mathtt{co}(\{x, y\})\rangle\, \mathtt{true}$$

thus $\mathcal{L}_{\mathsf{Canco}}(P) \neq \mathcal{L}_{\mathsf{Canco}}(Q)$. ◊

There is an obvious bisimulation equivalence ($\approx_{\mathsf{Canco}}$), based on that in Definition 2.8, which uses the commit sensitive transitions in Figure 5 and contains no intentional predicate on configurations. For the sake of clarity we spell it out.

**Definition 4.13** (Commit-sensitive bisimulation)**.** A binary relation $\mathcal{R}$ over configurations is a commit sensitive bisimulation when for all $\mathcal{C}_1 \,\mathcal{R}\, \mathcal{C}_2$,

(i) for all $\zeta \in \{\tau, k(a)\}$, if $\mathcal{C}_1 \stackrel{\zeta}{\Rrightarrow} \mathcal{C}_1'$ where $\zeta \,\sharp\, \mathcal{C}_2$, then $\mathcal{C}_2 \stackrel{\zeta}{\Rrightarrow} \mathcal{C}_2'$ for some $\mathcal{C}_2'$ such that $\mathcal{C}_1' \,\mathcal{R}\, \mathcal{C}_2'$

(ii) for all $K \subseteq \mathsf{exTrN}$, if $\mathcal{C}_1 \xrightarrow{\mathtt{co}(K)}\!\!\!\!\!\!\twoheadrightarrow \mathcal{C}_1'$, then $\mathcal{C}_2 \xrightarrow{\mathtt{co}(K)}\!\!\!\!\!\!\twoheadrightarrow \mathcal{C}_2'$ for some $\mathcal{C}_2'$ such that $\mathcal{C}_1' \,\mathcal{R}\, \mathcal{C}_2'$,

(iii) the converse of conditions (i) (ii).

We use $\approx_{\mathsf{Canco}}$ to denote the largest commit sensitive bisimulation over configurations; this is extended to processes in the standard manner. ◊

Note that this variation on bisimulations does not interrogate the past behaviour, as recorded in configurations, although it does use their equivalence relations in order to interpret the novel modal operator $\langle \mathtt{co}\, K\rangle$. Indeed it is essentially the default notion of bisimulation for the LTS generated by the transitions from Figure 5.

**Proposition 4.14.** *$\mathcal{C}_1 \approx_{\mathsf{Canco}} \mathcal{C}_2$ if and only if $\mathcal{L}_{\mathsf{Canco}}(\mathcal{C}_1) = \mathcal{L}_{\mathsf{Canco}}(\mathcal{C}_2)$.*

*Proof.* Similar to that of Propositions 4.4 and 4.5, although somewhat simpler. □

## 5. Distinguishability

A general logic $\mathcal{L}$ can be considered to be a set of formulae, together with a *satisfaction* relation between configurations and formulae, written $\mathcal{C} \models^{\mathcal{L}} \phi$, for $\phi \in \mathcal{L}$. We use $\mathcal{L}(\mathcal{C})$ to denote the set $\{\, \phi \in \mathcal{L} \mid \mathcal{C} \models^{\mathcal{L}} \phi \,\}$, and abbreviate $\mathcal{L}(\langle \emptyset; \varepsilon \bullet P\rangle)$ to $\mathcal{L}(P)$. Then we write $P \equiv_{\mathcal{L}} Q$ whenever $\mathcal{L}(P) = \mathcal{L}(Q)$; this defines an equivalence relation over processes.

**Definition 5.1** (Distinguishing power of logics)**.** For any two logics $\mathcal{L}_1$, $\mathcal{L}_2$ we write $\mathcal{L}_1 \preceq \mathcal{L}_2$ whenever $\equiv_{\mathcal{L}_2} \subseteq \equiv_{\mathcal{L}_1}$. ◊

The intuition here is that if $\mathcal{L}_1 \preceq \mathcal{L}_2$ then any two processes which can be distinguished by a formula in $\mathcal{L}_1$ can also be distinguished by a formula from $\mathcal{L}_2$. Suppose $P \models^{\mathcal{L}_1} \phi_1$ and $Q \not\models^{\mathcal{L}_1} \phi_1$ for some $\phi_1 \in \mathcal{L}_1$. Then $P \not\equiv_{\mathcal{L}_1} Q$ and so, since $\mathcal{L}_1 \preceq \mathcal{L}_2$, $P \not\equiv_{\mathcal{L}_2} Q$. This means that there exists a formula $\phi_2 \in \mathcal{L}_2$ such that $P \models^{\mathcal{L}_2} \phi_2$ and $Q \not\models^{\mathcal{L}_2} \phi_2$, or vice-versa.

We have already remarked, on page 24, that $\mathcal{L}_{\mathsf{Hasco}} \preceq \mathcal{L}_{\mathsf{Eq}}$. Despite Example 4.8, which involves arbitrary configurations, we will also show the converse which involves processes. A direct proof is not straightforward. Instead we first concentrate on relating $\mathcal{L}_{\mathsf{Hasco}}$ with $\mathcal{L}_{\mathsf{Canco}}$ via their associated bisimulations. Here, the central issue is to understand when the transitions $\mathcal{C}_1 \stackrel{\zeta}{\Rightarrow} \mathcal{C}_2$, used in the definition of $\mathcal{L}_{\mathsf{Hasco}}$, can be transformed into transitions used in the definition of $\mathcal{L}_{\mathsf{Canco}}$, $\mathcal{C}_1 \stackrel{\zeta}{\Rrightarrow} \mathcal{C}_2$.

**Lemma 5.2.** *Suppose* $\mathcal{C}_1 \stackrel{\zeta}{\twoheadrightarrow} \mathcal{C}_2$, *where* $\zeta \in \{\tau, k(a)\}$.

*(1) If* $\mathtt{Hasco}(\mathcal{C}_2) = \mathtt{Hasco}(\mathcal{C}_1)$ *then* $\mathcal{C}_1 \stackrel{\zeta}{\ggg} \mathcal{C}_2$.

*(2) Otherwise* $\zeta$ *is* $\tau$, $\mathtt{Hasco}(\mathcal{C}_2) = \mathtt{Hasco}(\mathcal{C}_1) \uplus A$ *for some set* $A$ *such that* $\mathcal{C}_1 \xrightarrow{\mathsf{co}(A)}\!\!\!\!\!\!\ggg \mathcal{C}_2$.

*Proof.* By inspection of Figure 4 and Figure 5. □

**Theorem 5.3.** $\mathcal{C}_1 \approx_{\mathsf{Hasco}} \mathcal{C}_2$ *implies* $\mathcal{C}_1 \approx_{\mathsf{Canco}} \mathcal{C}_2$.

*Proof.* We show that $\approx_{\mathsf{Hasco}}$ satisfies the requirements of Definition 4.13.

Let $\mathcal{C}_1 \approx_{\mathsf{Hasco}} \mathcal{C}_2$ and $\mathcal{C}_1 \stackrel{\zeta}{\ggg} \mathcal{C}_1'$. We show that $\mathcal{C}_2 \stackrel{\zeta}{\Rrightarrow} \mathcal{C}_2'$ such that $\mathcal{C}_1' \approx_{\mathsf{Hasco}} \mathcal{C}_2'$; this transfer condition will be sufficient to establish the more general one for the weak transitions $\mathcal{C}_1 \stackrel{\zeta}{\Rrightarrow} \mathcal{C}_1'$, and the result will follow by symmetry. There are two cases.

- Suppose $\zeta \neq \mathsf{co}(A)$. Then by Figures 4 and 5 and Lemma 4.11 (2), $\mathcal{C}_1 \stackrel{\zeta}{\twoheadrightarrow} \mathcal{C}_1'$ and $\mathtt{Hasco}(\mathcal{C}_1') = \mathtt{Hasco}(\mathcal{C}_1)$. Since $\mathcal{C}_1 \approx_{\mathsf{Hasco}} \mathcal{C}_2$, there is some transition $\mathcal{C}_2 \stackrel{\zeta}{\Rightarrow} \mathcal{C}_2'$ such that $\mathcal{C}_1' \approx_{\mathsf{Hasco}} \mathcal{C}_2'$, which in turn implies that $\mathtt{Hasco}(\mathcal{C}_2') = \mathtt{Hasco}(\mathcal{C}_2)$.

  So by Lemma 5.2 (1) $\mathcal{C}_2 \stackrel{\zeta}{\Rrightarrow} \mathcal{C}_2'$, which is the required matching transition.

- Suppose $\zeta = \mathsf{co}(A)$. Then $\mathcal{C}_1 \stackrel{\tau}{\twoheadrightarrow} \mathcal{C}_1'$ and since $\mathcal{C}_1 \approx_{\mathsf{Hasco}} \mathcal{C}_2$ we have a matching transition $\mathcal{C}_2 \stackrel{\tau}{\Rightarrow} \mathcal{C}_2'$ such that $\mathcal{C}_1' \approx_{\mathsf{Hasco}} \mathcal{C}_2'$. By Lemma 4.11 (3) we have $\mathtt{Hasco}(\mathcal{C}_1') = \mathtt{Hasco}(\mathcal{C}_1) \uplus A$, thus $\mathtt{Hasco}(\mathcal{C}_2') = \mathtt{Hasco}(\mathcal{C}_2) \uplus A$.

  We will show that the weak transition $\mathcal{C}_2 \stackrel{\tau}{\Rightarrow} \mathcal{C}_2'$ commits the names in $A$ with a single commit move. We distinguish the first commit move in this weak transition. Note that there is at least one commit move because $A$ is non-empty. There exist $C_2''$ and $C_2'''$, non-empty $A_1$ and $A_2$ such that $A_1 \uplus A_2 = A$ and $\mathcal{C}_2 \stackrel{\tau}{\Rightarrow} C_2'' \stackrel{\tau}{\twoheadrightarrow} C_2''' \stackrel{\tau}{\Rightarrow} \mathcal{C}_2'$ with $\mathtt{Hasco}(\mathcal{C}_2) = \mathtt{Hasco}(C_2'')$ and $\texttt{Hasco}(C_2''') = \texttt{Hasco}(C_2'') \uplus A_1$ and $\texttt{Hasco}(C_2') = \texttt{Hasco}(C_2''') \uplus A_2 = \texttt{Hasco}(C_2) \uplus A_1 \uplus A_2$.

If $A_2 = \emptyset$ then the proof is completed by deriving $\mathcal{C}_2 \stackrel{\tau}{\Rrightarrow} C_2'' \xrightarrow{\mathsf{co}(A)}\!\!\!\!\!\twoheadrightarrow C_2''' \stackrel{\tau}{\Rrightarrow} C_2'$ using Lemma 5.2.

Otherwise, $A_2 \neq \emptyset$ leads to a contradiction: because of the transition $\mathcal{C}_2 \stackrel{\tau}{\Rrightarrow} \mathcal{C}_2'''$ and $\mathcal{C}_1 \approx_{\mathsf{Hasco}} \mathcal{C}_2$, there exists $\mathcal{C}_1'''$ such that $\mathcal{C}_1 \stackrel{\tau}{\Rrightarrow} \mathcal{C}_1'''$ and $\mathcal{C}_1''' \approx_{\mathsf{Hasco}} \mathcal{C}_2'''$. Therefore, by the first condition of Definition 3.7, $\texttt{Hasco}(C_1''') = \texttt{Hasco}(C_2''') = \texttt{Hasco}(C_2) \uplus A_1$. We take $k \in A_1$ and $l \in A_2$. By Figure 5 and the transition $\mathcal{C}_1 \xrightarrow{\mathsf{co}(A)}\!\!\!\!\!\twoheadrightarrow \mathcal{C}_1'$, we have $k\ E_1\ l$, where $E_1$ is the equivalence relation of $\mathcal{C}_1$. By Lemma 3.3 (3) and the transition $\mathcal{C}_1 \stackrel{\tau}{\Rrightarrow} \mathcal{C}_1'''$, we get $k\ E_1'''\ l$, where $E_1'''$ is the equivalence relation of $\mathcal{C}_1'''$. By condition (ii) of well-formedness (Definition 3.1), $l \in \texttt{Hasco}(C_1''')$ which contradicts the assumption that $A_2$ is disjoint from $A_1$ and $\texttt{Hasco}(C_2)$. Thus this case is vacuously true. □

We now turn our attention to relating $\approx_{\mathsf{Canco}}$ and $\approx_{\mathsf{Eq}}$, which is considerably more straightforward. The following lemma shows how the predicate $\mathtt{Eq}$ interacts with the transitions of the form $\mathcal{C} \stackrel{\zeta}{\Rrightarrow} \mathcal{C}'$, used in the definition of $\approx_{\mathsf{Canco}}$.

**Lemma 5.4.**

*(1) Suppose $\mathcal{C} \stackrel{\zeta}{\Rrightarrow} \mathcal{C}'$, where $\zeta \in \{\tau, k(a)\}$. Then $\mathcal{C} \models l =_{\mathsf{co}} l'$ if and only if $\mathcal{C}' \models l =_{\mathsf{co}} l'$.*

*(2) Suppose $\mathcal{C} \xrightarrow{\mathsf{co}(A)}\!\!\!\!\!\twoheadrightarrow \mathcal{C}'$. Then*

- *(a) $\mathcal{C} \models l =_{\mathsf{co}} k$ implies $\mathcal{C}' \models l =_{\mathsf{co}} k$*
- *(b) $\mathcal{C}' \models l =_{\mathsf{co}} k$ for every $l, k \in A$*
- *(c) $\mathcal{C}' \models l =_{\mathsf{co}} k$ implies $\mathcal{C} \models l =_{\mathsf{co}} k$, or $l, k \in A$.*

*Proof.* (1) The forward direction follows immediately from Lemma 4.11 (2). Conversely suppose $\mathcal{C}' \models l =_{\mathsf{co}} l'$. By Lemma 4.11 (2), $\texttt{Hasco}(\mathcal{C}') = \texttt{Hasco}(\mathcal{C})$ and therefore $l, l' \in \texttt{Hasco}(\mathcal{C})$. Lemma 4.11 (4) ensures that $\mathcal{C} \models l =_{\mathsf{co}} l'$.

(2) Suppose $\mathcal{C} \xrightarrow{\mathsf{co}(A)}\!\!\!\!\!\twoheadrightarrow \mathcal{C}'$. By Lemma 4.11 (3), $\texttt{Hasco}(\mathcal{C}') = \texttt{Hasco}(\mathcal{C}) \uplus A$. Again parts (2)(a) and (2)(b) follow from Lemma 4.11 (2). For part (2)(c) suppose $\mathcal{C}' \models l =_{\mathsf{co}} l'$. Then $l, l' \in \texttt{Hasco}(\mathcal{C}) \uplus A$ and $l\,E\,l'$, where $E$ is the common equivalence relation in both configurations $\mathcal{C}, \mathcal{C}'$. If one of the names, say $l$, is in $\texttt{Hasco}(\mathcal{C})$ then by the well-formedness of configurations, Definition 3.1, so is $l'$, and therefore by definition $\mathcal{C} \models l =_{\mathsf{co}} l'$. Otherwise both are in $A$, as required. □

**Theorem 5.5.** *Suppose $\mathcal{C}_1 \,\mathtt{Eq}\, \mathcal{C}_2$. Then $\mathcal{C}_1 \approx_{\mathsf{Canco}} \mathcal{C}_2$ implies $\mathcal{C}_1 \approx_{\mathsf{Eq}} \mathcal{C}_2$.*

*Proof.* Let $\mathcal{R}$ be the relation defined by: $\mathcal{C}_1 \;\mathcal{R}\; \mathcal{C}_2$ when $\mathcal{C}_1 \,\mathtt{Eq}\, \mathcal{C}_2$ and $\mathcal{C}_1 \approx_{\mathsf{Canco}} \mathcal{C}_2$. We show that $\mathcal{R}$ satisfies the conditions of Definition 4.9.

By definition, condition (1) is satisfied. So we look at the Transfer property. Suppose $\mathcal{C}_1 \;\mathcal{R}\; \mathcal{C}_2$ and $\mathcal{C}_1 \xrightarrow{\zeta}\!\!\!\twoheadrightarrow \mathcal{C}_1'$ where $\zeta \in \{\tau, k(a)\}$ and is fresh from $\mathcal{C}_2$. We show that $\mathcal{C}_2 \overset{\zeta}{\Rightarrow}\!\!\!\twoheadrightarrow \mathcal{C}_2'$ for some $\mathcal{C}_2'$ such that $\mathcal{C}_1' \;\mathcal{R}\; \mathcal{C}_2'$. The more general transfer property, for $\mathcal{C}_1 \overset{\zeta}{\Rightarrow}\!\!\!\twoheadrightarrow \mathcal{C}_1'$, will follow by induction on the length of this weak transition, and condition (2)(ii) of Definition 4.9 will follow by symmetry. There are two cases:

- $\mathtt{Hasco}(\mathcal{C}_1') = \mathtt{Hasco}(\mathcal{C}_1)$: By Lemma 5.2(1) we have that $\mathcal{C}_1 \xrightarrow{\zeta}\!\!\!\twoheadrightarrow\!\!\!\!\rightarrow \mathcal{C}_1'$, and since $\mathcal{C}_1 \approx_{\mathsf{Canco}} \mathcal{C}_2$ there is some transition $\mathcal{C}_2 \overset{\zeta}{\Rightarrow}\!\!\!\twoheadrightarrow\!\!\!\!\rightarrow \mathcal{C}_2'$ such that $\mathcal{C}_1' \approx_{\mathsf{Canco}} \mathcal{C}_2'$. Since $\mathtt{Hasco}(\mathcal{C}_2') = \mathtt{Hasco}(\mathcal{C}_2)$ one can check that $\mathcal{C}_2 \overset{\zeta}{\Rightarrow}\!\!\!\twoheadrightarrow \mathcal{C}_2'$. So it only remains to show that $\mathcal{C}_1' \,\mathtt{Eq}\, \mathcal{C}_2'$. However this follows easily from the fact that $\mathtt{Hasco}(\mathcal{C}_1') = \mathtt{Hasco}(\mathcal{C}_1)$ and Lemma 5.4(1).

- $\mathtt{Hasco}(\mathcal{C}_1') = \mathtt{Hasco}(\mathcal{C}_1) \uplus A$, for some non-empty set $A$; in this case $\zeta$ is $\tau$. Here we use Lemma 5.2(2), obtaining a transition $\mathcal{C}_1 \xrightarrow{\mathsf{co}(A)}\!\!\!\twoheadrightarrow\!\!\!\!\rightarrow \mathcal{C}_1'$, and a matching transition from $\mathcal{C}_2$, $\mathcal{C}_2 \overset{\mathsf{co}(A)}{\Longrightarrow}\!\!\!\twoheadrightarrow\!\!\!\!\rightarrow \mathcal{C}_2'$ such that $\mathcal{C}_1' \approx_{\mathsf{Canco}} \mathcal{C}_2'$. Again it is easy to see that $\mathcal{C}_2 \overset{\tau}{\Rightarrow}\!\!\!\twoheadrightarrow \mathcal{C}_2'$ and so we have to establish $\mathcal{C}_1' \,\mathtt{Eq}\, \mathcal{C}_2'$.

  Suppose $\mathcal{C}_1' \models l =_{\mathsf{co}} k$. By Lemma 5.4(2)(c) either $\mathcal{C}_1 \models l =_{\mathsf{co}} k$ or $l, k \in A$. In the latter case Lemma 5.4(2)(b), together with repeated applications of part (1), ensures that $\mathcal{C}_2' \models l =_{\mathsf{co}} k$. In the former $\mathtt{Hasco}(\mathcal{C}_1) = \mathtt{Hasco}(\mathcal{C}_2)$ ensures that $\mathcal{C}_2 \models l =_{\mathsf{co}} k$, and by Lemma 5.4(2)(a), and repeated applications of part (1), we obtain the required $\mathcal{C}_2' \models l =_{\mathsf{co}} k$.

  The converse, $\mathcal{C}_2' \models l =_{\mathsf{co}} k$ implies $\mathcal{C}_1' \models l =_{\mathsf{co}} k$, is virtually identical. □

We now sum up the results of this section on distinguishability:

**Theorem 5.6.** $\mathcal{L}_{\mathsf{Canco}} \preceq \mathcal{L}_{\mathsf{Hasco}} \preceq \mathcal{L}_{\mathsf{Eq}} \preceq \mathcal{L}_{\mathsf{Canco}}$.

*Proof.* The first, $\mathcal{L}_{\mathsf{Canco}} \preceq \mathcal{L}_{\mathsf{Hasco}}$, follows from Theorem 5.3, together with Proposition 4.14, and Propositions 4.4 and 4.5. The second, $\mathcal{L}_{\mathsf{Hasco}} \preceq \mathcal{L}_{\mathsf{Eq}}$, has been explained on page 24, while the third, $\mathcal{L}_{\mathsf{Eq}} \preceq \mathcal{L}_{\mathsf{Canco}}$, follows from Theorem 5.5, together with Propositions 4.10 and 4.14. □

This result also means that the three bisimulations over processes coincide.

**Corollary 5.7.** $(\approx_{\mathsf{Hasco}}) = (\approx_{\mathsf{Eq}}) = (\approx_{\mathsf{Canco}})$.

## 6. Expressiveness

We have shown that the three logics $\mathcal{L}_{\mathsf{Hasco}}$, $\mathcal{L}_{\mathsf{Eq}}$ and $\mathcal{L}_{\mathsf{Canco}}$ are each individually sufficiently powerful to explain behavioural differences between processes.

They can also be compared with respect to their expressiveness; that is the complexity of the properties which they can describe. This is difficult to capture in general, but *relative expressiveness* is straightforward to formalise.

**Definition 6.1** (Relative expressiveness of logics)**.** For any two logics $\mathcal{L}_1$, $\mathcal{L}_2$ we write $\mathcal{L}_1 \preceq_{\text{exp}} \mathcal{L}_2$ if for every $\phi \in \mathcal{L}_1$ there exists some $\phi' \in \mathcal{L}_2$ such that for any process $P$, $P \models^{\mathcal{L}_1} \phi$ if and only if $P \models^{\mathcal{L}_2} \phi'$. We say that $\mathcal{L}_2$ is at least as expressive as $\mathcal{L}_1$. $\Diamond$

It is straightforward to see that $\mathcal{L}_{\mathsf{Eq}}$ is at least as expressive as $\mathcal{L}_{\mathsf{Hasco}}$. For $\phi \in \mathcal{L}_{\mathsf{Hasco}}$ let $\mathcal{L}_{\mathsf{Eq}}(\phi) \in \mathcal{L}_{\mathsf{Eq}}$ be the result of replacing all occurrences of $\texttt{Hasco}(v)$ with $v =_{\mathsf{co}} v$. As discussed on page 24, $\mathcal{C} \models_{\overline{\mathsf{hc}}} \phi$ iff $\mathcal{C} \models_{\overline{\mathsf{eq}}} \mathcal{L}_{\mathsf{Eq}}(\phi)$, for any configuration $\mathcal{C}$.

We now show that $\mathcal{L}_{\mathsf{Hasco}}$ is at least as expressive as $\mathcal{L}_{\mathsf{Canco}}$.

**Example 6.2.** Consider again the formula $\phi = \langle x(a)\rangle \langle y(b)\rangle \langle \mathsf{co}(\{x, y\})\rangle \, \texttt{true}$, from $\mathcal{L}_{\mathsf{Canco}}$, used in Example 4.12 to differentiate between processes $P_2, Q_2$. An equivalent formula in $\mathcal{L}_{\mathsf{Hasco}}$ is given by

$$\phi' = \langle x(a)\rangle \langle y(b)\rangle(\texttt{nHasco}(x, y) \wedge [\tau]\, \texttt{EquiCo}(x, y) \wedge \langle\tau\rangle\, \texttt{Hasco}(x, y))$$

where $\texttt{nHasco}(x, y)$ and $\texttt{EquiCo}(x, y)$ abbreviate $\neg\texttt{Hasco}(x) \wedge \neg\texttt{Hasco}(y)$ and $\texttt{Hasco}(x) \leftrightarrow \texttt{Hasco}(y)$, respectively. $\Diamond$

A general structural translation from $\mathcal{L}_{\mathsf{Canco}}$ into $\mathcal{L}_{\mathsf{Hasco}}$ has to take into account some properties of the environment in which the formula is being asserted. For instance in the above example, when translating the sub-formula $\langle y(b)\rangle \langle \mathsf{co}(\{x, y\})\rangle \, \texttt{true}$ we assume that some transaction bound to $x$ has performed some external transition, and has not yet committed. Tracking which transactions have not yet committed is carried out by systematic use of the predicate $\texttt{Hasco}(-)$ from $\mathcal{L}_{\mathsf{Hasco}}$ but we also need to know what transactions have been created. This is the role of the parameter $R$ in the definition of $[\![\cdot]\!]^{\mathsf{ch}}_R$ in Figure 6, which is assumed to be a subset of $\mathsf{exTrN} \cup \mathsf{Var}$. In the following lemma and theorem, however, $R$ is a subset of $\mathsf{exTrN}$.

**Lemma 6.3.** *Let $A$ and $R$ be subsets of* $\mathsf{exTrN}$*. Suppose* $\mathcal{C} \models_{\overline{\mathsf{hc}}} \texttt{nHasco}(A \cup R) \wedge [\tau]\, \texttt{EquiCo}(A)$ *where* $\texttt{IsR}(\mathcal{C}) \subseteq R$*. Then* $\mathcal{C} \xRightarrow{\tau} \mathcal{C}' \models_{\overline{\mathsf{hc}}} \texttt{Hasco}(A) \wedge \texttt{nHasco}(R \backslash A)$ *implies* $\mathcal{C} \xRightarrow{\mathsf{co}(A)} \mathcal{C}'$*.* $\Box$

**Theorem 6.4.** *Let* $R \subseteq \mathsf{exTrN}$ *be such that* $\texttt{IsR}(\mathcal{C}) \subseteq R$ *and* $\texttt{Hasco}(\mathcal{C}) \sharp R$*. Then for every closed* $\phi \in \mathcal{L}_{\mathsf{Canco}}$*,* $\mathcal{C} \models_{\overline{\mathsf{cc}}} \phi$ *if and only if* $\mathcal{C} \models_{\overline{\mathsf{hc}}} [\![\phi]\!]^{\mathsf{ch}}_R$*.*

*Proof.* By induction on the size of $\phi$, taking cases on its syntax. Here we consider the size of $\phi$ to be equal to the size of $\phi[k/x]$, for any transaction name $k$ and variable $x$. We start with the most difficult case, when it has the form $\langle \mathsf{co}(A)\rangle \, \phi'$.

**Translation** $[\![\cdot]\!]^{\mathsf{ch}}_{R} \in \mathcal{L}_{\mathsf{Canco}} \to \mathcal{L}_{\mathsf{Hasco}}$

$$[\![\langle \mathtt{co}(A)\phi\rangle]\!]^{\mathsf{ch}}_{R} = \langle\tau\rangle \Big(\mathtt{nHasco}(A \cup R) \wedge [\tau]\,\mathtt{EquiCo}(A) \wedge \langle\tau\rangle(\psi_{A,R} \wedge [\![\phi]\!]^{\mathsf{ch}}_{R\backslash A})\Big)$$

$$[\![\langle x(a)\rangle\,\phi]\!]^{\mathsf{ch}}_{R} = \langle x(a)\rangle \Big(\mathtt{nHasco}(R \cup \{x\}) \wedge [\![\phi]\!]^{\mathsf{ch}}_{R\cup\{x\}}\Big)$$

$$[\![\langle\tau\rangle\,\phi]\!]^{\mathsf{ch}}_{R} = \langle\tau\rangle \Big(\mathtt{nHasco}(R) \wedge [\![\phi]\!]^{\mathsf{ch}}_{R}\Big)$$

$$[\![\wedge_{i\in I}\phi_i]\!]^{\mathsf{ch}}_{R} = \bigwedge\nolimits_{i\in I}[\![\phi_i]\!]^{\mathsf{ch}}_{R}$$

$$[\![\neg\phi]\!]^{\mathsf{ch}}_{R} = \neg[\![\phi]\!]^{\mathsf{ch}}_{R}$$

where:

$$\mathtt{Hasco}(S) = \wedge_{s\in S}\mathtt{Hasco}(s)$$

$$\mathtt{EquiCo}(S) = \wedge_{s,s'\in S}(\mathtt{Hasco}(s) \leftrightarrow \mathtt{Hasco}(s'))$$

$$\mathtt{nHasco}(S) = \wedge_{s\in S}\neg\mathtt{Hasco}(s)$$

$$\psi_{A,R} = \mathtt{Hasco}(A) \wedge \mathtt{nHasco}(R\backslash A)$$

---

**Translation** $[\![\cdot]\!]^{\mathsf{ec}}_{R,E} \in \mathcal{L}_{\mathsf{Eq}} \to \mathcal{L}_{\mathsf{Canco}}$

$$[\![v_1 =_{\mathsf{co}} v_2]\!]^{\mathsf{ec}}_{R,E} = \mathtt{true} \text{ if } E \models v_1 =_{\mathsf{co}} v_2 \text{ and } \mathtt{false} \text{ otherwise}$$

$$[\![\langle x(a)\rangle\,\phi]\!]^{\mathsf{ec}}_{R,E} = \bigvee\nolimits_{\emptyset\neq A\subseteq R} \langle \mathtt{co}\ A\rangle[\![\langle x(a)\rangle\,\phi]\!]^{\mathsf{ec}}_{R\backslash A,E\cup A\times A} \ \vee\ \langle x(a)\rangle[\![\langle\tau\rangle\,\phi]\!]^{\mathsf{ec}}_{R\cup\{x\},E}$$

$$[\![\langle\tau\rangle\,\phi]\!]^{\mathsf{ec}}_{R,E} = \bigvee\nolimits_{\emptyset\neq A\subseteq R} \langle \mathtt{co}\ A\rangle[\![\langle\tau\rangle\,\phi]\!]^{\mathsf{ec}}_{R\backslash A,E\cup A\times A} \ \vee\ \langle\tau\rangle[\![\phi]\!]^{\mathsf{ec}}_{R,E}$$

$$[\![\bigwedge\nolimits_{i\in I}\phi_i]\!]^{\mathsf{ec}}_{R,E} = \bigwedge\nolimits_{i\in I}[\![\phi_i]\!]^{\mathsf{ec}}_{R,E}$$

$$[\![\neg\phi]\!]^{\mathsf{ec}}_{R,E} = \neg[\![\phi]\!]^{\mathsf{ec}}_{R,E}$$

Figure 6: Translations between logic formulae.

- Suppose $\mathcal{C} \models_{\mathsf{cc}} \langle\mathtt{co}(A)\rangle\,\phi'$. Then $\mathcal{C} \overset{\tau}{\Rrightarrow} \mathcal{C}_1 \xrightarrow{\mathtt{co}(A)} \mathcal{C}_2 \overset{\tau}{\Rrightarrow} \mathcal{C}_3$, where $\mathcal{C}_3 \models_{\mathsf{cc}} \phi'$. We show that $\mathcal{C}_1 \models_{\mathsf{hc}} \mathtt{nHasco}(A\cup R) \wedge [\tau]\,\mathtt{EquiCo}(A) \wedge \langle\tau\rangle(\psi_{A,R} \wedge [\![\phi]\!]^{\mathsf{ch}}_{R\backslash A})$, from which the required $\mathcal{C} \models_{\mathsf{hc}} [\![\langle\mathtt{co}(A)\rangle\,\phi']\!]^{\mathsf{ch}}_{R}$ will follow.

  By Lemma 4.11 (3), we have $\mathtt{Hasco}(\mathcal{C}) = \mathtt{Hasco}(\mathcal{C}_1)$ and $\mathtt{Hasco}(\mathcal{C}_3) = \mathtt{Hasco}(\mathcal{C}_2) = \mathtt{Hasco}(\mathcal{C}_1) \uplus A$. From the same lemma, $\mathtt{IsR}(\mathcal{C}_1) \subseteq \mathtt{IsR}(\mathcal{C})$ and $\mathtt{IsR}(\mathcal{C}_3) \subseteq \mathtt{IsR}(\mathcal{C}_2) = \mathtt{IsR}(\mathcal{C}_1) \setminus A$ and $A \subseteq \mathtt{IsR}(\mathcal{C}_1)$.

  Therefore, $A \cup R \sharp \mathtt{Hasco}(\mathcal{C}_1)$ and thus $\mathcal{C}_1 \models_{\mathsf{hc}} \mathtt{nHasco}(A \cup R)$.

  Next suppose $\mathcal{C}_1 \overset{\tau}{\Rrightarrow} \mathcal{C}'$. Because $\mathcal{C}_1 \xrightarrow{\mathtt{co}(A)} \mathcal{C}_2$, by Lemma 4.11 (3), $A \subseteq E$, where $E$ is the common equivalence relation of $\mathcal{C}_1$ and $\mathcal{C}_2$. By Lemma 4.11 (2), $A \subseteq E'$, where $E'$ is the equivalence relation of $\mathcal{C}'$. By property (ii) of Definition 3.1, we establish $\mathcal{C}' \models_{\mathsf{hc}} \mathtt{EquiCo}(A)$; it follows that $\mathcal{C}_1 \models_{\mathsf{hc}} [\tau]\,\mathtt{EquiCo}(A)$.

  Finally we show that $\mathcal{C}_3 \models_{\mathsf{hc}} \psi_{A,R}$ and $\mathcal{C}_3 \models_{\mathsf{hc}} [\![\phi']\!]^{\mathsf{ch}}_{R\backslash A}$. We observe $\mathtt{Hasco}(\mathcal{C}_3) \sharp (R \setminus A)$ and $\mathtt{IsR}(\mathcal{C}_3) \subseteq R \setminus A$, which implies $\mathcal{C}_3 \models_{\mathsf{hc}} \psi_{A,R}$ and by the induction hypothesis $\mathcal{C}_3 \models_{\mathsf{hc}} [\![\phi']\!]^{\mathsf{ch}}_{R\backslash A}$.

- Conversely suppose $\mathcal{C} \models_{\mathsf{hc}} [\![\langle\mathtt{co}(A)\rangle\,\phi']\!]^{\mathsf{ch}}_{R}$. This means that $\mathcal{C} \overset{\tau}{\Rrightarrow} \mathcal{C}_1 \overset{\tau}{\Rrightarrow} \mathcal{C}_2$ where $\mathcal{C}_1 \models_{\mathsf{hc}} \mathtt{nHasco}(A \cup R) \wedge [\tau]\, \mathtt{EquiCo}(A)$ and $\mathcal{C}_2 \models_{\mathsf{hc}} \psi^2_{A,R} \wedge [\![\phi']\!]^{\mathsf{ch}}_{R\setminus A}$.

By Lemma 6.3 for $\mathcal{C}_1$, we obtain $\mathcal{C}_1 \overset{\mathsf{co}(A)}{\Longrightarrow} \mathcal{C}_2$. Because $\mathcal{C}_1 \models_{\mathsf{hc}} \mathtt{nHasco}(A \cup R)$ we can also argue that $\mathtt{Hasco}(\mathcal{C}_1) = \mathtt{Hasco}(\mathcal{C})$ and so by Lemma 5.2 we have $\mathcal{C} \overset{\tau}{\Longrightarrow} \mathcal{C}_1$, which in turn means that $\mathcal{C} \overset{\mathsf{co}(A)}{\Longrightarrow} \mathcal{C}_2$.

By Lemma 4.11 (3), $\mathtt{Hasco}(\mathcal{C}_2) = \mathtt{Hasco}(\mathcal{C}) \uplus A$, and therefore $\mathtt{Hasco}(\mathcal{C}_2) \sharp R \setminus A$. From the same lemma, $\mathtt{IsR}(\mathcal{C}_2) \subseteq \mathtt{IsR}(\mathcal{C}) \subseteq R$. So we can apply the induction hypothesis to obtain $\mathcal{C}_2 \models_{\mathsf{cc}} \phi'$. It therefore follows that $\mathcal{C} \models_{\mathsf{cc}} \langle \mathsf{co}(A) \rangle\, \phi'$, as required.

Now consider the case when $\phi$ is $\langle x(a) \rangle\, \phi$.

- Suppose $\mathcal{C} \models_{\mathsf{cc}} \langle x(a) \rangle\, \phi'$. By definition there exists a cofinite set $K$ such that for all $k \in K$, $\mathcal{C} \overset{k(a)}{\Longrightarrow} \mathcal{C}'$ such that $\mathcal{C}' \models_{\mathsf{cc}} \phi'[k/x]$. By Lemma 4.11 (2), $\mathtt{Hasco}(\mathcal{C}') = \mathtt{Hasco}(\mathcal{C})$ and $k \sharp \mathtt{Hasco}(\mathcal{C}')$. Therefore, $\mathcal{C}' \models_{\mathsf{hc}} \mathtt{nHasco}(R \cup \{k\})$ and $\mathtt{Hasco}(\mathcal{C}') \sharp (R \cup \{k\})$. Moreover, from the same lemma, $\mathtt{IsR}(\mathcal{C}') \subseteq \mathtt{IsR}(\mathcal{C}) \cup \{k\} = R \cup \{k\}$. It follows by repeated application of Lemma 5.2 that $\mathcal{C} \overset{k(a)}{\Longrightarrow} \mathcal{C}'$. By the induction hypothesis we get $\mathcal{C} \models_{\mathsf{hc}} [\![\phi'[k/x]]\!]^{\mathsf{ch}}_{R\cup\{k\}}$. This is true for every $k \in K$, and therefore by definition $\mathcal{C} \models_{\mathsf{hc}} [\![\langle x(a) \rangle\, \phi']\!]^{\mathsf{ch}}_{R}$.
- Conversely suppose $\mathcal{C} \models_{\mathsf{hc}} [\![\langle x(a) \rangle\, \phi']\!]^{\mathsf{ch}}_{R}$, and that there is some cofinite set $K$ such that for all $k \in K$, $\mathcal{C} \overset{k(a)}{\Longrightarrow} \mathcal{C}'$ for some $\mathcal{C}'$ such that $\mathcal{C}' \models_{\mathsf{hc}} \mathtt{nHasco}(R \cup \{k\})$ and $\mathcal{C}' \models_{\mathsf{hc}} [\![\phi]\!]^{\mathsf{ch}}_{R\cup\{x\}}[k/x] = [\![\phi[k/x]]\!]^{\mathsf{ch}}_{R\cup\{k\}}$. Using Lemma 3.3 (3) together with the facts that $\mathtt{Hasco}(\mathcal{C}) \sharp R$ and $\mathcal{C}' \models_{\mathsf{hc}} \mathtt{nHasco}(R \cup \{k\})$ and $\mathtt{IsR}(\mathcal{C}) \subseteq R$ one can calculate that $\mathtt{Hasco}(\mathcal{C}') = \mathtt{Hasco}(\mathcal{C})$. Then repeated application of Lemma 5.2 gives $\mathcal{C} \overset{k(a)}{\Longrightarrow} \mathcal{C}'$. Lemma 4.11 (2) ensures that $\mathtt{IsR}(\mathcal{C}') \subseteq R \cup \{k\}$ and therefore by induction $\mathcal{C}' \models_{\mathsf{cc}} \phi'[k/x]$.

  This argument works for all $k$ in $K$ and therefore by definition $\mathcal{C} \models_{\mathsf{cc}} \langle x(a) \rangle\, \phi$.

The case when $\phi$ has the form $\langle \tau \rangle\, \phi'$ is similar, while the remaining cases follow directly by structural induction. □

**Corollary 6.5.** *For every closed $\phi \in \mathcal{L}_{\mathsf{Canco}}$, $P \models_{\mathsf{cc}} \phi$ if and only if $P \models_{\mathsf{hc}} [\![\phi]\!]^{\mathsf{ch}}_{\emptyset}$.*

We now proceed to show that $\mathcal{L}_{\mathsf{Canco}}$ is at least as expressive as $\mathcal{L}_{\mathsf{Eq}}$, by providing yet another structural translation $[\![\cdot]\!]^{\mathsf{ec}}_{R,E}$, shown in Figure 6. This again uses the set of running transactions $R \subseteq \mathsf{exTrN} \cup \mathsf{Var}$, but also requires to keep track of the equivalence relation $E \subseteq (\mathsf{exTrN} \cup \mathsf{Var}) \times (\mathsf{exTrN} \cup \mathsf{Var})$, which equates the transaction names that committed simultaneously by the commit modalities. Compared to the previous translation, this one can give rise to a considerably larger blow-up in the size of the resulting formulae. This is because for each $\mathcal{L}_{\mathsf{Eq}}$ diamond operator, it creates an $\mathcal{L}_{\mathsf{Canco}}$ disjunction that exhaustively explores all scenarios of possible transactional commits that can occur simultaneously.

**Example 6.6.** Consider the $\mathcal{L}_{\mathsf{Eq}}$ formula $\phi = \langle x(a)\rangle \langle y(b)\rangle (x =_{\mathsf{co}} y)$ which distinguishes the processes $P_2$, $Q_2$ from the introduction. This can be translated to an equivalent $\mathcal{L}_{\mathsf{Canco}}$ formula according to the above translation $[\![\phi]\!]^{\mathsf{ec}}_{\emptyset,\emptyset}$:

$$\begin{aligned}
[\![\phi]\!]^{\mathsf{ec}}_{\emptyset,\emptyset} &= \langle x(a)\rangle [\![\langle\tau\rangle \langle y(b)\rangle (x =_{\mathsf{co}} y)]\!]^{\mathsf{ec}}_{\{x\},\emptyset} \\
&= \langle x(a)\rangle \big( \langle \mathtt{co}\{x\}\rangle [\![\langle\tau\rangle \langle y(b)\rangle (x =_{\mathsf{co}} y)]\!]^{\mathsf{ec}}_{\emptyset,E_x} \vee \langle\tau\rangle [\![\langle y(b)\rangle (x =_{\mathsf{co}} y)]\!]^{\mathsf{ec}}_{\{x\},\emptyset}\big) \\
&= \langle x(a)\rangle \big( \langle \mathtt{co}\{x\}\rangle \langle\tau\rangle [\![\langle y(b)\rangle (x =_{\mathsf{co}} y)]\!]^{\mathsf{ec}}_{\emptyset,E_x} \\
&\vee \langle\tau\rangle \langle \mathtt{co}\{x\}\rangle [\![\langle y(b)\rangle (x =_{\mathsf{co}} y)]\!]^{\mathsf{ec}}_{\emptyset,E_x} \vee \langle\tau\rangle \langle y(b)\rangle [\![\langle\tau\rangle (x =_{\mathsf{co}} y)]\!]^{\mathsf{ec}}_{\{x,y\},\emptyset}\big) \\
&= \langle x(a)\rangle \big( \langle \mathtt{co}\{x\}\rangle \langle\tau\rangle \langle y(b)\rangle [\![\langle\tau\rangle (x =_{\mathsf{co}} y)]\!]^{\mathsf{ec}}_{\{y\},E_x} \\
&\vee \langle\tau\rangle \langle \mathtt{co}\{x\}\rangle \langle y(b)\rangle [\![\langle\tau\rangle (x =_{\mathsf{co}} y)]\!]^{\mathsf{ec}}_{\{y\},E_x} \vee \langle\tau\rangle \langle y(b)\rangle [\![\langle\tau\rangle (x =_{\mathsf{co}} y)]\!]^{\mathsf{ec}}_{\{x,y\},\emptyset}\big)
\end{aligned}$$

Where $E_x = \{x = x\}$. This formula is semantically equivalent to

$$\langle x(a)\rangle \Big( \langle \mathtt{co}\{x\}\rangle \langle y(b)\rangle [\![\langle\tau\rangle (x =_{\mathsf{co}} y)]\!]^{\mathsf{ec}}_{\{y\},E_x} \vee \langle y(b)\rangle [\![\langle\tau\rangle (x =_{\mathsf{co}} y)]\!]^{\mathsf{ec}}_{\{x,y\},\emptyset}\Big)$$

Continuing the translation we have:

$$\begin{aligned}
[\![\langle\tau\rangle (x =_{\mathsf{co}} y)]\!]^{\mathsf{ec}}_{\{y\},E_x} &= \langle \mathtt{co}\{y\}\rangle \langle\tau\rangle [\![(x =_{\mathsf{co}} y)]\!]^{\mathsf{ec}}_{\emptyset,E} \vee \langle\tau\rangle [\![(x =_{\mathsf{co}} y)]\!]^{\mathsf{ec}}_{\{y\},E_x} \\
[\![\langle\tau\rangle (x =_{\mathsf{co}} y)]\!]^{\mathsf{ec}}_{\{x,y\},\emptyset} &= \langle \mathtt{co}\{x,y\}\rangle \langle\tau\rangle [\![(x =_{\mathsf{co}} y)]\!]^{\mathsf{ec}}_{\emptyset,E_{xy}} \\
&\vee \langle \mathtt{co}\{x\}\rangle \langle \mathtt{co}\{y\}\rangle \langle\tau\rangle [\![(x =_{\mathsf{co}} y)]\!]^{\mathsf{ec}}_{\emptyset,E} \\
&\vee \langle \mathtt{co}\{x\}\rangle \langle\tau\rangle [\![(x =_{\mathsf{co}} y)]\!]^{\mathsf{ec}}_{\{y\},E_x} \\
&\vee \langle \mathtt{co}\{y\}\rangle \langle \mathtt{co}\{x\}\rangle \langle\tau\rangle [\![(x =_{\mathsf{co}} y)]\!]^{\mathsf{ec}}_{\emptyset,E} \\
&\vee \langle \mathtt{co}\{y\}\rangle \langle\tau\rangle [\![(x =_{\mathsf{co}} y)]\!]^{\mathsf{ec}}_{\{x\},E_y} \\
&\vee \langle\tau\rangle [\![(x =_{\mathsf{co}} y)]\!]^{\mathsf{ec}}_{\{x,y\},\emptyset}
\end{aligned}$$

where $E$ and $E_{xy}$ are the identity and universe relations over $\{x, y\}$, respectively, and $E_y = \{y = y\}$. Only $[\![(x =_{\mathsf{co}} y)]\!]^{\mathsf{ec}}_{\emptyset,E_{xy}}$ translates to `true`, the rest translate to `false`. $\Diamond$

In the following theorem, when $\mathcal{C} = \langle E; H \bullet P\rangle$, $E_{\mathtt{Hasco}}(\mathcal{C})$ denotes the restriction of $E$ to the committed external names in $\mathcal{C}$. That is, $E_{\mathtt{Hasco}}(\mathcal{C}) = E \cap (\mathtt{Hasco}(C) \times \mathtt{Hasco}(C))$.

**Theorem 6.7.** *Let $R$ be a finite set and $S$ an equivalence relation over* $\mathsf{exTrN}$ *such that* $\mathtt{IsR}(\mathcal{C}) \subseteq R$ *and* $E_{\mathtt{Hasco}}(\mathcal{C}) \subseteq S$. *Then for every closed* $\phi \in \mathcal{L}_{\mathsf{Eq}}$, $\mathcal{C} \models_{\mathsf{eq}} \phi$ *if and only if* $\mathcal{C} \models_{\mathsf{cc}} [\![\phi]\!]^{\mathsf{ec}}_{R,S}$.

*Proof.* By lexicographic induction on the size of $\phi$ and the cardinality of $R$, taking cases on the syntax of $\phi$. Here we consider the size of $\phi$ to be equal to the size of $\phi[k/x]$, for any transaction name $k$ and variable $x$. We also consider the size of $\langle x(a)\rangle \phi$ to be strictly larger than the size of $\langle\tau\rangle \phi$.

The only nontrivial cases are the two diamond operators; here we only analyse the more involved case when $\phi$ is of the form $\langle x(a)\rangle \phi'$.

- Suppose $\mathcal{C} \models_{\mathsf{eq}} \langle x(a)\rangle\, \phi$. From the semantics of the satisfaction relation, we know that there is a cofinite set $K \subseteq \mathsf{exTrN}$ such that for all $k \in K$, $\mathcal{C} \overset{\tau}{\Rightarrow} \mathcal{C}_1 \xrightarrow{k(a)} \mathcal{C}_2 \overset{\tau}{\Rightarrow} \mathcal{C}_3$, where $\mathcal{C}_3 \models_{\mathsf{eq}} \phi[k/x]$. We distinguish two cases for the weak transition $\mathcal{C} \overset{\tau}{\Rightarrow} \mathcal{C}_1$:

  - The committed names in $\mathcal{C}$ and $\mathcal{C}_1$, and all intermediate configurations are the same; i.e., $\mathtt{Hasco}(\mathcal{C}) = \mathtt{Hasco}(\mathcal{C}_1)$. In this case, we show that $\mathcal{C} \models_{\mathsf{cc}} [\![\langle x(a)\rangle\, \phi]\!]^{\mathsf{ec}}_{R,S}$ because $\mathcal{C} \models_{\mathsf{cc}} \langle x(a)\rangle [\![\phi]\!]^{\mathsf{ec}}_{R\cup\{x\},S}$.

    By repeated applications of Lemma 5.2 (1), we get $\mathcal{C} \overset{\tau}{\Rrightarrow} \mathcal{C}_1$. By Lemma 3.3 (3), $\mathtt{IsR}(\mathcal{C}_1) \subseteq \mathtt{IsR}(\mathcal{C}) \subseteq R$, and by Lemma 3.3 (4), $E_{\mathtt{Hasco}}(\mathcal{C}_1) = E_{\mathtt{Hasco}}(\mathcal{C}) \subseteq S$. By Figures 4 and 5, $\mathcal{C}_1 \xrightarrow{k(a)} \mathcal{C}_2$, and $\mathtt{IsR}(\mathcal{C}_2) = \mathtt{IsR}(\mathcal{C}_1) \cup \{k\} \subseteq R \cup \{k\}$ and $E_{\mathtt{Hasco}}(\mathcal{C}_2) = E_{\mathtt{Hasco}}(\mathcal{C}_1) \subseteq S$. Moreover, by definition, $\mathcal{C}_2 \models_{\mathsf{eq}} \langle\tau\rangle\, \phi[k/x]$. Therefore, by the induction hypothesis,

    $$\mathcal{C}_2 \models_{\mathsf{cc}} [\![\langle\tau\rangle\, \phi[k/x]]\!]^{\mathsf{ec}}_{R\cup\{k\},S} = \left([\![\langle\tau\rangle\, \phi]\!]^{\mathsf{ec}}_{R\cup\{x\},S}\right)[k/x]$$

    Since this holds for all $k \in K$, by definition, $\mathcal{C} \models_{\mathsf{cc}} \langle x(a)\rangle [\![\langle\tau\rangle\, \phi]\!]^{\mathsf{ec}}_{R\cup\{x\},S}$.

  - When the committed names in $\mathcal{C}$ and $\mathcal{C}_1$ are different, we show $\mathcal{C} \models_{\mathsf{cc}} [\![\langle x(a)\rangle\, \phi]\!]^{\mathsf{ec}}_{R,S}$ because there exists $\emptyset \neq A \subseteq R$ such that $\mathcal{C} \models_{\mathsf{cc}} \langle \mathsf{co}\, A\rangle [\![\langle x(a)\rangle\, \phi]\!]^{\mathsf{ec}}_{R\setminus A,\, S\cup A\times A}$.

    We analyse the weak transition $\mathcal{C} \overset{\tau}{\Rightarrow} \mathcal{C}_1$ as follows:

    $$\mathcal{C} \overset{\tau}{\Rightarrow} \mathcal{C}_0 \xrightarrow{\tau} \mathcal{C}_0' \overset{\tau}{\Rightarrow} \mathcal{C}_1$$

    where, by Lemma 5.2, $\mathtt{Hasco}(\mathcal{C}) = \mathtt{Hasco}(\mathcal{C}_0)$ and $\mathtt{Hasco}(\mathcal{C}_0') = \mathtt{Hasco}(\mathcal{C}_0) \uplus A$, for some $A$ such that $\mathcal{C}_0 \xrightarrow{\mathsf{co}\, A} \mathcal{C}_0'$. In other words, $\mathcal{C}_0 \xrightarrow{\tau} \mathcal{C}_0'$ is the first commit transition in the sequence of $\tau$-transitions from $\mathcal{C}$ to $\mathcal{C}_1$.

    By repeated applications of Lemma 5.2 (1), $\mathcal{C} \overset{\tau}{\Rrightarrow} \mathcal{C}_0$. Thus, $\mathcal{C} \overset{\mathsf{co}\, A}{\Rrightarrow} \mathcal{C}_0'$. By Lemma 3.3 (3), $\mathtt{IsR}(\mathcal{C}_0') \subseteq \mathtt{IsR}(\mathcal{C}) \subseteq R$, and by Lemma 3.3 (4), $E_{\mathtt{Hasco}}(\mathcal{C}_0) = E_{\mathtt{Hasco}}(\mathcal{C}) \subseteq S$. By Lemma 4.11 (3), $\mathtt{IsR}(\mathcal{C}_0') \subseteq R \setminus A$ and $E_{\mathtt{Hasco}}(\mathcal{C}_0') = E_0 \cap (\mathtt{Hasco}(\mathcal{C}_0) \uplus A)^2$, where $E_0$ is the equivalence relation in $\mathcal{C}_0$. Because none of the names in $A$ is related with a committed name in $\mathcal{C}_0$, $E_{\mathtt{Hasco}}(\mathcal{C}_0') = E_0 \cap (\mathtt{Hasco}(\mathcal{C}_0)^2 \uplus A^2) = E_{\mathtt{Hasco}}(\mathcal{C}_0) \uplus (E_0 \cap A^2) \subseteq S \cup (A \times A)$. Moreover, $\mathcal{C}_0' \models_{\mathsf{eq}} \langle x(a)\rangle\, \phi$. Therefore, by the induction hypothesis, $\mathcal{C}_0' \models_{\mathsf{cc}} [\![\langle x(a)\rangle\, \phi]\!]^{\mathsf{ec}}_{R\setminus A,\, S\cup A\times A}$ and by definition $\mathcal{C} \models_{\mathsf{cc}} \langle \mathsf{co}\, A\rangle [\![\langle x(a)\rangle\, \phi]\!]^{\mathsf{ec}}_{R\setminus A,\, S\cup A\times A}$.

- Conversely, suppose $\mathcal{C} \models_{\mathsf{cc}} [\![\langle x(a)\rangle\, \phi]\!]^{\mathsf{ec}}_{R,S}$. There are two cases:

  - $\mathcal{C} \models_{\mathsf{cc}} \langle x(a)\rangle [\![\phi]\!]^{\mathsf{ec}}_{R\cup\{x\},S}$. Here, by definition, for all $k$ from some cofinite set $K$, $\mathcal{C} \overset{k(a)}{\Rrightarrow} \mathcal{C}_1$ with $\mathcal{C}_1 \models_{\mathsf{cc}} [\![\phi[k/x]]\!]^{\mathsf{ec}}_{R\cup\{k\},S}$. As above, from the definitions of the two transition systems (Figures 4 and 5), $\mathcal{C} \overset{k(a)}{\Longrightarrow} \mathcal{C}_1$ and $\mathtt{IsR}(\mathcal{C}_2) = \mathtt{IsR}(\mathcal{C}_1) \cup \{k\} \subseteq R \cup \{k\}$ and $E_{\mathtt{Hasco}}(\mathcal{C}_2) = E_{\mathtt{Hasco}}(\mathcal{C}_1) \subseteq S$. Therefore, by the induction hypothesis, $\mathcal{C}_1 \models_{\mathsf{eq}} \phi[k/x]$. This is true for every $k \in K$, thus by definition, $\mathcal{C} \models_{\mathsf{eq}} \langle x(a) \rangle\, \phi$.

– $\mathcal{C} \models_{\mathsf{cc}} \langle \mathtt{co}\, A \rangle [\![ \langle x(a) \rangle\, \phi ]\!]^{\mathsf{ec}}_{R \setminus A,\, S \cup A \times A}$ for some non-empty $A \subseteq \mathtt{Hasco}(\mathcal{C})$. In this case, $\mathcal{C} \overset{\mathtt{co}\, A}{\Longrightarrow} \mathcal{C}'_0$ with $\mathcal{C}'_0 \models_{\mathsf{cc}} [\![ \langle x(a) \rangle\, \phi ]\!]^{\mathsf{ec}}_{R \setminus A,\, S \cup A \times A}$. As before, $\mathcal{C} \overset{\tau}{\Longrightarrow} \mathcal{C}'_0$ and $\mathtt{IsR}(\mathcal{C}'_0) \subseteq R \setminus A$ and $E_{\mathtt{Hasco}}(\mathcal{C}'_0) \subseteq R \cup (A \times A)$. Therefore, by the induction hypothesis, $\mathcal{C}'_0 \models_{\mathsf{eq}} \langle x(a) \rangle\, \phi$, and by definition, $\mathcal{C} \models_{\mathsf{eq}} \langle x(a) \rangle\, \phi$. □

**Corollary 6.8.** *For every closed $\phi \in \mathcal{L}_{\mathsf{Eq}}$, $P \models_{\mathsf{eq}} \phi$ if and only if $P \models_{\mathsf{cc}} [\![\phi]\!]^{\mathsf{ec}}_{\emptyset,\emptyset}$.*

As a summary of the results in this section, we obtain the following theorem stating that all three logics have the same relative expressiveness.

**Theorem 6.9.** $\mathcal{L}_{\mathsf{Eq}} \preceq_{exp} \mathcal{L}_{\mathsf{Canco}} \preceq_{exp} \mathcal{L}_{\mathsf{Hasco}} \preceq_{exp} \mathcal{L}_{\mathsf{Eq}}$.

*Proof.* The first two expressiveness inclusions follow from Corollaries 6.8 and 6.5, respectively. The third inclusion is straightforward, as discussed at the beginning of this section. □

Note that this result involves processes; Example 4.8 shows that it does not hold for arbitrary configurations.

## 7. Conclusions

We have extended the classical theory associating the property logic HML with *CCS* processes to the language *TCCS*$^m$ for defining communicating transactions. We are not aware of other work on property logics for transactions that communicate.

We have provided three extensions to standard HML, each containing modal operators for examining future behaviour. Two of the logics have operators for examining past behaviour, encoded in a novel notion of extended configuration, which remembers the status of transactions, and the relationship between their names. In addition we have used a novel *nominal* interpretation of the forward looking modal operators of standard HML, based on the ideas of [12, 21].

For future work we intend to evaluate the usefulness of these three logics for giving coherent explanations for the difference in the behaviour of processes. This will be done by developing algorithms which take descriptions of two transactions, and either return a bisimulation containing the pair, or a distinguishing formula from one of our logics. Such algorithms already exist, and are implemented, for finite-state *CCS* and other process description languages, [4, 13, 2]. One major obstacle here is to develop a useful notion of finite-state transactions. The semantics for *TCCS*$^m$, given in Figure 4, associates an infinite LTS to even the simplest transactions, such as $\mathtt{rec}X.[\![a.\,\mathtt{co} \blacktriangleright X]\!]$. This is because of the freshness requirements in the LTS rules. Relaxing such freshness conditions may allow us to associate finite LTSs with a significant subset of $TCCS^m$. Ideas from [24] may be useful in this.

Previous work [20] gives a uniform framework for HML-like logics over *nominal transition systems*. With Proposition 3.5 it is possible to show that our LTSs generated by the transitions in both Figure 4 and Figure 5 satisfy the requirements of nominal transition systems. Thus, in principle, variations of our novel property logics could be formulated using the framework of [20]. The use of this framework may alleviate some of the proof burden in Section 4; however these results (Propositions 4.4, 4.5, 4.10 and 4.14) are fairly straightforward to derive from first principles. The major results of this paper, those in Sections 5 and 6, are not addressed by [20]. Moreover, these alternative property logics are quite different in style to those developed here. For example satisfaction would be defined over alpha-equivalence classes of configurations, and attention would have to be restricted to *finitely-supported* formulae. Our property logics tolerate formulae with infinite names and support. Nevertheless it would be interesting to compare the two approaches. Again this would be best carried out in terms of the algorithms already alluded to above.

## Appendix A. Properties of Process Transitions

We must first derive some properties of the reduction semantics given in Figure 1. To this end, for any name substitution $\sigma = (\widetilde{k} \mapsto l)$ and name permutation $\pi$ let $\sigma_\pi$ be the name substitution $(\pi(\widetilde{k}) \mapsto \pi(l))$.

**Lemma A.1.** *Let $E$ be an equivalence relation and $\sigma$ a name substitution satisfying $range(\sigma) \sharp E$.*

1. $\pi \cdot \sigma = \sigma_\pi \cdot \pi$
2. $k\,\sigma(E)\,k'$ *if and only if, either*
   - *i)* $k\,E\,k'$, *or*
   - *ii)* $k, k' \in E(dom(\sigma)) \cup dom(\sigma) \cup range(\sigma)$,

   *where* $E(dom(\sigma))$ *denotes* $\{\, l \mid l\,E\,l' \text{ for some } l' \in dom(\sigma) \,\}$
3. $\sigma_\pi(E_\pi) = (\sigma(E))_\pi$.[3]

*Proof.* The proof of both (1) and (2) is by calculation. Then for (3) the characterisation in (2) can be used to prove $\sigma_\pi(E_\pi) \subseteq (\sigma(E))_\pi$ and $(\sigma(E))_\pi \subseteq \sigma_\pi(E_\pi)$. □

**Lemma A.2.** *Suppose $P \xrightarrow{l(\mu)}_\sigma Q$; then:*

1. *The substitution $\sigma$ has the form $\widetilde{k} \mapsto l$, where $l \sharp P$ and $\widetilde{k} \sharp Q$ and $dom(\sigma) \subseteq ftn(P) \backslash ftn(Q)$.*

[3] See paragraph under Definition 3.4 for $E_\pi$.

2. *If $P$ is well-formed than so is $Q$.*

3. *For any permutation $\pi$, $\pi(P) \xrightarrow{\pi(l)(\mu)}_{\sigma_\pi} \pi(Q)$, where $\sigma_\pi = (\pi(\widetilde{k}) \mapsto \pi(l))$.*

4. *$ftn(Q) \subseteq ftn(P) \cup \{l\}$.*

*Proof.* By rule induction on $P \xrightarrow{l(\mu)}_{\sigma} Q$. In part (3) Lemma A.1 (1) is useful. □

**Lemma A.3.** *Suppose $P \xrightarrow{\beta} Q$, where $\beta \in \{\mathtt{co}\,k, \mathtt{ab}\,k, \mathtt{new}\,k\}$.*

1. *If $P$ is well-formed then so is $Q$.*

2. *For any permutation $\pi$, $\pi(P) \xrightarrow{\pi(\beta)} \pi(Q)$,*

*Proof.* By rule induction on $P \xrightarrow{\beta} Q$. □

**Lemma A.4.** *Let $P \xrightarrow{\mathsf{ab}\,k} P'$.*

1. *If $P \xrightarrow{\mathsf{ab}\,k} Q$ then $P' = Q$.*

2. *If $P \xrightarrow{\beta} Q$ and $\beta \sharp k$ then $P' \xrightarrow{\beta} Q'$ and $Q \xrightarrow{\mathsf{ab}\,k} Q'$, for some $Q'$.*

3. *If $P \xrightarrow{\alpha}_{\sigma} Q$ and $dom(\sigma) \sharp k$ then $P' \xrightarrow{\alpha}_{\sigma} Q'$ and $Q \xrightarrow{\mathsf{ab}\,k} Q'$, for some $Q'$.*

4. *If $P \xrightarrow{\alpha}_{\sigma} Q$ and $\sigma = (k, \widetilde{l}) \mapsto m$ then $P' \xrightarrow{\mathsf{ab}\,\widetilde{l}} Q'$ and $Q \xrightarrow{\mathsf{ab}\,m} Q'$, for some $Q'$.*

*Proof.* Each property is proved by structural induction on $P$. □

The first three cases of this lemma hold also for commit transitions, but they are not needed here.

## Appendix B. Comparing Bisimulations

Here we develop the proof of Theorem 3.9, which shows that $\approx_{\mathsf{Hasco}}$ (Definition 3.7), restricted to processes, coincides with the original bisimulation of [16], $\approx$ (Definition 2.8). The two directions of this proof are significantly different and are shown in the following subsections. The theorem follows from Theorems B.7 and B.14.

*Appendix B.1. Preliminaries*

Permutations have been defined in Section 3.2 and applied to extended configurations (Definition 3.1) and their transitions (Figure 4). Here we also apply them to configurations (Definition 2.5) but, since their transition semantics does not distinguish between transaction names from $\mathsf{inTrN}$ and $\mathsf{exTrN}$, when applied to configurations we drop the requirement (2) in Definition 3.4. We will also assume a result corresponding to Proposition 3.5 for the transitions $\mathcal{C} \xrightarrow{\zeta} \mathcal{C}'$ from Figure 3.

**Definition B.1.** We say the configuration $\langle H \bullet P \rangle$ is well-formed if

(i) $H(i) = k(\star)$ implies $k \sharp P$

(ii) $H(i) = k(a)$ implies $k \in P$. ◊

*Appendix B.2. Relating histories and extended histories*

The proof of Theorem 3.9 relies on the definition of a relation $\prec$ between configurations and extended configurations which in some sense is preserved by their respective transition semantics. This is first defined in the current subsection, and some of its properties are developed. It is then used in the following sub-section to relate $\approx_{\mathsf{Hasco}}$ and $\approx$.

**Definition B.2** (Comparing configurations)**.** Let $H$ be a history as given in Definition 2.5 and $E; K$ an extended history as given in Definition 3.1. We write $H \prec E; K$ if $dom(H) = dom(K)$ and for all $i \in dom(H)$

(a) if $H(i) = k(a)$ then $K(i) = k'(a)$ for some $k'$ such that $(k, k') \in E$

(b) if $H(i) = k(\star)$ then $K(i) = k'(\mathsf{ab})$ for any name $k'$

(c) if $H(i) = \mathsf{ab}$ then $K(i) = k'(\mathsf{ab})$ for any name $k'$

(d) if $H(i) = a$ then $K(i) = k'(\mathsf{co})$ for any name $k'$

This is extended to configurations by letting $\langle H \bullet P \rangle \prec \langle E; K \bullet P \rangle$ whenever $H \prec E; K$. ◊

**Lemma B.3.** *Suppose $H \prec E; K$. Then*

*(i)* $(\pi \cdot H) \prec \pi(E; K)$*, for any permutation* $\pi$

*(ii)* $(\sigma \cdot H) \prec \sigma(E); K$*, for any substitution* $\sigma$

*(iii) if* $\langle H \bullet P \rangle$ *and* $\langle E; K \bullet P \rangle$ *are well-formed configurations and* $k \in ftn(P)$ *then*

- $H \setminus_{\mathsf{co}} k \prec (E; K) \setminus_{\mathsf{co}} k$
- $H \setminus_{\mathsf{ab}} k \prec (E; K) \setminus_{\mathsf{ab}} k$

*Proof.* Parts (i) and (ii) can be established by straightforward calculations. In part (iii) the cases for co and ab are identical. We consider briefly the former; for convenience let $E; K'$ represent $(E; K) \setminus_{\mathsf{co}} k$.

As an example suppose $H(i) = l(a)$, in which case $K(i) = k'(a)$ for some $(l, k') \in E$. If $l$ coincides with $k$ then $(H \setminus_{\mathsf{co}} k)(i) = a$. But by definition $K'(i) = k'(\mathsf{co})$, as required, since $(k, k') \in E$. On the other hand if $l$ is different from $k$ then $(H \setminus_{\mathsf{co}} k)(i)$ remains as $l(a)$. By the well-formedness of $\langle H \bullet P \rangle$ $l$ must be in $ftn(P)$. We are assuming that $k$ is also in $ftn(P)$, and therefore by the well-formedness of $\langle E; K \bullet P \rangle$ it follows that $(k, k') \notin E$ – see condition (iv) of Definition 3.1. Consequently, as required $K'(i)$ also remains the same, as $k'(\mathsf{co})$.

As another example suppose $H(i) = l(\star)$, in which case $K(i)$ has the form $k'(\mathsf{ab})$, and $K'(i)$ remains the same. By well-formedness of $\langle H \bullet P \rangle$ we know that $l$ must be different than $k$, and therefore $(H \setminus_{\mathsf{co}} k)(i)$ also remains as $l(\star)$.

The remaining possibilities for $H(i)$ are trivial. □

**Lemma B.4.** *Suppose $\mathcal{C} \prec \mathcal{D}$.*

*(i)* $\mathcal{D} \xrightarrow{k(a)}\!\!\!\!\rightarrow \mathcal{D}'$ *implies challenger move* $\mathcal{C} \xrightarrow{k} \mathcal{C}'$ *for some* $\mathcal{C}'$ *such that* $\mathcal{C}' \prec \mathcal{D}'$*.*

*(ii) Conversely,* $\mathcal{C} \xrightarrow{k} \mathcal{C}'$*, where* $k \in \mathsf{exTrN}$ *and* $k \sharp \mathcal{D}$*, implies* $\mathcal{D} \xrightarrow{k(a)}\!\!\!\!\rightarrow \mathcal{D}'$ *for some* $a, \mathcal{D}'$*, such that* $\mathcal{C}' \prec \mathcal{D}'$*.*

*(iii)* $\mathcal{D} \xrightarrow{\tau}\!\!\!\!\rightarrow \mathcal{D}'$ *implies* $\mathcal{C} \xrightarrow{\tau} \mathcal{C}'$ *for some* $\mathcal{C}'$ *such that* $\mathcal{C}' \prec \mathcal{D}'$*.*

*(iv) Conversely,* $\mathcal{C} \xrightarrow{\tau} \mathcal{C}'$*, where* $eftn(\mathcal{C}') \subseteq eftn(\mathcal{C})$*, implies* $\mathcal{D} \xrightarrow{\tau}\!\!\!\!\rightarrow \mathcal{D}'$ *such that* $\mathcal{C}' \prec \mathcal{D}'$*.*

*Proof.* By a case analysis of the rules in Figure 3 and Figure 4, using the relevant part of Lemma B.3. We look at two examples.

Suppose $\mathcal{D} = \langle E; K \bullet P \rangle \xrightarrow{\tau}\!\!\!\!\rightarrow \langle \sigma(E); K \bullet Q \rangle = \mathcal{D}'$ because $P \xrightarrow{k(\tau)}_{\sigma} Q$; an instance of (iii). It follows from the rule LTS$k(\tau)$ in Figure 3 that $\mathcal{C} = \langle H \bullet P \rangle \xrightarrow{\tau} \langle \sigma(H) \bullet Q \rangle = \mathcal{C}'$, and by part (ii) of Lemma B.3 we have the required $\mathcal{C}' \prec \mathcal{D}'$.

Suppose $\mathcal{C} = \langle H \bullet P \rangle \xrightarrow{\tau} \langle H \setminus_{\mathsf{ab}} k \bullet Q \rangle = \mathcal{C}'$ because $P \xrightarrow{\mathsf{ab}\,k} Q$, an instance of (iv). Using the rule eLTSab in Figure 4 we have $\mathcal{D} = \langle E; K \bullet P \rangle \xrightarrow{\tau}\!\!\!\!\rightarrow \langle (E; K) \setminus_{\mathsf{ab}} k \bullet Q \rangle = \mathcal{D}'$. By examining the Transactional Reconfiguration Transition rules in Figure 1 one can see that $k$ must appear in $P$. So this time part (iii) of Lemma B.3 gives the required $\mathcal{C}' \prec \mathcal{D}'$. □

*Appendix B.3.* $(\approx_{\mathsf{Hasco}})$ *implies* $(\approx)$

For the forward direction of Theorem 3.9 we need to strengthen the consistency of extended histories.

**Definition B.5** (Very consistent extended histories)**.** Two extended histories $\Delta_1 = E_1; H_1$ and $\Delta_2 = E_2; H_2$ are *very consistent*, written $\Delta_1$ vcons $\Delta_2$, if

(i) $\Delta_1$ $\texttt{Hasco}$ $\Delta_2$

(ii) for all $i \in I$, $H_1{}^{\texttt{trn}}(i) = H_2{}^{\texttt{trn}}(i)$

(iii) for all $i \in I$, if $H_1(i) = k(a)$ and $H_2(i) = l(b)$ then $a = b$.

This is generalised to extended configurations in the standard manner. $\diamond$

**Proposition B.6.** $(\approx_{\mathsf{Hasco}} \cap \mathsf{vcons})$ *is a* $\texttt{Hasco}$ *bisimulation.*

*Proof.* Straightforward, since the transfer property used in Definition 3.7 demands that both extended configurations use the same fresh transaction names and the same action names. $\square$

**Theorem B.7.** $P \approx_{\mathsf{Hasco}} Q$ *implies* $P \approx Q$.

*Proof.* We prove a more general result. Let $\mathcal{R}$ be the relation over configurations defined by $\mathcal{C}_1 \;\mathcal{R}\; \mathcal{C}_2$ whenever

(i) $\mathcal{C}_1$ and $\mathcal{C}_2$ are well-formed

(ii) $\mathcal{C}_1$ and $\mathcal{C}_2$ are consistent

(iii) $\mathcal{C}_1 \prec \cdot (\approx_{\mathsf{Hasco}} \cap \mathsf{vcons}) \cdot \succ \mathcal{C}_2$

We show that $\mathcal{R}$ satisfies the requirements of a bisimulation, in Definition 2.8, from which the result follows.

Suppose $\mathcal{C}_1 \;\mathcal{R}\; \mathcal{C}_2$ and $\mathcal{C}_1 \xrightarrow{\zeta} \mathcal{C}_1'$ is a challenger move; we must find $\mathcal{C}_2 \stackrel{\zeta}{\Rightarrow} \mathcal{C}_2'$ such that $\mathcal{C}_1' \;\mathcal{R}\; \mathcal{C}_2'$.

We look at the case where $\zeta$ is $k$; the case for $\tau$ is similar and omitted. In order to apply Lemma B.4 (ii) let $\pi$ be a renaming which maps $k$ to some element in $\mathsf{exTrN}$ but is invariant over $\mathit{ftn}(\mathcal{C}_1)$; $\pi(k)$ should also be chosen so as to be fresh from $\mathcal{C}_1$, $\mathcal{C}_2$, $\mathcal{D}_1$, where $\mathcal{D}_1$ is such that $\mathcal{C}_1 \prec \mathcal{D}_1 \;(\approx_{\mathsf{Hasco}} \cap \mathsf{vcons}) \cdot \succ \mathcal{C}_2$. It will also be convenient later to ensure $\pi$ is also invariant over $\mathit{ftn}(\mathcal{C}_2)$. Then using a variation of Proposition 3.5 for the semantics in Figure 3 we have $\mathcal{C}_1 \xrightarrow{\pi(k)} \pi(\mathcal{C}_1')$.

Now Lemma B.4 (ii) can be applied, and using Prop. B.6 and Lemma B.4 (i), we can find a transition $\mathcal{C}_2 \stackrel{\pi(k)}{\Longrightarrow} \mathcal{C}_2''$ such that

$$\pi(\mathcal{C}_1') \prec \cdot (\approx_{\mathsf{Hasco}} \cap \mathsf{vcons}) \cdot \succ \mathcal{C}_2''$$

Using Lemma B.3 and Proposition 3.8 we have that $\mathcal{C}_1' \prec \cdot (\approx_{\mathsf{Hasco}} \cap \mathsf{vcons}) \cdot \succ \pi^{-1}(\mathcal{C}_2'')$ and since Proposition 3.5 for the semantics in Figure 3 ensures that $\mathcal{C}_2 \stackrel{k}{\Rightarrow} \pi^{-1}(\mathcal{C}_2'')$, we have found a $\mathcal{C}_2'$ such that $\mathcal{C}_1' \prec \cdot (\approx_{\mathsf{Hasco}} \cap \mathsf{vcons}) \cdot \succ \mathcal{C}_2'$, namely $\pi^{-1}(\mathcal{C}_2'')$.

It is straightforward to show that the transitions in Figure 3 also preserve well-formedness and therefore $\mathcal{C}_1'$ and $\mathcal{C}_2'$ are also well-formed (condition (i)).

So in order to establish that $\mathcal{C}_1' \;\mathcal{R}\; \mathcal{C}_2'$ it is sufficient to prove that they are consistent (condition (ii)). To this end, let $\mathcal{C}_1$, $\mathcal{C}_2$ take the forms $\langle H_1 \bullet P_1 \rangle$, $\langle H_2 \bullet P_2 \rangle$, respectively, and $\mathcal{C}_1'$, $\mathcal{C}_2'$ the corresponding forms $\langle H_1' \bullet P_1' \rangle$, $\langle H_2' \bullet$ $P_2'\rangle$. Suppose that for some index $i$, $H_1'(i) = a$ for some action $a$; we prove $H_2'(i) = a$. The proof of the converse property is symmetric, and therefore omitted.

First suppose $H_1(i) = a$. Then since $\mathcal{C}_1$ and $\mathcal{C}_2$ are consistent we must have that $H_2(i) = a$, and therefore, since $\mathcal{C}_2 \overset{k}{\twoheadrightarrow} \mathcal{C}_2'$, $H_2'(i) = a$. Otherwise $H_1(i)$ must have the form $k_1(a)$ for some (unimportant) $k_1$. At this point there are two steps in the argument. First, using

$$\mathcal{C}_1' \prec \cdot \; \mathsf{vcons} \; \cdot \succ \mathcal{C}_2'$$

it follows that $H_2'(i)$ must be equal to $b$, for some action $b$. Moreover, because $\mathcal{C}_2 \overset{k}{\twoheadrightarrow} \mathcal{C}_2''$, $H_2(i)$ must be of the form $k_2(b)$ for some (unimportant) $k_2$. The second step now uses

$$\mathcal{C}_1 \prec \cdot \; \mathsf{vcons} \; \cdot \succ \mathcal{C}_2$$

to conclude that $b$ must coincide with $a$. □

*Appendix B.4. (Un-)committable history indices*

To prove the backward direction of Theorem 3.9 we need to deal with uncommittable history indices; that is, indices recording actions $k(a)$ that do not appear commit in all subsequent configurations. The transfer condition of $\approx$ allows these actions to be matched with any other uncommittable action $k(b)$. However, the transfer condition of $\approx_{\mathsf{Hasco}}$ requires that $a = b$.

To bridge this difference between the two bisimulations we define the notion of committable and uncommittable positions in the history, which is an adaptation of committable actions from [16, Definition 4.3].

**Definition B.8** ((Un-)Committable History Index)**.** Suppose $\mathcal{C} = \langle H \bullet P\rangle$ is a configuration and $i$ and index such that $H(i) = k(a)$, for some $k, a$. We will call $i$

- *committable* in $\mathcal{C}$, if there exists sequence of actions $\zeta_1, \ldots, \zeta_n$ and configuration $\langle H' \bullet P'\rangle$ such that

$$\mathcal{C} \xrightarrow{\zeta_1} \ldots \xrightarrow{\zeta_n} \langle H' \bullet P'\rangle \qquad \text{and} \qquad H'(i) = a$$

- and *uncommittable* in $\mathcal{C}$ otherwise.

◊

Obviously, if an index is uncommittable in a configuration, in any future configuration it either appears uncommittable or aborted.

**Lemma B.9.** *Let $i$ be uncommittable in $\mathcal{C}$ and $\mathcal{C} \xrightarrow{\zeta} \mathcal{C}' = \langle H' \bullet P'\rangle$. Then either $i$ is uncommittable in $\mathcal{C}'$, or $H'(i) = \mathsf{ab}$.*

The following lemma states that aborting the transaction names at uncommittable indices has no effect in the observable behaviour of configurations.

**Lemma B.10.** *Let $i$ be uncommittable in $\mathcal{C} = \langle H \bullet P \rangle$. Then there exists $\mathcal{C}' = \langle H' \bullet P' \rangle$ such that $\mathcal{C} \xrightarrow{\tau} \mathcal{C}'$ and $H'(i) = \mathsf{ab}$ and $\mathcal{C} \approx \mathcal{C}'$.*

*Proof.* By the definition of uncommittability (Definition B.8), we get $H(i) = k(a)$, for some $k, a$. Because aborts are non-deterministic transitions (Figure 1), we can deduce $P \xrightarrow{\mathsf{ab}\,k} P'$, for some $P'$. Thus, we can apply rule LTS$\mathsf{ab}$ from Figure 3, and get $\mathcal{C} \xrightarrow{\tau} \mathcal{C}' = \langle H' \bullet P' \rangle$ for $H' = H \setminus_{\mathsf{ab}} k$. Note that $H(i) = \mathsf{ab}$.

It remains to show $C \approx C'$. We do this by showing that the following relation is a weak bisimulation according to Definition 2.8:

$$[R_1]\frac{\mathcal{C} \;\mathsf{Id}\; \mathcal{C}'}{\mathcal{C} \;\mathcal{R}\; \mathcal{C}'} \qquad [R_2]\frac{\mathcal{C} \;\mathcal{R}\; \langle H' \bullet P' \rangle \qquad H'(i) = \mathsf{ab} \qquad k \sharp H', P'}{\mathcal{C} \;\mathcal{R}\; \langle H'[i \mapsto k(\star)] \bullet P' \rangle}$$

$$[R_3]\frac{\begin{array}{c}\langle H \bullet P \rangle \xrightarrow{\tau} \langle H' \bullet P' \rangle \\ i \text{ is uncommittable in } \langle H \bullet P \rangle \qquad H(i) = k(a) \qquad H'(i) = \mathsf{ab}\end{array}}{\langle H \bullet P \rangle \;\mathcal{R}\; \langle H' \bullet P' \rangle}$$

where $\mathsf{Id}$ is the identity relation on configurations.

We consider $\mathcal{C} \;\mathcal{R}\; \mathcal{C}'$ and prove that the conditions of the definition of bisimulation (Definition 2.8) are satisfied.

We proceed by induction on the derivation of $\mathcal{C} \;\mathcal{R}\; \mathcal{C}'$. The proof in the case of $[R_1]$ is trivial, and in $[R_2]$ follows by the induction hypothesis and a simple property $\langle H' \bullet P' \rangle \approx \langle H'[i \mapsto k(\star)] \bullet P' \rangle$ when $H'(i) = \mathsf{ab}$ and $k \sharp H', P'$.

*Case* $[R_3]$. In this case, in addition to $\mathcal{C} \;\mathcal{R}\; \mathcal{C}'$, we also have:

$$\begin{array}{c}\mathcal{C} = \langle H \bullet P \rangle \xrightarrow{\tau} \mathcal{C}' = \langle H' \bullet P' \rangle \\ i \text{ is uncommittable in } \langle H \bullet P \rangle \qquad H(i) = k(a) \qquad H'(i) = \mathsf{ab}\end{array}$$

By case analysis on the derivation of $\mathcal{C} \xrightarrow{\tau} \mathcal{C}'$ we get $H' = H \setminus_{\mathsf{ab}} k$ and $P \xrightarrow{\mathsf{ab}\,k} P'$ and the transition $\mathcal{C} \xrightarrow{\tau} \mathcal{C}'$ is an abort transition produced by rule LTS$\mathsf{ab}$ of Figure 3. We examine each of the conditions of Definition 2.8:

- $\mathcal{C}$ and $\mathcal{C}'$ are consistent because the histories $H$ and $H \setminus_{\mathsf{ab}} k$ have the same committed indices.
- If $\mathcal{C}' \xrightarrow{\zeta} \mathcal{C}_1'$ then $\mathcal{C} \xrightarrow{\tau} \mathcal{C}' \xrightarrow{\zeta} \mathcal{C}_1'$ and $\mathcal{C}_1' \;\mathcal{R}\; \mathcal{C}_1'$ by $[R_1]$.
- If $\mathcal{C} \xrightarrow{\zeta} \mathcal{C}_1$ is a challenger transition with $\zeta \sharp \mathcal{C}'$ then we have

$$\begin{array}{ccc}\mathcal{C} = \langle H \bullet P \rangle & \xrightarrow{\quad\zeta\quad} & \mathcal{C}_1 \\ \Big\downarrow{\scriptstyle \tau} & & \\ \mathcal{C}' = \langle H \setminus_{\mathsf{ab}} k \bullet P' \rangle & & \end{array}$$

and we need to show that for some $\mathcal{C}_1'$:

$$\begin{array}{ccc} & & \mathcal{C}_1 \\ & & \Big|\ \mathcal{R} \\ \mathcal{C}' & \stackrel{\zeta}{\Longrightarrow} & \mathcal{C}_1' \end{array}$$

We proceed by cases on the rule of Figure 3 that produced the $\zeta$-transition:

- Cases LTS$\tau$, LTS`new` and LTS`ab` follow from Lemma A.4. Here we show only the case for LTS`ab`, where $\zeta = \tau$, $C_1 = \langle H \setminus_{\mathsf{ab}} k' \bullet Q_1\rangle$ and $P \xrightarrow{\mathsf{ab}\, k'} Q_1$.
  If $k' = k$ then, by Lemma A.4 (1), $\mathcal{C}_1 = \mathcal{C}'$ and we take $\mathcal{C}_1' = \mathcal{C}$ and derive $\mathcal{C}' \stackrel{\tau}{\Rightarrow} \mathcal{C}_1'$ and $\mathcal{C}_1 \,\mathcal{R}\, \mathcal{C}_1'$. If $k' \neq k$ then, by Lemma A.4 (2), we get $P' \xrightarrow{\mathsf{ab}\, k'} Q_1'$ and $Q_1 \xrightarrow{\mathsf{ab}\, k} Q_1'$, for some $Q_1'$. Therefore, by LTS`ab`, we deduce $\mathcal{C}' \xrightarrow{\tau} \mathcal{C}_1' = \langle H \setminus_{\mathsf{ab}} k \setminus_{\mathsf{ab}} k' \bullet Q_1'\rangle$ and, because

  $$H \setminus_{\mathsf{ab}} k \setminus_{\mathsf{ab}} k' = H \setminus_{\mathsf{ab}} k' \setminus_{\mathsf{ab}} k$$

  also $\mathcal{C}_1 \xrightarrow{\tau} \mathcal{C}_1'$. Thus, by Lemma B.9 and $[R_3]$, we derive $\mathcal{C}_1 \,\mathcal{R}\, \mathcal{C}_1'$.
- In the case of LTS`co` we have $\zeta = \tau$, $\mathcal{C}_1 = \langle H \setminus_{\mathsf{co}} k' \bullet Q_1\rangle$ and $P \xrightarrow{\mathsf{co}\, k'} Q_1$. Because index $i$ is uncommittable in $\mathcal{C}$, it must be that $k \neq k'$. Therefore this case is proved as above from Lemma A.4 (2), Lemma B.9 and the equation

  $$H \setminus_{\mathsf{ab}} k \setminus_{\mathsf{co}} k' = H \setminus_{\mathsf{co}} k' \setminus_{\mathsf{ab}} k$$

- In case LTS$k(\tau)$ we have $\zeta = \tau$, $\mathcal{C}_1 = \langle \sigma(H) \bullet Q_1\rangle$, $m \sharp H$ and $P \xrightarrow{m(\tau)}_{\sigma} Q_1$.

  If $dom(\sigma) \sharp k$ then Lemma A.4 (3) and equation $\sigma(H) \setminus_{\mathsf{ab}} k = \sigma(H \setminus_{\mathsf{ab}} k)$ give us $\mathcal{C}' \xrightarrow{\tau} \mathcal{C}_1' = \langle \sigma(H \setminus_{\mathsf{ab}} k) \bullet Q_1'\rangle$, for some $Q_1'$, and the abort transition $\mathcal{C}_1 \xrightarrow{\tau} \mathcal{C}_1'$. Thus, from Lemma B.9 and $[R_3]$, $\mathcal{C}_1 \,\mathcal{R}\, \mathcal{C}_1'$.

  If $k \in dom(\sigma)$ then $\sigma = (k, \widetilde{l} \mapsto m)$, for some $\widetilde{l}$. By Lemma A.4 (4), $P' \xrightarrow{\mathsf{ab}\, \widetilde{l}} Q_1'$, for some $Q_1'$, and $Q_1 \xrightarrow{\mathsf{ab}\, m} Q_1'$. Therefore, by LTS`ab`, $\mathcal{C}' \xrightarrow{\tau} \mathcal{C}_1' = \langle H \setminus_{\mathsf{ab}} k \setminus_{\mathsf{ab}} \widetilde{l} \bullet Q_1'\rangle$ and $\mathcal{C}_1 \xrightarrow{\tau} \mathcal{C}_1'' = \langle \sigma(H) \setminus_{\mathsf{ab}} m \bullet Q_1'\rangle$. Moreover, $\mathcal{C}_1' = \mathcal{C}_1''$ because of the equation $H \setminus_{\mathsf{ab}} k \setminus_{\mathsf{ab}} \widetilde{l} = \sigma(H) \setminus_{\mathsf{ab}} m$. It remains to show $\mathcal{C}_1 \,\mathcal{R}\, \mathcal{C}_1'$ which follows from $[R_3]$ and Lemma B.9, because $i$ is uncommittable.
- In Case LTS$k(a)$ we have $\zeta = m$, $\mathcal{C}_1 = \langle \sigma(H), (j \mapsto m(a)) \bullet Q_1\rangle$, $m \sharp H$ and $P \xrightarrow{m(a)}_{\sigma} Q_1$.

  If $dom(\sigma) \sharp k$ then the proof is the same as in the case of LTS$k(\tau)$.

If $k \in dom(\sigma)$ then $\sigma = (k \mapsto m)$. By Lemma A.4 (4), $P' \xrightarrow{\mathsf{ab}\,\varepsilon} P'$ and $Q_1 \xrightarrow{\mathsf{ab}\,m} P'$. By LTS$\star$, $\mathcal{C}' \overset{m}{\Rightarrow} \mathcal{C}_1' = \langle (H \setminus_{\mathsf{ab}} k), (j \mapsto m(\star)) \bullet P' \rangle$. By LTSab, $\mathcal{C}_1 \xrightarrow{\tau} \mathcal{C}_1'' = \langle (\sigma(H) \setminus_{\mathsf{ab}} m), (j \mapsto \mathsf{ab}) \bullet P' \rangle$. It remains to show $\mathcal{C}_1 \mathrel{\mathcal{R}} \mathcal{C}_1'$. From $[R_3]$ and Lemma B.9, $\mathcal{C}_1 \mathrel{\mathcal{R}} \mathcal{C}_1''$. By the equation

$$(\sigma(H) \setminus_{\mathsf{ab}} m) = H \setminus_{\mathsf{ab}} k$$

we obtain $\mathcal{C}_1'' = \langle (H \setminus_{\mathsf{ab}} k), (j \mapsto \mathsf{ab}) \bullet P' \rangle$, and by $[R_2]$ we obtain $\mathcal{C}_1 \mathrel{\mathcal{R}} \mathcal{C}_1'$. □

*Appendix B.5.* ($\approx$) *implies* ($\approx_{\mathsf{Hasco}}$)

To prove the forward direction of Theorem 3.9 in Appendix B.3 we have strengthened the consistency requirement of $\approx_{\mathsf{Hasco}}$ by defining *very consistent* extended histories (vcons). Similarly here, to prove the backward direction of the theorem we need to strengthen the consistency requirement of $\approx$, by defining *action consistent* histories.

**Definition B.11** (Action Consistent Histories)**.** Histories $H_1$ and $H_2$ are action-consistent when for all $i \in dom(H_1) \cup dom(H_2)$, and all $k$, $l$, $a$ and $b$:

$$H_1(i) = k(a) \text{ and } H_2(i) = l(b) \text{ implies } a = b$$

Configurations $C_1$ and $C_2$ are action-consistent, when their histories are. We write acons for the largest relation over action-consistent configurations. ◇

**Lemma B.12.** *Let* $\mathcal{C}_1 \xrightarrow{\tau} \mathcal{C}_2$. *Then* $\mathcal{C}_1$ acons $\mathcal{C}_2$.

*Proof.* By inspection of the rules of Figure 1 and observing that $\tau$ transitions either commit, abort, or rename history indices. □

The following proposition shows that the action-consistent subset of weak bisimilarity is also a weak bisimulation.

**Proposition B.13.** *The relation* $\mathcal{R} = (\approx \cap \mathsf{acons})$ *is a weak bisimulation according to Definition 2.8.*

*Proof.* Let $\mathcal{C}_1 \mathrel{\mathcal{R}} \mathcal{C}_2$. Then $\mathcal{C}_1$ and $\mathcal{C}_2$ satisfy the first condition of bisimulation (Definition 2.8).

We need to prove the transfer condition: if $\zeta \in \{\tau, k\}$, and $\mathcal{C}_1 \xrightarrow{\zeta} \mathcal{C}_1'$ is a challenger transition and $\zeta \sharp \mathcal{C}_2$, then there exists $\mathcal{C}_2'$ such that $\mathcal{C}_2 \overset{\zeta}{\Rightarrow} \mathcal{C}_2'$ and $\mathcal{C}_1' \mathrel{\mathcal{R}} \mathcal{C}_2'$.

We assume $\zeta \in \{\tau, k\}$, and challenger transition $\mathcal{C}_1 \xrightarrow{\zeta} \mathcal{C}_1'$ where $\zeta \sharp \mathcal{C}_2$. Because $\mathcal{C}_1 \approx \mathcal{C}_2$ we get $\mathcal{C}_2'$ such that $\mathcal{C}_2 \overset{\zeta}{\Rightarrow} \mathcal{C}_2'$ and $\mathcal{C}_1' \approx \mathcal{C}_2'$.

In the case where $\zeta = \tau$ we have from Lemma B.12: $\mathcal{C}_1'$ acons $\mathcal{C}_1$ acons $\mathcal{C}_2$ acons $\mathcal{C}_2'$. By transitivity, $\mathcal{C}_1'$ acons $\mathcal{C}_2'$ and thus $\mathcal{C}_1' \mathrel{\mathcal{R}} \mathcal{C}_2'$.

In the case where $\zeta = k$ the transition $\mathcal{C}_1 \xrightarrow{\zeta} \mathcal{C}_1'$ is derived by rule LTS$k(a)$ of Figure 3. Thus $\mathcal{C}_1 = \langle H_1 \bullet P_1 \rangle$ and $\mathcal{C}_1' = \langle \sigma_1(H_1), (i \mapsto k(a)) \bullet Q_1 \rangle$. Moreover, the transition $\mathcal{C}_2 \overset{\zeta}{\Rightarrow} \mathcal{C}_2'$ is derived by rule LTS$k(a)$ or LTS$\star$ and therefore, $\mathcal{C}_2 = \langle H_2 \bullet P_2 \rangle$ and either $\mathcal{C}_2' = \langle \sigma_2(H_2), (i \mapsto k(b)) \bullet Q_2 \rangle$ or $\mathcal{C}_2' = \langle H_2, (i \mapsto k(\star)) \bullet P_2 \rangle$.

We distinguish two cases, whether $i$ is committable in $\mathcal{C}_1$:

- Index $i$ is committable in $\mathcal{C}_1$: In this case $\mathcal{C}_2'$ must have been produced by the LTS$k(a)$ transition and $\mathcal{C}_2' = \langle \sigma_2(H_2), (i \mapsto k(b)) \bullet Q_2 \rangle$, and moreover, $a = b$. If this is not the case then $\mathcal{C}_1$ and $\mathcal{C}_1'$ will be distinguished by weak bisimulation after $i$ is committed in a subsequent configuration.

- Index $i$ is uncommittable in $\mathcal{C}_1$: Because $\mathcal{C}_1 \approx \mathcal{C}_2$, $i$ is also uncommittable in $\mathcal{C}_2$. By Lemma B.10, there exists $\mathcal{C}_2'' = \langle H_2'', (i \mapsto \mathsf{ab}) \bullet Q_2'' \rangle$ such that $\mathcal{C}_2' \xrightarrow{\tau} \mathcal{C}_2''$ and $\mathcal{C}_2' \approx \mathcal{C}_2''$.

  Because $H_1$ $\mathsf{acons}$ $H_2$ and Lemma B.12 we have

  $$\sigma_1(H_1) \;\mathsf{acons}\; \sigma_2(H_2) \;\mathsf{acons}\; H_2 \;\mathsf{acons}\; H_2''$$

  Therefore, $\sigma_1(H_1), (i \mapsto k(a))$ $\mathsf{acons}$ $H_2'', (i \mapsto \mathsf{ab})$ and $\mathcal{C}_1'$ $\mathsf{acons}$ $\mathcal{C}_2''$. Hence, $\mathcal{C}_2 \xrightarrow{k} \mathcal{C}_2''$ and $\mathcal{C}_1' \;\mathcal{R}\; \mathcal{C}_2''$, as needed by the proof. □

**Theorem B.14.** *$P \approx Q$ implies $P \approx_{\mathsf{Hasco}} Q$.*

*Proof.* Let $\mathcal{R}$ be the relation over extended configurations defined by $\mathcal{C}_1 \;\mathcal{R}\; \mathcal{C}_2$ whenever

(i) $\mathcal{C}_1$ and $\mathcal{C}_2$ are extended configurations

(ii) $\mathcal{C}_1 \,\mathtt{Hasco}\, \mathcal{C}_2$

(iii) $\mathcal{C}_1 \succ \cdot (\approx \cap \;\mathsf{acons}) \cdot \prec \mathcal{C}_2$ and $\mathcal{C}_1$ $\mathsf{vcons}$ $\mathcal{C}_2$

We show that $\mathcal{R}$ satisfies the requirements of a bisimulation, in Definition 3.7, from which the result follows.

Suppose $\mathcal{C}_1 \;\mathcal{R}\; \mathcal{C}_2$ and $\mathcal{C}_1 \overset{\zeta}{\twoheadrightarrow} \mathcal{C}_1'$; we must find $\mathcal{C}_2 \overset{\zeta}{\Rrightarrow} \mathcal{C}_2'$ such that $\mathcal{C}_1' \;\mathcal{R}\; \mathcal{C}_2'$.

Consider $\mathcal{D}_1$,$\mathcal{D}_2$ is such that $\mathcal{C}_1 \succ \mathcal{D}_1$ $(\approx \cap \;\mathsf{acons})$ $\mathcal{D}_2 \prec \mathcal{C}_2$. By Lemma B.4 (i) and (iii) we get the challenger move $\mathcal{D}_1 \xrightarrow{\zeta} \mathcal{D}_1'$ such that $\mathcal{C}_1' \prec \mathcal{D}_1'$. By Proposition B.13, $\mathcal{D}_2 \overset{\zeta}{\Rightarrow} \mathcal{D}_2'$ such that $\mathcal{D}_1'$ $(\approx \cap \;\mathsf{acons})$ $\mathcal{D}_2'$.

Before applying Lemma B.4 (ii) and (iv) we need to ensure that $\mathit{eftn}(\mathcal{D}_2') \subseteq \mathit{eftn}(\mathcal{D}_2)$. We take permutation $\pi$ which maps all the names in $\mathit{eftn}(\mathcal{D}_2') \setminus \mathit{eftn}(\mathcal{D}_2)$ into fresh names in $\mathsf{inTrN}$. Using a variation of Proposition 3.5 for the semantics in Figure 3 we have $\mathcal{D}_2 \overset{\zeta}{\Rightarrow} \pi(\mathcal{D}_2')$. Now Lemma B.4 (ii) and (iv) can be applied, and get $\mathcal{C}_2 \overset{\zeta}{\Rightarrow} \pi(\mathcal{C}_2')$ such that $\mathcal{C}_2' \prec \pi(D_2')$. By equivariance of $\approx$ [16, Lemma 3.9], $\mathcal{C}_1' \succ \cdot (\approx \cap \;\mathsf{acons}) \cdot \prec \mathcal{C}_2'$. Moreover, $\mathcal{C}_1'$ $\mathsf{vcons}$ $\mathcal{C}_2'$ follows by the rules of Figure 4. Condition (i) follows by Lemma 3.3 (2).

Thus, to establish $\mathcal{C}_1' \;\mathcal{R}\; \mathcal{C}_2'$, it is sufficient to prove $\mathcal{C}_1 \,\mathtt{Hasco}\, \mathcal{C}_2'$. Because of symmetry we only need to show $\mathtt{Hasco}(\mathcal{C}_1) \subseteq \mathtt{Hasco}(\mathcal{C}_2)$. Let $\mathcal{C}_1$, $\mathcal{C}_2$ take the forms $\langle E_1; H_1 \bullet P_1 \rangle$, $\langle E; H_2 \bullet P_2 \rangle$, respectively, and $\mathcal{C}_1'$, $\mathcal{C}_2'$ the corresponding forms $\langle E_1'; H_1' \bullet P_1' \rangle$, $\langle E_2'; H_2' \bullet P_2' \rangle$. Suppose that for some $k$, $k \in \texttt{Hasco}(H_1')$.

First consider $k \in \texttt{Hasco}(H_1)$. Then because $\mathcal{C}_1 \,\texttt{Hasco}\, \mathcal{C}_2$ we have $k \in \texttt{Hasco}(\mathcal{C}_2)$, and therefore, by Lemma 3.3 (3), $k \in \texttt{Hasco}(\mathcal{C}_2')$.

Otherwise there exists index $i$ such that $H_1'(i) = k(\texttt{co})$ and $H_1(i) = k(a)$, for some $a$. Therefore, by $\mathcal{C}_1$ $\textsf{vcons}$ $\mathcal{C}_2$, $H_2(i) = k(b)$. Moreover, by definitions of $\prec$ and $\approx$, $H_2'(i) = l(\texttt{co})$. Because extended transitions do not rename the histories, it must be $k = l$. This implies $k \in \texttt{Hasco}(\mathcal{C}_2')$, which completes the proof. $\square$